%% file: tac20_bha_r1.tex
\documentclass[journal,onecolumn,draftclsnofoot]{IEEEtran}
\usepackage{float}
\usepackage{amsmath}
\usepackage{amsthm}
\usepackage{amssymb}
\usepackage{graphicx}
\usepackage{color}
\usepackage{comment}
\usepackage[arxiv]{optional} 

\newcommand\numberthis{\addtocounter{equation}{1}\tag{\theequation}}

\usepackage{mathtools}

\allowdisplaybreaks[1]

\makeatletter

\floatstyle{ruled}
\newfloat{algorithm}{tbp}{loa}
\providecommand{\algorithmname}{Algorithm}
\floatname{algorithm}{\protect\algorithmname}

\theoremstyle{plain}
\newtheorem{thm}{\protect\theoremname}
\theoremstyle{definition}
\newtheorem{defn}{\protect\definitionname}
\theoremstyle{plain}
\newtheorem{lem}{\protect\lemmaname}
\newtheorem{algo}{Fixed-Point Equation}

\IEEEoverridecommandlockouts
\usepackage[caption=false,font=footnotesize]{subfig}
\usepackage{cite}

\newcommand{\tx}{\tilde{x}}
\newcommand{\tV}{\tilde{V}}
\newcommand{\mN}{\mathcal{N}}
\newcommand{\mK}{\mathcal{K}}
\newcommand{\mV}{\mathcal{V}}
\newcommand{\mX}{\mathcal{X}}
\newcommand{\mA}{\mathcal{A}}
\newcommand{\mP}{\mathcal{P}}
\renewcommand{\Pr}{\mathbb{P}}

\makeatletter
\renewcommand{\fnum@algorithm}{}
\makeatother

\optv{2col}{\title{Informational Cascades with Non-myopic Agents}}

\optv{arxiv}{\title{Do Informational Cascades Happen with Non-myopic Agents?}}

\author{Ilai Bistritz, Nasimeh Heydaribeni and Achilleas Anastasopoulos
\thanks{This work was supported in part by NSF Grant ECCS-1608361.}
\thanks{Some of the results of this paper were presented in  \cite{bistritz2018characterizing,heydaribeni2019informational}.}

\thanks{Ilai Bistritz is with the Department of Electrical Engineering, Stanford University, Stanford, CA, USA
      {\tt\small bistritz@stanford.edu}}%
\thanks{Nasimeh Heydaribeni and Achilleas Anastasopoulos are with the Department of Electrical Engineering and
Computer Science, University of Michigan, Ann Arbor, MI, USA
        {\tt\small \{heydari,anastas\}@umich.edu}}%
}

\makeatother

\providecommand{\definitionname}{Definition}
\providecommand{\lemmaname}{Lemma}
\providecommand{\theoremname}{Theorem}

\DeclareMathOperator*{\argmax}{arg\,max}
\begin{document}
\maketitle
\begin{abstract}
We consider an environment where players need to decide whether
to buy a certain product (or adopt a technology) or not. The product is either good or bad, but its true value is unknown to the players.
Instead, each player has her own private information on its quality.
Each player can observe the previous actions of other
players and estimate the quality of the product.
A classic result in the literature shows that in similar settings informational cascades occur where learning stops for the whole network and players repeat the actions of their predecessors.
In contrast to this literature, in this work, players get more than one opportunity to act.
In each turn, a player is chosen uniformly at random from all players and can decide to buy the product and leave the market or wait. Her utility is the total expected
discounted reward, and thus myopic strategies may not constitute equilibria.
We provide a characterization of perfect Bayesian equilibria (PBE)
with forward-looking strategies through a fixed-point equation of
dimensionality that grows only quadratically with the number of players.
Using this tractable fixed-point equation, we show the existence of a PBE and characterize PBE with threshold strategies.
Based on this characterization we study informational cascades in two regimes.
First, we show that for a discount factor $\delta$ strictly smaller than one, informational cascades happen with high probability as the number of players $N$ increases.
Furthermore, only a small portion of the total information in the system is revealed before a cascade occurs.
Secondly, and more surprisingly, we show that for a fixed $N$, and for a sufficiently large $\delta<1$, when the product is bad, there exists an equilibrium where an informational cascade can happen only after at least half of the players revealed their private information, and consequently, the probability for a ``bad cascade" where all the players buy the product vanishes exponentially with $N$. 
Finally, when $\delta=1$ and the product is bad, there exists an equilibrium where informational cascades do not happen at all.

\end{abstract}

\IEEEpeerreviewmaketitle{}

\section{Introduction}\label{sec:intro}

When a new product/technology is deployed one cannot be certain about
its quality in the early stages of the deployment.
Many people together may form a more accurate prediction about
its quality, but in a strategic environment players act selfishly and may not want to share their private information about the product/technology.
Hence, other players' opinions (private information about the product quality)
are revealed only indirectly through their actions, i.e., whether they bought the product (adopted the technology) or not.
This means that from the perspective of a strategic player,
waiting to see what other people have done may provide
more certainty about the quality of the product. On the other hand,
many products or trends which turn out to be beneficial are better
to be adopted as early as possible.
This interaction can be formalized as a dynamic game with asymmetric
information and a discounted reward. Players want to avoid buying a bad product,
so they may postpone their decision to buy/adopt until
more information is revealed, while at the same time they want to buy/adopt a good product as soon as possible. This scenario generalizes the classical problem of sequential Bayesian learning to a setting with forward-looking players and no predefined order of play.

Sequential learning has been extensively explored in the literature,
with a special focus on a phenomenon known as an \emph{informational cascade}.
In two seminal papers~\cite{Banerjee1992,Bikhchandani1992} the authors
investigated the occurrence of fads in a social network, which was
later generalized in~\cite{smith2000pathological}.
Alternative learning models that have been studied in the literature include
\cite{Acemoglu2011} where players only observe a random set of past actions,
\cite{LeSuBe17} where players observe the past actions through a noisy process,
\cite{Celen2004} where players observe only their immediate predecessor, and
\cite{ScSuSu19} where players are allowed to ask questions to a bounded subset of their predecessors.

The common assumption in all of these models is that players act only once in
the game and there are informational externalities only, which allows
for relatively easy computation of game equilibrium strategies.
Some other works where all players act in each period but are myopic
by design include~\cite{bala1998learning,mossel2013making,mossel2014asymptotic,mossel2015strategic,Harel,Jadbabaie,dasaratha2018social}.
In~\cite{ccelen2004distinguishing,guarino2011aggregate,herrera2013biased,guarino2013social},
different models of Bayesian learning were studied where players
do not observe the entire action history of the past players, but a
``coarser'' history.
There are also
works on non-Bayesian learning models where players do not update
their beliefs in a Bayesian sense~\cite{degroot1974reaching,bala1998learning,ellison1993rules,ellison1995word,molavi2017foundations,jadbabaie2012non},
or do so only with some probability~\cite{Peres}. A survey of such
models can be found in~\cite{nedic2016tutorial}.

An informational cascade is a phenomenon where no player has an incentive
to reveal her private information, hence learning stops in the system.
This is an interesting case of herd behavior that happens even with
fully rational players. While information cascades do not necessarily
happen in all systems (see~\cite{Lee1993,Hann-Caruthers2018}), they
represent a universal phenomenon in sequential Bayesian learning where
players act once in a sequence that is predefined before the game starts.
In such systems, when the turn of a certain player arrives, she has
no choice but to
either buy the product if it seems profitable to her at the moment
or forever forgo the opportunity. Hence, it is natural
to ask whether cascades occur because this one-shot opportunity
was forced upon the players. It is conceivable that if players had the freedom to choose
to wait and gather more information about the product, a herd behavior,
especially a wrong one, could be avoided. This question provides
the motivation for studying information cascades in more complex environments.
In~\cite{VaAn16_learning_arxiv}, informational cascades were defined for a general
dynamic scenario. However, no evidence for their occurrence was provided.

From a technical perspective, the sequential one-shot framework introduced in~\cite{Banerjee1992,Bikhchandani1992} and followed in most of the subsequent literature, lends itself to relatively simple equilibrium analysis, since players do not have to account for how much their estimation on the value of the product is going to improve by
waiting. This is simply because players are given a single opportunity
to act, and cannot wait. In this case, players form a posterior belief on the value of the product based on their public and private signals.
Consequently, the equilibrium consists of strategies that maximize each player's instantaneous reward based on this posterior belief.

In this paper we consider a setting with a finite number of players with no predefined order of action.
An exogenous process determines who enters the marketplace at each time epoch. Once a player is chosen, she is given the opportunity to buy the product (and leave the marketplace forever) or wait and have the opportunity to be called again at future times.
In this setting strategic players  take the future into account since they have multiple interactions with the environment. As a result, our players are typically non-myopic.
This problem can be formulated as a dynamic game with asymmetric information.


In general, one appropriate solution concept for dynamic games with
asymmetric information is the perfect Bayesian equilibrium (PBE)~\cite{FuTi91b}. Finding
a PBE is a crucial first step for establishing whether an informational cascade
occurs. Finding a PBE in a general dynamic scenario with asymmetric information is an extremely challenging
task. In~\cite{VaAn16,VaSiAn19}, the independence of players' types
was exploited to introduce a sequential decomposition methodology to find PBE
involving strategies with time-invariant domain.  This sequential decomposition methodology was based on the common information approach in team problems \cite{nayyar2013decentralized} where the strategies are broken into two partial strategies and dynamic programming equations are used to generate the partial strategies to be applied to the private part of the history. The common information approach in games \cite{VaAn16,VaSiAn19,ouyang2016dynamic,HeAn20arxiv,HeAn19}  is what we use in order to characterize PBE in this paper.

The first contribution of this paper is to characterize a class of PBE where strategies depend on the private observation, as well as the public history of previous actions summarized into a sufficient statistic, the size of which does not increase with time. As a result, equilibrium strategies have a time-invariant domain, and are characterized through the solution of a fixed-point equation (FPE). Furthermore, the domain of the value functions in the FPE we characterize is finite. The finite dimension of the FPE holds even though, for a system with $N$ players, the belief by definition is a probability distribution over a set of size $2^{N+1}$ (all possible realizations of the quality of the product and players' private observation), and thus it is itself an infinite-dimensional object.

Although this sequential decomposition and the ensuing FPE reduce considerably the problem of finding a PBE, the FPE is still quite cumbersome since it has an exponential dimension in the number of players $N$ (the dimensions of the domain of the value functions). Hence, solving the FPE to find PBE is infeasible for large-scale systems. 
%
The second contribution of this paper is to show that by exploiting the structure of our model, we can further simplify the FPE such that the dimension of the domain of the value functions only grows quadratically with $N$.
This simplification and the resulting summarizing variables have a very intuitive explanation that relates this model to the original sequential model of~\cite{Banerjee1992,Bikhchandani1992} and highlights the fundamental differences between the two models. This quadratic-dimension FPE can be solved numerically in practice even for relatively large $N$. We present numerical results indicating that more collaborative equilibria emerge in this setting if players are sufficiently patient. In particular, players are willing to reveal their information even though they are quite certain that the value of the product is good and they would have bought it if they were acting myopically.

\optv{arxiv}{
The third contribution of this paper is to prove existence for the solution of the FPE and to characterize the structure of the solutions. Structural properties of the equilibrium strategies that apply to all of the solutions of the FPE are investigated.
Specifically, the existence of a specific type of strategies, i.e., threshold policies, is proved.
}

The final contribution of this paper is to study whether informational cascades can occur in this model. We study two settings.
In the first setting, the discount factor, $\delta$, is strictly below one.  We show that in this case, the probability of a cascade approaches one as the number of players, $N$, approaches infinity. Moreover, the number of players who have revealed their information before the cascade occurs is small, which formalizes their inefficiency.
%
The second setting involves a fixed number of players $N$ with the discount factor approaching one.
A surprising result emerges in this setting: when the product is bad, there exists a PBE where at least $\frac{N}{2}$ players reveal their information before the wrong cascade, when players buy the product, can occur. Since each revealing player is wrong with probability $p<\frac{1}{2}$, this implies that the probability for a wrong cascade vanishes with $N$. Furthermore, when the discount factor is exactly one and the product is bad, we show that there exists a PBE where a bad informational cascade does not happen at all. 


The rest of this paper is organized as follows.
In Section~\ref{sec:model} we present the model and formulate
the game of non-myopic players. 
In Section~\ref{sec:pbe} we characterize PBE through
a FPE on appropriate beliefs.
In Section~\ref{sec:PBE_quadratic} we summarize
the information contained in the aforementioned beliefs and provide characterization through FPEs
with quadratic dimension in $N$.
Existence results and further characterization of equilibrium strategies are presented in Section~\ref{sec:existence}.
In Section~\ref{sec:cascades} we analyze informational cascades and we show that quite inefficient informational cascades happen with high probability (for large
$N$) for discount factors strictly smaller than one.
Furthermore, we show the surprising result that bad informational cascades can be avoided completely when the product is bad.
Some numerical results are presented in Section~\ref{sec:numerical}, while conclusions are drawn in Section~\ref{sec:conclusions}.
Most of the proof of the Theorems are relegated to the Appendices.
\optv{2col}{
Some of the proofs are omitted due to space limitations and are presented in~\cite{BiHeAn19arxiv}.}

\subsection{Notation}
We use upper case letters for scalar and vector random
variables. We use lower case letters for scalars and bold lower case letters for vectors. We denote the indicator function by $\textbf{1}_a(b)$, such that $\textbf{1}_a(b)=1$ if $a=b$ and $\textbf{1}_a(b)=0$ otherwise.
The space of distributions on a general set $\mA$ is denoted as $\mP(\mA)$.%

\section{Problem Formulation}\label{sec:model}
Consider an infinite horizon dynamic game with $N$ players in the set $\mathcal{N}$. Time is
discrete and the current turn is denoted by $t$, starting from $t=0$.
At each turn, a player is chosen uniformly at random to act, independently
between turns. Only a single player acts in each turn. The random index
of the acting player at time $t$ is denoted $N_{t}$, and its realization
is $n_{t}$.

There is a product with a random state $V\in\mathcal{V}=\left\{ -1,1\right\} $
where $V=-1$ means that the product is bad and $V=1$ means that
the product is good\footnote{In our model we assume the product has infinite many copies, or alternatively the product is a technology that can be adopted by all without scarcity constraints.}. We define $Q\left(v\right)=\Pr\left(V=v\right)$. In the following we assume for simplicity of exposition that $Q(1)=Q(-1)=0.5$.

Each player has her own private information on the product. The private
information of player $n$ is the random variable $X^{n}\in\mathcal{X}\triangleq\left\{ -1,1\right\} $,
with distribution
\begin{equation}
Q\left(x^{n}|v\right)=\Pr\left(X^{n}=x^{n}\,|\,V=v\right)=\left\{
\begin{array}{cc}
1-p & x^{n}=v\\
p & x^{n}\neq v
\end{array}\right.
\end{equation}
where $p\in(0,1/2)$. Define the vector of private information as
$\boldsymbol{X}=\left(X_{1},...,X_{N}\right)$. The private information is independent
between players conditioned on the true value of $V$, so
\begin{equation}
\Pr(\boldsymbol{X}=(x^1,\ldots,x^N)|V=v)=\prod_{n=1}^{N}Q(x^{n}|v).
\end{equation}
Player $n$'s action at turn $t$, denoted by $a_{t}^{n},$ is equal
to 1 if player $n$ buys the product at time $t$ and 0
otherwise. Below, we restrict the action sets such that
only player $n_{t}$ can buy the product at time $t$, and she can do that only once.

Denote $\boldsymbol{a}_{0:t-1}=\left(\boldsymbol{a}_{0},...,\boldsymbol{a}_{t-1}\right)$ and $n_{0:t}=\left(n_{0},...,n_{t}\right)$, where $\boldsymbol{a}_{t}=(a_{t}^{n})_{n\in\mathcal{N}}$ is the action profile at time $t$. The total history
of the game at time $t$ is
\begin{equation}
\boldsymbol{h}_{t}=\left(v,\boldsymbol{x},\boldsymbol{a}_{0:t-1},n_{0:t}\right)\ensuremath{\in\mathcal{H}}_{t}.
\end{equation}
We assume each player can observe all the previous actions taken by
the other players, as well as their identities. Hence the common history at time $t$ is
\begin{equation}
\boldsymbol{h}_{t}^{c}=\left(\boldsymbol{a}_{0:t-1},n_{0:t}\right)\ensuremath{\in\mathcal{H}}_{t}^{c}.
\end{equation}
The common history of actions provide the player with additional information about the quality
of the product. Together with her private information, they form the
information set of player $n$ at time $t$, denoted by
\begin{equation}
\boldsymbol{h}_{t}^{n}=\left(x^{n},\boldsymbol{a}_{0:t-1},n_{0:t}\right)\in\mathcal{H}_{t}^{n}.
\end{equation}
We define $\boldsymbol{b}_{t}=(b_{t}^{n})_{n\in\mathcal{N}}$ with $b_{t}^{n}$
equal to 1 if and only if player $n$ has already bought the product before time $t$. Clearly, $\boldsymbol{b}_t$ can be determined recursively through the publicly observed action profile history $\boldsymbol{a}_{0:t-1}$ and thus it is part of the common history of the players.

A player's pure strategy is a sequence of functions from the information sets of the game to the action space (i.e., a decision whether
to buy or not). In this work, we consider pure strategies. Formally,
player $n$'s strategy is $\boldsymbol{s}^{n}=(s_{t}^{n})_{t=0}^{\infty}$, with
\begin{equation}
s_{t}^{n}:\mathcal{H}_{t}^{n}\rightarrow\mathcal{A}^{n}\left(b_{t}^{n},n_{t}\right)
\label{eq:strategies}
\end{equation}
where
\begin{equation}
\mathcal{A}^{n}\left(b_{t}^{n},n_{t}\right)=\Biggl\{\begin{array}{cc}
\left\{ 0,1\right\}  & \text{if } b_{t}^{n}=0\,,n_{t}=n\\
\left\{ 0\right\}  & \text{else}
\end{array}\label{eq:actionset}
\end{equation}
so that any player $n$ can buy the product only once, and $a^n_{t}=0$
for all $t$ afterwards. In all the turns when player
$n$ does not act ($n_t\neq n$), she is restricted not to buy (``play zero'').

Note that for player $n$, the unknown variables in $\boldsymbol{h}_{t}$ are $X^{-n}$
and $V$. Hence, we define the private belief of player $n$ on the
history of the game as $\mu_{t}^{n}:\mathcal{H}_{t}^{n}\rightarrow\mathcal{P}(\mathcal{X}^{-n}\times \mathcal{V})$
and denote the sequence of private beliefs by $\boldsymbol{\mu}^n=(\mu^n_t)_{t\geq 0}$.
Taking the expectation with respect to this belief and the strategies
in \eqref{eq:strategies}, we define the expected reward-to-go of player $n$ at time $t$
as

\begin{equation}\label{eq:totalreward}
R^{n}\left(\boldsymbol{s}_{t:\infty},\mu_{t}^{n},\boldsymbol{h}_{t}^{n}\right)=\mathbb{E}^{s,\mu_{t}^{n}}\left\{ \sum_{t'=t}^{\infty}\delta^{t'-t}VA_{t'}^{n}\,|\,\boldsymbol{h}_{t}^{n}\right\} ,
\end{equation}

where $0\leq\delta\leq1$ is the discount factor.
Note that at most a single term in the sum \eqref{eq:totalreward} can be non-zero, since $VA_{t'}^{n}=V$ only in the first time that player $n$
buys the product, and 0 otherwise.

The strategies in \eqref{eq:strategies} are functions of $x^{n}$, $\boldsymbol{a}_{0:t-1}$
and $n_{0:t}.$ While $\boldsymbol{a}_{0:t-1}$ and $n_{0:t}$ are observed by
all players, $x^{n}$ is only known to player $n$. Throughout the
paper, it will be useful to decompose those strategies into their common and private components as follows.
\begin{defn} \label{def:Policy-1}
Player $n$ at time $t$ observes $\boldsymbol{h}_{t}^{c}$
and takes an action $a^n_t=\gamma^n_{t}\left(x^{n}\right)$, where $\gamma^n_{t}:\mathcal{X}\rightarrow\mathcal{A}^{n}\left(b_{t}^{n},n_{t}\right)$
is the partial function from her private information to her action.
These partial functions are generated through some policy\footnote{Throughout the paper we use square brackets for mappings that produce functions.}
\begin{equation}
\psi^n_{t}:\mathcal{H}_{t}^{c}\rightarrow\left\{ \mathcal{X}\rightarrow\mathcal{A}^{n}\right\}
\qquad \forall n \in\mathcal{N}
\end{equation}
which operates on $\boldsymbol{h}_{t}^{c}$ and returns a mapping from $x^{n}$
to an action $a^n_t$, so $\gamma^n_{t}=\psi^n_{t}[\boldsymbol{h}_{t}^{c}]$ and
$a^n_{t}=\psi^n_{t}[\boldsymbol{h}_{t}^{c}](x^{n})$.
\end{defn}

The above decomposition is a trivial consequence of the fact that any function $\mathcal{H}_{t}^{c}\times \mathcal{X}\rightarrow \mathcal{A}^{n}$ is equivalent to a function $\mathcal{H}_{t}^{c}\rightarrow  \{ \mathcal{X}\rightarrow \mathcal{A}^{n} \}$.
In the first form, the strategy is a direct function of both the public history $\boldsymbol{h}^c_t$ and the private signal $x^n$, so that $a^n_t=s^n_t(\boldsymbol{h}^c_t,x^n)$.
In the second form, the strategy is decomposed into two steps: in the first step the public history produces a partial function $\gamma^n_{t}=\psi^n_{t}[\boldsymbol{h}_{t}^{c}]$, and in the second step this partial function is evaluated at the private signal to generate the final action $a^n_{t}=\gamma^n_t(x^n)=\psi^n_{t}[\boldsymbol{h}_{t}^{c}](x^{n})$.
Note that there are only four possible deterministic gamma functions
$\gamma^n_{t}$: wait for any $x^{n}$ (denoted by $\boldsymbol{0}$),
buy for any $x^{n}$ (denoted by $\boldsymbol{1}$),
buy according to $x^{n}$ (denoted by $\boldsymbol{I}$)
and buy according to $-x^{n}$. The last one is clearly
dominated by one of the other three so it is never considered. Hence,
we are left with three possible partial strategies, namely,
$\gamma^n_t\in\left\{ \boldsymbol{0},\boldsymbol{1},\boldsymbol{I}\right\}$.
Furthermore, since every non-acting player is essentially waiting (i.e., playing $\gamma^n_t=\boldsymbol{0}$ for $n\neq n_t$), in the following
we will drop the superscript $^n$ and only refer to the acting player's partial function
as $\gamma_t=\psi_t[\boldsymbol{h}^c_t]$.

We conclude this section by remarking that players' strategies and particularly their partial function $\gamma_t$ are responsible for the revelation of the private information $x^n$ to the rest of the community. Indeed, if a player plays according to $\gamma_t=\boldsymbol{I}$ then she reveals her private information $x^n$ through her action $a^n_t$. Conversely, if she either plays according to $\gamma_t=\boldsymbol{0}$, or $\boldsymbol{1}$, her private information is not revealed. We note that ``revealing" is a special case of ``signaling", where the exact private information of a player can be inferred as opposed to only some Bayesian estimation of it \cite{vasal2021signaling}.%

\section{Characterization of Structured Perfect Bayesian Equilibria}\label{sec:pbe}

\subsection{Perfect Bayesian Equilibrium}

Our main goal is to study if an informational cascade occurs in the above setting. An
informational cascade is defined as a state of the game where learning
stops since actions no longer reveal new information. To do so, we
first have to study the equilibrium strategies of this game. Since
this is a dynamic game with asymmetric information, an appropriate
solution concept is the PBE~\cite{FuTi91b}, defined
as follows.

\begin{defn}\label{def:PBE}
A PBE with pure strategies is a pair $\left(\boldsymbol{s}^{*},\boldsymbol{\mu}^{*}\right)$
of
\begin{itemize}
\item a strategy profile $\boldsymbol{s}^{*}=(\boldsymbol{s}^{*n})_{n\in\mathcal{N}}$,
\item a belief profile sequence $\boldsymbol{\mu}^{*}=(\boldsymbol{\mu}^{*n})_{n\in\mathcal{N}}$,
\end{itemize}
such that sequential rationality holds, i.e.,  for each $n\in\mN$, $t\geq 0$ and $\boldsymbol{h}_{t}^{n}\in\mathcal{H}_{t}^{n}$,
and each strategy $\boldsymbol{s}^{n}$
\begin{equation}
R^{n}\left(s_{t:\infty}^{*n},s_{t:\infty}^{*-n},\mu^{*n}_t,\boldsymbol{h}_{t}^{n}\right)\geq R^{n}\left(s_{t:\infty}^{n},s_{t:\infty}^{*-n},\mu^{*n}_t,\boldsymbol{h}_{t}^{n}\right),
\end{equation}
and the beliefs satisfy Bayesian updating whenever  $\Pr^{s^*}(\boldsymbol{h}^n_t|\boldsymbol{h}^n_{t-1})>0$.
\end{defn}

In this paper, we are interested in PBE that depend on the history of the game only through a summary in the form of the belief of the players about $V$ and $X$. Hence, we formulate FPE for which the set of solutions is the set of these PBE,  which are known as structured PBE \cite{VaSiAn19}. Structured PBEs represent a more reasonable behavior since strategies that depend on sequentially updatable beliefs are more tractable than strategies that require tracking the whole history. 

We remark that strategies and beliefs should be defined for all information sets, even those that occur with zero probability under equilibrium strategies (off-equilibrium paths). \optv{2col}{In our setting,  a deviation of the equilibrium strategies is known to other players only when non-signalling strategies are played. Since the beliefs are not updated for the on-equilibrium strategies at such points, we assume the beliefs are not changed even if a player deviates. However, there are some deviations that are only known to the acting player and we note that she should not change her belief about neither $v$ nor others' private signals when she is playing, no matter what she plays and whether she deviates or not. Detailed explanations on how we deal with off-equilibrium beliefs can be found in \cite{BiHeAn19arxiv}.}%
\optv{arxiv}{ In our setting, there are both public and private off-equilibrium paths. The public off-equilibrium paths (i.e., paths where all players can confirm that there was a deviation from equilibrium) are those for which $a^{n_{t-1}}_{t-1}=0$, but $s^{*n_{t-1}}(x^{n_{t-1}},h^c_{t-1})=1$, for all $x^{n_{t-1}}$ or similarly, $a^{n_{t-1}}_{t-1}=1$, but $s^{*n_{t-1}}(x^{n_{t-1}},\boldsymbol{h}^c_{t-1})=0$, for all $x^{n_{t-1}}$. In both of these situations, we have $\Pr^{s^*}(\boldsymbol{h}^n_t|\boldsymbol{h}^n_{t-1})=0$ and we pose no restriction on the belief updating.
 As will be shown in Lemma~\ref{lem:beliefupdate}, in both of these cases, the beliefs are not updated for on-equilibrium actions, and so we choose to not update them even if the actions are not according to the equilibrium strategies.
The beliefs at the continuation of the game from these points on, however, will be updated according to Bayes' rule if $\Pr^{s^*}(\boldsymbol{h}^n_t|\boldsymbol{h}^n_{t-1})>0$.
 The private off-equilibrium paths (i.e., paths where all players other than the acting player do not have a way to confirm if a deviation from equilibrium occurred) are when $s^{*n_{t-1}}(x^{n_{t-1}}=1,\boldsymbol{h}^c_{t-1})=1$ and $s^{*n_{t-1}}(x^{n_{t-1}}=-1,\boldsymbol{h}^c_{t-1})=0$ (playing $\gamma_t^{n_t}=\boldsymbol{I}$) and the acting player played $a^{n_{t-1}}_{t-1}=1$ with a private signal $x^{n_{t-1}}=-1$ or played $a^{n_{t-1}}_{t-1}=0$ with a private signal $x^{n_{t-1}}=1$, and she has not yet revealed her private information. In this situation, no player other than player $n_{t-1}$ is aware of the deviation because both actions are possible. We impose the restriction on player $n_{t-1}$'s belief to not be updated at the time of her deviation, although other players update their beliefs about $x^{n_{t-1}}$ and consequently $v$. Intuitively, a player can not learn anything more by her own actions but she can induce different beliefs in others. One can refer to~\cite{FuTi91b,Fudenberg1991} in order to justify this constraint on the off-equilibrium beliefs. Specifically, one of the conditions posed on off-equilibrium beliefs for PBE is referred to as ``no signaling what you don't know"~\cite[p.~332]{Fudenberg1991}. This condition indicates that if one considers two different action profiles in which a specific player's action is the same, the belief about that player's type should be updated similarly for both action profiles. This implies that in our setting, the acting player should not change her belief about any other player's private signal because they are not playing. On the other hand, learning about $v$ happens through players' private signals. If the belief about others' private signals does not change, the belief about $v$ should not change either. So the acting player should not change her belief about neither $v$ nor others' private signals when she is playing, no matter what she plays and whether she deviates or not.
}


\subsection{Characterization of Structured PBE}

We now present a methodology for characterizing PBE where
the strategy for the acting player $n_{t}$ depends on the common
history only through the common belief on the variables $V,X$ (as
well as the variable $\boldsymbol{B}_{t}$). In particular, we define the common
belief $\pi_{t}\in\mathcal{P}\left(\mathcal{X}^{N}\times\mathcal{V}\right)$
where $\pi_{t}(\boldsymbol{x},v):=\Pr^s\left(X=\boldsymbol{x},V=v|\boldsymbol{a}_{0:t-1},\boldsymbol{b}_{1:t},n_{0:t}\right)
=\Pr^{\psi}\left(X=\boldsymbol{x},V=v|\boldsymbol{a}_{0:t-1},\boldsymbol{b}_{1:t},n_{0:t},\gamma_{0:t-1}\right)$. For $t=0$, we set $\pi_0(\boldsymbol{x},v)=Q(v)\prod_n Q(x^n|v)$.
We first show that the belief $\pi_{t}$ can be updated using only public information and that the update depends on  $\psi_t$ only through $\gamma_t$.
Note that the dependence of the update equation on $\gamma_t$ is the manifestation of ``signaling'' in our model. When the equilibrium strategy is $\gamma_t=\boldsymbol{I}$, acting player's action reveals her private information and changes the beliefs of other players about $V$ and $X$.
\begin{lem}\label{lem:beliefupdate}
There exists a function $F$ such that the belief $\pi_{t}$ can be updated as $\pi_{t+1}=F(\pi_{t},\gamma_{t},a^{n_t}_{t},n_{t})$.
In particular, if $\gamma_{t}\neq\boldsymbol{I}$, the belief is not
updated.
\end{lem}
\begin{IEEEproof}
By simple application of Bayes' rule we have
\optv{2col}{
\begin{subequations}
\begin{align}
&\pi_{t+1}\left(\boldsymbol{x},v\right) \nonumber \\
&=\Pr^s\left(\boldsymbol{x},v|\boldsymbol{a}_{0:t},\boldsymbol{b}_{1:t+1},n_{0:t+1}\right)\\
&=\Pr^{\psi}\left(\boldsymbol{x},v|\boldsymbol{a}_{0:t},\boldsymbol{b}_{1:t+1},n_{0:t+1},\gamma_{0:t}\right)\\
&=\Pr^{\psi}\left(\boldsymbol{x},v|\boldsymbol{a}_{0:t},\boldsymbol{b}_{1:t},n_{0:t},\gamma_{0:t}\right)\\
&=\frac{\Pr^{\psi}\left(\boldsymbol{x},v,\boldsymbol{a}_{t}|\boldsymbol{a}_{0:t-1},\boldsymbol{b}_{1:t},n_{0:t},\gamma_{0:t}\right)}{\Pr^{\psi}\left(\boldsymbol{a}_{t}|\boldsymbol{a}_{0:t-1},\boldsymbol{b}_{1:t},n_{0:t},\gamma_{0:t}\right)}\\
&=\frac{\Pr^{\psi}\left(\boldsymbol{a}_{t}|\boldsymbol{x},v,\boldsymbol{a}_{0:t-1},\boldsymbol{b}_{1:t},n_{0:t},\gamma_{0:t}\right)}{\Pr^{\psi}\left(\boldsymbol{a}_{t}|\boldsymbol{a}_{0:t-1},\boldsymbol{b}_{1:t},n_{0:t},\gamma_{0:t}\right)} \times \nonumber \\
 &\qquad \qquad \Pr^{\psi}\left(\boldsymbol{x},v|\boldsymbol{a}_{0:t-1},\boldsymbol{b}_{1:t},n_{0:t},\gamma_{0:t}\right)\\
&=\frac{\textbf{1}_{\gamma_{t}(x^{n_{t}})}\left(a^{n_t}_{t}\right)\pi_{t}\left(\boldsymbol{x},v\right)}{\sum_{\boldsymbol{x'},v'}\textbf{1}_{\gamma_{t}(x^{'n_{t}})}\left(a^{n_t}_{t}\right)\pi_{t}\left(\boldsymbol{x'},v'\right)}.
\end{align}
\end{subequations}
}
\optv{arxiv}{
	\begin{subequations}
		\begin{align}
		\pi_{t+1}\left(\boldsymbol{x},v\right) &=\Pr^s\left(\boldsymbol{x},v|\boldsymbol{a}_{0:t},\boldsymbol{b}_{1:t+1},n_{0:t+1}\right)\\
		&=\Pr^{\psi}\left(\boldsymbol{x},v|\boldsymbol{a}_{0:t},\boldsymbol{b}_{1:t+1},n_{0:t+1},\gamma_{0:t}\right)\\
		&=\Pr^{\psi}\left(\boldsymbol{x},v|\boldsymbol{a}_{0:t},\boldsymbol{b}_{1:t},n_{0:t},\gamma_{0:t}\right)\\
		&=\frac{\Pr^{\psi}\left(\boldsymbol{x},v,\boldsymbol{a}_{t}|\boldsymbol{a}_{0:t-1},\boldsymbol{b}_{1:t},n_{0:t},\gamma_{0:t}\right)}{\Pr^{\psi}\left(\boldsymbol{a}_{t}|\boldsymbol{a}_{0:t-1},\boldsymbol{b}_{1:t},n_{0:t},\gamma_{0:t}\right)}\\
		&=\frac{\Pr^{\psi}\left(\boldsymbol{a}_{t}|\boldsymbol{x},v,\boldsymbol{a}_{0:t-1},\boldsymbol{b}_{1:t},n_{0:t},\gamma_{0:t}\right)\Pr^{\psi}\left(\boldsymbol{x},v|\boldsymbol{a}_{0:t-1},\boldsymbol{b}_{1:t},n_{0:t},\gamma_{0:t}\right)}{\Pr\left(\boldsymbol{a}_{t}|\boldsymbol{a}_{0:t-1},\boldsymbol{b}_{1:t},n_{0:t},\gamma_{0:t}\right)}   \\
		&=\frac{\textbf{1}_{\gamma_{t}(x^{n_{t}})}\left(a^{n_t}_{t}\right)\pi_{t}\left(\boldsymbol{x},v\right)}{\sum_{\boldsymbol{x'},v'}\textbf{1}_{\gamma_{t}(x^{'n_{t}})}\left(a^{n_t}_{t}\right)\pi_{t}\left(\boldsymbol{x'},v'\right)}.
		\end{align}
	\end{subequations}
}
Note that if $\gamma_t$ is a constant function (i.e., $\gamma_t\neq \boldsymbol{I}$) the quantity $\textbf{1}_{\gamma_{t}(x^{n_{t}})}(a^{n_t}_t)$ cancels from numerator and denominator of the above expression, thus resulting in $\pi_{t+1}=\pi_t$.
Furthermore, whenever the denominator is zero (off-equilibrium paths) we set $\pi_{t+1}=\pi_{t}$. Additionally, while $\pi_{t}(\boldsymbol{x},v)$ depends on $\boldsymbol{a}_{0:t}$, $n_{0:t+1}$ and $\boldsymbol{b}_{1:t+1}$, the update function $F$ only depends on $\pi_{t},\gamma_{t},a_{t}^{n_t}$ and $n_t$. By definition of the game, $a_t^{m}=0$ for all $m \neq n_t$, and so factors of the form $\textbf{1}_{\boldsymbol{0}}(a^m_t)$ are all one in both numerator and denominator in the last equality.
\end{IEEEproof}

The private beliefs of players on $v$ on equilibrium paths are obtained by conditioning the public belief on $V$ on players' private signal, $X^n$. More specifically, player $n$'s private belief on equilibrium path is $\mu^n_t(v)=\pi_{t}(v|x^n)=\frac{\pi_t(x^n,v)}{\pi_t(x^n)}$, where  $\pi_t(x^n,v)$ and $\pi_t(x^n)$ are marginal beliefs of $\pi_t(\boldsymbol{x},v)$.

A player is only interested in the previous actions since they carry information about $V$. However, not every action reveals the private information of the acting player. For that to
happen, the action that the player took must be determined by her
private information. This motivates characterizing the beliefs using the following finite dimensional variables:%
\begin{defn}
	\label{def:Revealing}Let $\tilde{x}_{t}^{n}\in\{0,-1,1\}$ be the revealed information of player $n$ at the beginning of period $t$,
	so $\tilde{x}_{t}^{n}=0$ if the player has
	not yet revealed her private information before time $t$, while $\tilde{x}_{t}^{n}=\pm1$
	if the player has already revealed her private signal and the value is as indicated. The quantity $\tilde{x}_{t}^{n}$ remains unchanged for non-acting players while it is recursively updated for the acting player as
	\begin{equation}\label{eq:xtilde}
	\widetilde{x}_{t+1}^{n}
 =f(\tx^{n}_{t},\gamma_{t},a^{n}_{t})
 =\left\{ \begin{array}{cc}
	2a_{t}^{n}-1 & \gamma_{t}=\boldsymbol{I},\tilde{x}_{t}^{n}=0\\
	\widetilde{x}_{t}^{n} & o.w.,
	\end{array}\right.
	\end{equation}
	with the initial condition $\widetilde{x}_{0}^{n}=0$. Note that $\widetilde{x}_{t+1}^{n}$
	is a function of $\tilde{x}_{0:t}^{n}$, $\boldsymbol{a}_{0:t}$ and $n_{0:t}$,
	or equivalently of $\gamma_{0:t}$, $\boldsymbol{a}_{0:t}$ and $n_{0:t}$.
	We also define the function $\tilde{F}$ such that  
	\begin{align*}
	   \ensuremath{\boldsymbol{\tilde{x}}_{t+1}
	    &=\left(\boldsymbol{\tx}_{t+1}^{-n_{t}},\tx_{t+1}^{n_{t}}\right)
	    =\tilde{F}(\boldsymbol{\tilde{x}}_{t},\gamma_{t},a_{t}^{n_{t}},n_{t})\\
	    &\triangleq\left(\boldsymbol{\tx}_{t}^{-n_{t}},f(\tx_{t}^{n_{t}},\gamma_{t},a_{t}^{n_{t}})\right)}
	   \numberthis
	\end{align*}
to summarize the recursive update of the entire vector $\boldsymbol{\tilde{x}}_t=(\tx^1_t,\ldots,\tx^N_t)$. 
Note that only the acting player's component of this vector is updated.%
Furthermore,  $\tilde{x}^n_t$ can be derived from the belief $\pi_t$  since if $\pi_t(x^n)=\textbf{1}_{k}(x^n)$ for $k \in \{-1,1\}$ then  $\tilde{x}^n_t=k$ and otherwise  $\tilde{x}^n_t=0$.
\end{defn}

 Following the discussion after Definition \ref{def:PBE} and by using $\boldsymbol{\tilde{x}_t}$, we characterize the off-equilibrium private beliefs as follows:%
\begin{equation}\label{eq:private}
\frac{\mu_t^n(v=1)}{\mu_t^n(v=-1)}
 =\frac{\pi_t(v=1)}{\pi_t(v=-1)} \left(\frac{1-p}{p}\right)^{-\tilde{x}_{t}^n+x^n}.
\end{equation}
Intuitively, equation \eqref{eq:private} says that if a player has not yet revealed her information ($\tilde{x}_{t}^n=0$), then her private likelihood about $V$ is the public likelihood amplified by the private factor $(\frac{1-p}{p})^{x^n}$.
If however she has already revealed her information and she is on equilibrium $\tilde{x}_{t}^n=x^n$
then her private belief is the same as the public belief, which includes her private information since it was revealed.
Finally, if she has already revealed her information and she is off-equilibrium $\tilde{x}_{t}^n=-x^n$
then her private likelihood has to correct for the erroneous public belief through the factor $(\frac{1-p}{p})^{-\tilde{x}_{t}^n}$ and then amplified by the true factor $(\frac{1-p}{p})^{x^n}$.


The following lemma shows that the common belief decomposes into a
belief on $v$ and a belief on $x$, and that each part can be updated
recursively. Specifically, it proves that
the private information variables $X^1,\ldots,X^N$ are conditionally
independent given $v\,,\boldsymbol{h}_{t}^{c}$ and that the common belief can be expressed in terms of $\boldsymbol{\widetilde{x}}_t$ from Definition~\ref{def:Revealing}.
\begin{lem}\label{lem:BeliefDecomposition}
The public belief $\pi_{t}\left(\boldsymbol{x},v\right)=\Pr\left(X=\boldsymbol{x},V=v|\boldsymbol{h}_{t}^{c}\right)$
can be decomposed as follows
\begin{equation}\label{eq:cond_ind}
\pi_{t}\left(\boldsymbol{x},v\right)=\pi_{t}\left(v\right)\prod_{m=1}^{N}\pi_{t}\left(x^{m}|v\right)
\end{equation}
where $\pi_{t}\left(v\right)\triangleq\Pr\left(V=v\,|\,\boldsymbol{h}_{t}^{c}\right)$
and $\pi_{t}\left(x^{m}|v\right)\triangleq\Pr\left(X^{m}=x^{m}\,|\,v,\boldsymbol{h}_{t}^{c}\right)$.
Furthermore,
\begin{equation}\label{eq:conditional}
\pi_{t}\left(x^{m}|v\right)=
\begin{cases}
\textbf{1}_{\widetilde{x}_{t}^{m}}(x^{m}), & \widetilde{x}_{t}^{m}\neq0\\
Q\left(x^{m}|v\right), & \widetilde{x}_{t}^{m}=0
\end{cases}
\end{equation}
and the belief on $V$ can be updated as
\begin{equation}\label{eq:marginal_up}
\frac{\pi_{t+1}(1)}{\pi_{t+1}(-1)}=\frac{\pi_{t}(1)}{\pi_{t}(-1)}  \times
\begin{cases}
q^{2a^{n_t}_{t}-1},
  & \hspace{-0.2cm}\gamma_{t}=\boldsymbol{I}\text{ and } \tilde{x}^{n_t}_t=0\\
1,
  & o.w.,
\end{cases}
\end{equation}
with $q=\frac{1-p}{p}$.
Finally, the belief on $V$ can be explicitly expressed as
\begin{equation}\label{eq:beliefv}
 \frac{\pi_{t}(1)}{\pi_{t}(-1)}=q^{\sum_{n}\widetilde{x}_{t}^{n}}.
\end{equation}
\end{lem}
\begin{IEEEproof}
See Appendix~\ref{lem:BeliefDecomposition_proof}.
\end{IEEEproof}

We would like to characterize equilibrium strategies for the acting player  $a^{n_t}_{t}=\psi_{t}[\boldsymbol{h}_{t}^{c}](x^{n_{t}})$
for which the ever-increasing common history $\boldsymbol{h}^c_t=(\boldsymbol{a}_{0:t-1},n_{0:t})$ is summarized into the time-invariant
quantities $(n_t,\pi_{t},\boldsymbol{b}_t)\in\mN\times \mP(\mX^N\times \mV) \times\{0,1\}^N$, i.e., equilibrium strategies of the form $a^{n_t}_{t}=\theta[n_t,\pi_{t},\boldsymbol{b}_t](x^{n_{t}})$. In other words, we seek equilibrium strategies where the partial functions are of the form $\gamma_{t}=\theta[n_t,\pi_{t},\boldsymbol{b}_t]$. Thanks to Lemma \ref{lem:BeliefDecomposition} we know that the beliefs can be summarized using $\boldsymbol{\tilde{x}}$. Hence, with a slight abuse of notation, we can write $\gamma_{t}=\theta[n_t,\boldsymbol{\tilde{x}}_t,\boldsymbol{b}_t]$.%

Using the above structural results for the beliefs, we can construct our finite-dimensional FPE.%

\begin{algo}[Finite dimensional]\label{alg:Fixed-Point finite}

For every $n\in\mathcal{N}$, $\boldsymbol{\tilde{x}}\in\{-1,0,1\}^N$,
$\boldsymbol{b}\in\{0,1\}^{N}$ we evaluate $\gamma^{*}=\theta\left[n,\boldsymbol{\tilde{x}},\boldsymbol{b}\right]$
as follows
\begin{itemize}
\item If $b^{n}=1$ then $\gamma^{*}=\textbf{0}$.

\item If $b^{n}=0$ then $\gamma^{*}$ is the solution of the following
system of equations, $\forall x^{n}\in\mathcal{X}$
\optv{2col}{
\begin{subequations}\label{eq:fpe2}
\begin{align}\label{eq:fpe2a}
&\gamma^{*}\left(x^{n}\right)=\arg\max\{
\underbrace{\frac{q^{\sum_{m}\widetilde{x}^{m}-\widetilde{x}^n+x^{n}} -1 }{ q^{\sum_{m}\widetilde{x}^{m}-\widetilde{x}^n+x^{n}} +1}}_{1=\text{``buy''}}, \nonumber \\
&\quad \underbrace{\frac{\delta}{N}\sum_{n'=1}^{N}V^{n}\left(x^{n},n',\tilde{F}\left(\boldsymbol{\tilde{x}},\gamma^{*},0,n\right),\boldsymbol{b}\right)}_{0=\text{``don't buy''}}   \}
\end{align}
where the value functions for all $m\in\mathcal{N}$ satisfy
\begin{align}\label{eq:fpe2b}
&V^{m}\left(x^{m},n,\boldsymbol{\tilde{x}},\boldsymbol{b}\right)=\nonumber \\
&\left\{
\begin{aligned}
&0, \hspace{3.2cm} b^{m}=1\\
&\frac{\delta}{N}\sum_{n'=1}^{N}V^{m}\left(x^{m},n',\tilde{F}\left(\boldsymbol{\tilde{x}},\gamma^{*},0,m\right),\boldsymbol{b}\right), \\ & \hspace{3.6cm} b^{m}=0,n=m,\gamma^{*}\left(x^{m}\right)=0\\
%
&\frac{q^{\sum_{m'}\widetilde{x}^{m'}-\widetilde{x}^{m}+x^{m}} -1}{q^{\sum_{m'}\widetilde{x}^{m'}-\widetilde{x}^{m}+x^{m}} +1}
 , \\ & \hspace{3.6cm}  b^{m}=0,n=m,\gamma^{*}\left(x^{m}\right)=1\\
&\frac{\delta}{N}\sum_{n'=1}^{N}\mathbb{E}\left[V^{m}(x^{m},n',\tilde{F}(\boldsymbol{\tilde{x}},\gamma^{*},\gamma^{*}(X^{n}),n),(\boldsymbol{b}^{-n},B^{n}))\right],\\ & \hspace{3.6cm}  b^{m}=0,n\neq m,
\end{aligned}\right.
\end{align}
where expectation in \eqref{eq:fpe2b} is wrt the RVs $X^{n}$ and $B^n$
with
\begin{align}\label{eq:fpe2c}
\Pr(&X^{n}=x^{n},B^{n}=b^{\prime n}|x^{m},n,\boldsymbol{\tilde{x}},\boldsymbol{b}) \nonumber\\
=&\Pr\left(B^{n}=b^{\prime n}|X^{n}=x^{n},x^{m},n,\boldsymbol{\tilde{x}},\boldsymbol{b}\right) \nonumber\\
 &\quad \Pr\left(X^{n}=x^{n}|x^{m},n,\boldsymbol{\tilde{x}},\boldsymbol{b}\right),
\end{align}
where
\begin{align}\label{eq:fpe2d}
&\Pr\left(B^{n}=1|X^{n}=x^{n},x^{m},n,\boldsymbol{\tilde{x}},\boldsymbol{b}\right) \nonumber \\
 &=\left\{ \begin{array}{ll}
1 & \text{, if }b^{n}=1\text{ or }\gamma^{*}\left(x^{n}\right)=1\\
0 & \text{, else,}
\end{array}\right.
\end{align}
and
\begin{align}\label{eq:fpe2e}
&\Pr(X^{n}=x^{n}|x^{m},n,\boldsymbol{\tilde{x}},\boldsymbol{b}) \nonumber \\
&=\left\{
 \begin{array}{ll}
\textbf{1}_{\tilde{x}^n}(x^n) & \text{, if } \tilde{x}^n\neq 0\\
\frac{Q(x^n|-1)+Q(x^n|1)q^{\sum_{m'}\widetilde{x}^{m'}-\widetilde{x}^m+x^{m}
}}{1+q^{\sum_{m'}\widetilde{x}^{m'}-\widetilde{x}^m+x^{m}
}} & \text{, if } \tilde{x}^n = 0.
\end{array}
 \right.
\end{align}
\end{subequations}}
\optv{arxiv}{
	\begin{subequations}\label{eq:fpe2}
		\begin{align}\label{eq:fpe2a}
		&\gamma^{*}\left(x^{n}\right)=\arg\max\left\{
		\underbrace{\frac{q^{\sum_{m}\widetilde{x}^{m}-\widetilde{x}^n+x^{n}} -1 }{ q^{\sum_{m}\widetilde{x}^{m}-\widetilde{x}^n+x^{n}} +1}}_{1=\text{``buy''}},  \underbrace{\frac{\delta}{N}\sum_{n'=1}^{N}V^{n}\left(x^{n},n',\tilde{F}\left(\boldsymbol{\tilde{x}},\gamma^{*},0,n\right),\boldsymbol{b}\right)}_{0=\text{``don't buy''}}   \right\}
		\end{align}
		where the value functions for all $m\in\mathcal{N}$ satisfy
		\begin{align}\label{eq:fpe2b}
		V^{m}&\left(x^{m},n,\boldsymbol{\tilde{x}},\boldsymbol{b}\right)=\nonumber \\&\left\{
		\begin{array}{ll}
		0, &  b^{m}=1\\
		\frac{\delta}{N}\sum_{n'=1}^{N}V^{m}\left(x^{m},n',\tilde{F}\left(\boldsymbol{\tilde{x}},\gamma^{*},0,m\right),\boldsymbol{b}\right),  &  b^{m}=0,n=m,\gamma^{*}\left(x^{m}\right)=0\\
		%
		\frac{q^{\sum_{m'}\widetilde{x}^{m'}-\widetilde{x}^{m}+x^{m}} -1}{q^{\sum_{m'}\widetilde{x}^{m'}-\widetilde{x}^{m}+x^{m}} +1}
		, &   b^{m}=0,n=m,\gamma^{*}\left(x^{m}\right)=1\\
		\frac{\delta}{N}\sum_{n'=1}^{N}\mathbb{E}\left[V^{m}(x^{m},n',\tilde{F}(\boldsymbol{\tilde{x}},\gamma^{*},\gamma^{*}(X^{n}),n),\boldsymbol{b}^{-n}B^{n})\right], &  b^{m}=0,n\neq m,
		\end{array}\right.
		\end{align}
		where expectation in \eqref{eq:fpe2b} is wrt the RVs $X^{n}$ and $B^n$
		with
		\begin{align}\label{eq:fpe2c}
		\Pr(&X^{n}=x^{n},B^{n}=b^{\prime n}|x^{m},n,\boldsymbol{\tilde{x}},\boldsymbol{b})=\Pr\left(B^{n}=b^{\prime n}|X^{n}=x^{n},x^{m},n,\boldsymbol{\tilde{x}},\boldsymbol{b}\right)\Pr\left(X^{n}=x^{n}|x^{m},n,\boldsymbol{\tilde{x}},\boldsymbol{b}\right),
		\end{align}
		where
		\begin{align}\label{eq:fpe2d}
		\Pr\left(B^{n}=1|X^{n}=x^{n},x^{m},n,\boldsymbol{\tilde{x}},\boldsymbol{b}\right)=\left\{ \begin{array}{ll}
		1 & \text{, if }b^{n}=1\text{ or }\gamma^{*}\left(x^{n}\right)=1\\
		0 & \text{, else,}
		\end{array}\right.
		\end{align}
		and
		\begin{align}\label{eq:fpe2e}
		\Pr(X^{n}=x^{n}|x^{m},n,\boldsymbol{\tilde{x}},\boldsymbol{b})=\left\{
		\begin{array}{ll}
		\textbf{1}_{\tilde{x}^n}(x^n) & \text{, if } \tilde{x}^n\neq 0\\
		\frac{Q(x^n|-1)+Q(x^n|1)q^{\sum_{m'}\widetilde{x}^{m'}-\widetilde{x}^m+x^{m}
		}}{1+q^{\sum_{m'}\widetilde{x}^{m'}-\widetilde{x}^m+x^{m}
		}} & \text{, if } \tilde{x}^n = 0.
		\end{array}
		\right.
		\end{align}
	\end{subequations}}
\hfill$\blacksquare$
\end{itemize}
\end{algo}

Once the mapping $\theta\left[\cdot\right]$ has been found through
the FPE~\ref{alg:Fixed-Point finite}, the PBE strategies and beliefs are
generated through the following forward recursion.

\begin{itemize}
\item \textbf{Initialization}: Let $\boldsymbol{\tilde{x}}_0= \boldsymbol{0}\in\mathbb{R}^{N}$.
\begin{subequations}\label{eq:forward}
\item \textbf{For} $t=0,1,2,\ldots$, $\forall n\in\mathcal{N}$, $\boldsymbol{h}_{t}^{c}\in\mathcal{H}_{t}^{c}$,
$x^{n}\in\mathcal{X}$:

\begin{enumerate}
\item Compute
\begin{align}
s_{t}^{*n}\left(\boldsymbol{h}_{t}^{n}\right) & :=\left\{ \begin{array}{ll}
\theta\left[n_{t},\boldsymbol{\tilde{x}}_{t}^{*}\left[\boldsymbol{h}_{t}^{c}\right],\boldsymbol{b}_{t}\right]\left(x^{n}\right) & n=n_{t}\\
0 & o.w.
\end{array}\right.\label{eq:forwardb}
\end{align}

\item 
Compute $\pi_{t}^{*}$ according to Lemma \ref{lem:BeliefDecomposition}.
\item
Generate the private beliefs $\mu_{t}^{*n}$ from $\pi_{t}^{*}$ by
as
\begin{equation}
\mu_{t}^{*n}\left(\boldsymbol{x}^{-n},v\right)= \pi^*_t(\boldsymbol{x}^{-n}|v)\mu^{*n}_t(v) \label{eq:forwardd}
\end{equation}
where
\begin{equation}
\frac{\mu^{*n}_t(v=1)}{\mu^{*n}_t(v=-1)}
 =\frac{\pi^*_t(v=1)}{\pi^*_t(v=-1)}
 \left(\frac{1-p}{p}\right)^{-\tilde{x}^n+x^n}.
\end{equation}

\item Let $a_t^{n_t}=s_{t}^{*n}\left(\boldsymbol{h}_{t}^{n}\right)$. For every $n_{t+1}\in \mathcal{N}$, let  $\boldsymbol{h}_{t+1}^{c}=(\boldsymbol{h}_{t}^{c},a_t^{n_t},n_{t+1})$ and compute:\begin{align*}
\boldsymbol{\tilde{x}}_{t+1}^{*}&\left[\boldsymbol{h}_{t+1}^{c}\right]  :=\\&\tilde{F}\left(\boldsymbol{\tilde{x}}_{t}^{*}\left[\boldsymbol{h}_{t}^{c}\right],\theta\left[n_{t},\boldsymbol{\tilde{x}}_{t}^{*}\left[\boldsymbol{h}_{t}^{c}\right],\boldsymbol{b}_{t}\right],a^{n_t}_{t},n_{t}\right).\numberthis
\label{eq:forwarde}
\end{align*}
\end{enumerate}
\end{subequations}
\end{itemize}


The following theorem establishes that the above construction
generates a PBE.%
\begin{thm} \label{thm:PBE}
Whenever FPE 1 has a solution, the forward
construction described in \eqref{eq:forward} generates a PBE.
\end{thm}

\begin{IEEEproof}
See Appendix \ref{thm:PBE_proof}.
\end{IEEEproof}

FPE~\ref{alg:Fixed-Point finite}, and in particular in~\eqref{eq:fpe2a}
is akin to a dynamic programming FPE in an infinite-horizon stopping-time problem.
There is however a significant difference: although player $n$ is deciding about her strategy which will lead to an action by maximizing the reward between buying and waiting,
we use the equilibrium strategy $\gamma^*$ in the update  function of the belief $\pi$.
The reason for this twist is shown in the proof of Theorem~ \ref{thm:PBE}.
This proof shows that player $n$ faces an MDP only if every other player plays according to $\gamma^*$, and also, most crucially, if the update of $\pi$ is according to $\gamma^*$. Hence, if these two requirements hold, the best response of player $n$ will give us the PBE strategies, $\gamma^*$. Therefore, we have a FPE that contains $\gamma^*$ in both the left- and right-hand side of the equation.
In other words, $\gamma^*_t$ is an equilibrium strategy only if it is the best response assuming that the belief update $\pi_{t+1}=F(\pi_{t},\gamma^*_{t},a^{n_t}_{t},n_{t})$ (or equivalently $\boldsymbol{\tilde{x}}_{t+1}=\tilde{F}(\boldsymbol{\tilde{x}}_{t},\gamma^*_{t},a^{n_{t}}_{t},n_{t})$) is evaluated using the equilibrium strategy.

We now provide intuition for the expressions in~\eqref{eq:fpe2b}. The first equation describes the case where a player has already bought the product so there is no additional expected reward. The second equation refers to the case where the acting player chooses to wait and so the future reward is averaged over all acting players at time $t+1$ with the beliefs being updated according to the equilibrium strategy $\gamma^*$ and the action $0$.
The third equation refers to the case where the acting player chooses to buy the product and thus it receives the expected value estimated by her private belief. Finally, the last equation refers to non-acting players who evaluate their future rewards by taking expectation over all possible acting players at the next stage, as well as the private information of the currently acting player and whether she will buy the product or not.

The domain of the value functions $V^m(\cdot)$ in FPE \ref{alg:Fixed-Point finite} is finite, with size $2\times N\times 3^{N}\times 2^N$. For practical systems with a large number of users $N$, the exponential dimension of FPE \ref{alg:Fixed-Point finite} renders the computation of the PBE infeasible. In the next section we show that using the structure of the problem, these equations can be simplified considerably, resulting in quadratic dimension in $N$. Then, the efficient computation of the PBE would allow characterizing informational cascades in large systems where the implication of a cascade can be dramatic.

\section{Computing a PBE though a quadratic-dimensional FPE}\label{sec:PBE_quadratic}

In this section, we exploit the structure of the problem to simplify FPE~\ref{alg:Fixed-Point finite}.
\optv{arxiv}{
This simplification is done in two steps. The first step results in a FPE with value functions having domain that grows polynomially with $N$, and in particular as $\sim N^4$.
However, we only present this result in Appendix~\ref{sec:PBE_polynomial} for completeness. The second step results in an even more drastic simplification with}\optv{2col}{This simplification is dramatic, since it results in} strategies and value functions having domain that grows only quadratically with $N$\optv{2col}{ as opposed to the exponential growth of FPE~\ref{alg:Fixed-Point finite}}.
The key observation here is that the indexing of the players has no effect on the future reward a player estimates she would get by waiting. Since $\boldsymbol{\widetilde{x}}$
contains this information, it can be reduced to the following two quantities:
\begin{defn}
Define the aggregated state information as
\begin{equation}
y_{t}=\sum_{n=1}^{N}\widetilde{x}_{t}^{n}\in\mathcal{Y}=\{-N,\ldots,N\}.\label{eq:y}
\end{equation}
Further, define the indicator that player $n$ has revealed her private
information as $r_{t}^{n}=|\tilde{x}_{t}^{n}|$. Using $z_{t}^{n}=\max\left\{ r_{t}^{n},b_{t}^{n}\right\} $,
define the number of players who cannot reveal their private information
after turn $t$ by
\begin{equation}
w_{t}=\sum_{n=1}^{N}z_{t}^{n}\in\mathcal{W}=\{0,\ldots,N\}.\label{eq:w}
\end{equation}
These are the players that have already revealed their private information or have already bought the product and cannot buy it again.
\end{defn}
Since the value function and strategy of players with $b^n=1$ are evidently 0 and $\gamma^*=\boldsymbol{0}$, respectively, we only argue for the players with $b^n=0$ and drop $b^n$ from the state variables. We define the functions
$U_{a}:\mathcal{X}\times\left\{ 0,1\right\}\times\mathcal{Y}\times\mathcal{W}\rightarrow\mathbb{R}$
and
$U_{na}^{\tilde{r}}:\mathcal{X}\times\left\{ 0,1\right\}\times\mathcal{Y}\times\mathcal{W} \rightarrow\mathbb{R}$
$\forall\tilde{r}\in\{0,1\}$ as follows.
$U_{a}\left(x,r,y,w\right)$ is the value function of the acting player $n$ whose private information
is $x^{n}=x$, she has revealed if $r=1$ and the aforementioned state variables are $(y_t,w_t)=(y,w)$.
Similarly, $U_{na}^{\tilde{r}}\left(x,z,y,w\right)$ is the value function of a non-acting
player $m$, whose private information is $x^{m}=x$, she has revealed if $\tilde{r}=1$ with an acting player $n$ who can reveal
her private information if $z=0$, and $y,w$ as before.

Finally, define the update functions $G^{r},G^{z},G^y,G^w$ as follows
\begin{subequations}
\begin{align}
 &G^{r}(r,\gamma)  =
 \left\{\begin{array}{ll}
 1, &  r= 0 \text{ and } \gamma = \textbf{I}\\
 r, & \text{else}
 \end{array} \right. \\
%
%
&G^{z}(z,\gamma,a)=
 \left\{\begin{array}{ll}
 1, &  z= 0 \text{ and } (a=1 \text{ or } \gamma = \textbf{I})\\
 z, & \text{else}
 \end{array}\right. \\
&G^y(z,y,\gamma,a) =
 \left\{\begin{array}{ll}
 y+(2a-1), &  z= 0 \text{ and } \gamma = \textbf{I}\\
 y, & \text{else}
 \end{array}\right.\\
 &G^w(z,w,\gamma,a) =w+G^z(z,\gamma,a)-z.
\end{align}
\end{subequations}

We now can formulate the alternative FPE~\ref{alg:Fixed-Point WY}.


\begin{algo}[Quadratic dimension]\label{alg:Fixed-Point WY}

For every $r\in\{0,1\}$, $y\in\mathcal{Y}$, $w\in\mathcal{W}$,
we evaluate $\gamma^{*}=\phi\left[r,y,w\right]$
as follows
\begin{itemize}
\begin{subequations}\label{eq:fp_quadratic}
\item $\gamma^{*}$ is the solution of
\begin{equation}
\gamma^{*}(x)=\arg\max\left\{
\underbrace{\frac{q^{y+r+x}-1}{q^{y+r+x}+1}}_{1=\text{buy}},
\underbrace{A}_{0=\text{don't buy}}
\right\}
\forall x\in\mathcal{X},
\label{eq:fp_quadratica}
\end{equation} where

\optv{2col}{
\begin{align}\label{eq:fp_quadraticb}
 A=&\frac{\delta}{N}U_{a}\left(x,r',y',w'\right)
\nonumber \\
 &+\frac{\delta}{N}(N-w-1+r)U_{na}^{r'}(x,0,y',w') \nonumber \\
 &+
  \frac{\delta}{N}\left(w-r\right)U_{na}^{r'}\left(x,1,y',w'\right).
\end{align}}
\optv{arxiv}{
	\begin{align}\label{eq:fp_quadraticb}
	A=\frac{\delta}{N}U_{a}\left(x,r',y',w'\right)&+
	\frac{\delta}{N}(N-w-1+r)U_{na}^{r'}(x,0,y',w') +
	\frac{\delta}{N}\left(w-r\right)U_{na}^{r'}\left(x,1,y',w'\right).
	\end{align}}
\optv{2col}{where the next state variables are
	\begin{align}
	&r'=G^{r}(r,\gamma^{*})\\
	   & y'=G^{y}\left(r,y,\gamma^{*},0\right)\\
	   &w'=G^{w}\left(r,w,\gamma^{*},0\right).
	\end{align}}\optv{arxiv}{where the next state variables are
	\begin{align}
	&r'=G^{r}(r,\gamma^{*})\\
	   & y'=G^{y}\left(r,y,\gamma^{*},0\right)\\
	   &w'=G^{w}\left(r,w,\gamma^{*},0\right).
	\end{align}}
The value functions satisfy
\begin{align}\label{eq:fp_quadraticc}
U_{a} & \left(x,r,y,w\right)=\left\{ \begin{array}{cc}
A & \gamma^{*}(x)=0\\
\frac{q^{y+r+x}-1}{q^{y+r+x}+1} & \gamma^{*}(x)=1
\end{array}\right.
\end{align}
and for all $\tilde{r}\in\{0,1\}$
\optv{2col}{
\begin{align}\label{eq:fp_quadraticd}
 &U_{na}^{\tilde{r}}(x,z,y,w) \nonumber \\
 &=\frac{\delta}{N}\mathbb{E}\left\{ U_{a}(x,\tilde{r},\tilde{Y},\tilde{W})
 \right\} +\frac{\delta}{N}\mathbb{E}\left\{ U_{na}^{\tilde{r}}(x,\tilde{Z},\tilde{Y},\tilde{W})
 \right\}  \nonumber \\
 &+\frac{\delta}{N}(w-z-\tilde{r})\mathbb{E}\left\{ U_{na}^{\widetilde{r}}(x,1,\tilde{Y},\tilde{W})
 \right\}+ \nonumber \\
 &\frac{\delta}{N}\left(N-w-2+z+\tilde{r}\right) 
 \mathbb{E}\left\{ U_{na}^{\tilde{r}}(x,0,\tilde{Y},\tilde{W})
 \right\},
\end{align}}
\optv{arxiv}{
	\begin{align}\label{eq:fp_quadraticd}
	U_{na}^{\tilde{r}}\left(x,z,y,w\right) &=\frac{\delta}{N}\mathbb{E}\left\{ U_{a}\left(x,\tilde{r},\tilde{Y},\tilde{W}\right)
	\right\}+\frac{\delta}{N}\mathbb{E}\left\{ U_{na}^{\tilde{r}}\left(x,\tilde{Z},\tilde{Y},\tilde{W}\right)
	\right\}  \nonumber \\
	&+\frac{\delta}{N}\left(w-z-\tilde{r}\right)\mathbb{E}\left\{ U_{na}^{\widetilde{r}}\left(x,1,\tilde{Y},\tilde{W}\right)
	\right\}  \nonumber \\ &+\frac{\delta}{N}\left(N-w-2+z+\tilde{r}\right) \mathbb{E}\left\{ U_{na}^{\tilde{r}}\left(x,0,\tilde{Y},\tilde{W}\right)
	\right\},
	\end{align}
}
where the (random) next state variables from the point of view of a non-acting player are: 
\begin{align}
& \tilde{Z}=G^{z}\left(z,\gamma^{*},\gamma^{*}\left(X^{n}\right)\right)\\
   & \tilde{Y}=G^{y}\left(z,y,\gamma^{*},\gamma^{*}\left(X^{n}\right)\right)\\
   & \tilde{W}=G^{w}\left(z,w,\gamma^{*},\gamma^{*}\left(X^{n}\right)\right)
\end{align}
and the expectation is wrt the RV $X^{n}$, where
\optv{2col}{\\
	$\Pr(X^{n}=x^{n}|\widetilde{r},x,w,y)=\frac{Q(x^{n}|-1)+Q(x^{n}|1)q^{y+r+x}}{1+q^{y+r+x}}.$}
\optv{arxiv}{
$$\Pr(X^{n}=x^{n}|\widetilde{r},x,w,y)=\frac{Q(x^{n}|-1)+Q(x^{n}|1)q^{y+r+x}}{1+q^{y+r+x}}.$$}
Specifically, for $z=1$ the above becomes
\optv{2col}{
\begin{align}\label{eq:fp_quadratice}
 &U_{na}^{\tilde{r}}\left(x,1,y,w\right)=
 \frac{\delta}{N} U_{a}\left(x,\tilde{r},y,w\right) \nonumber  \\
 &+\frac{\delta}{N}\left(w-z-\tilde{r}+1\right) U_{na}^{\widetilde{r}}\left(x,1,y,w\right) \nonumber \\
 &+\frac{\delta}{N}\left(N-w-2+z+\tilde{r}\right) U_{na}^{\tilde{r}}\left(x,0,y,w\right).
\end{align}}
\optv{arxiv}{
	\begin{equation}\label{eq:fp_quadratice}
U_{na}^{\tilde{r}}\left(x,1,y,w\right)=
	\frac{\delta}{N} U_{a}\left(x,\tilde{r},y,w\right) +\frac{\delta}{N}\left(w-z-\tilde{r}+1\right) U_{na}^{\widetilde{r}}\left(x,1,y,w\right) 
	+\frac{\delta}{N}\left(N-w-2+z+\tilde{r}\right) U_{na}^{\tilde{r}}\left(x,0,y,w\right).
	\end{equation}}
\end{subequations}%
\hfill $\blacksquare$
\end{itemize}
\end{algo}

The intuitive explanation for FPE~\ref{alg:Fixed-Point WY} is as follows.
Equation~\eqref{eq:fp_quadratica} quantifies the decision between buying now or waiting, given
the quality of information about $V$ evaluated through $y$.
Specifically, the reward-to-go for waiting in~\eqref{eq:fp_quadraticb} averages out the rewards
obtained by whether the acting player will also be acting at the next epoch (first term),
or whether she will be non-acting and the acting player can reveal her private information or not (the two terms with $z=0,1$).
Similarly, a non-acting player updates her value function in~\eqref{eq:fp_quadraticd} by averaging out four possibilities for the next epoch: whether she will be the acting player (first term),  whether she will be non-acting but the acting player will be the same as in the current epoch (second term), and whether she will be non-acting and the acting player will be some other than herself and the current acting player (last two terms). Specifically, if the current acting player has either bought the product or revealed her private information ($z=1$) the second term in this equation is absorbed into the third one as shown in~\eqref{eq:fp_quadratice}.

The next Theorem shows that by finding a solution to FPE~\ref{alg:Fixed-Point WY}, we obtain a solution
to  FPE~\ref{alg:Fixed-Point finite}. Since equations \eqref{eq:fp_quadratic}
have quadratic dimension in $N$, this significantly
reduces the complexity of solving FPE~\ref{alg:Fixed-Point finite}.
Specifically, given the solution $U^{*}$ of FPE~\ref{alg:Fixed-Point WY} (together
with $\phi$) we construct the following strategies and value functions.
\begin{align}\label{eq:mapping}
\gamma^{*}=\theta\left[n,\boldsymbol{\tilde{x}},b^n,\boldsymbol{b}^{-n}\right]=
 \begin{cases}
 \phi\left[r^{n},y,w\right],     &  b^n=0 \\
 \textbf{0},   & b^n=1
 \end{cases}
\end{align}
\optv{2col}{
\begin{align}\label{eq:quadratic_finite}
%
&\tilde{V}^{m}\left(\cdot,n,\boldsymbol{\tilde{x}},b^m,b^{-m}\right)\nonumber\\
&=\begin{cases}
U_{a}\left(\cdot,\left|\tilde{x}^{n}\right|,y,w\right), &   b^m=0, m=n \\
U_{na}^{\left|\tilde{x}^{m}\right|}\left(\cdot,\max\{|\tilde{x}^n|,b^n\},y,w\right), & b^m=0, m\neq n \\
0, & b^m=1,
\end{cases}
\end{align}}%
\optv{arxiv}{
	\begin{align}\label{eq:quadratic_finite}
	%
\tilde{V}^{m}\left(\cdot,n,\boldsymbol{\tilde{x}},b^m,\boldsymbol{b}^{-m}\right)=\begin{cases}
	U_{a}\left(\cdot,\left|\tilde{x}^{n}\right|,y,w\right), &   b^m=0, m=n \\
	U_{na}^{\left|\tilde{x}^{m}\right|}\left(\cdot,\max\{|\tilde{x}^n|,b^n\},y,w\right), & b^m=0, m\neq n \\
	0, & b^m=1,
	\end{cases}
	\end{align}
}%
where we note that $y,w$ and $r^n$ are all determined by $\boldsymbol{\tilde{x}}$ and $n$
through \eqref{eq:y} and \eqref{eq:w}. We will show that these value
functions are solutions of the original FPE~\ref{alg:Fixed-Point finite}.
\begin{thm} \label{thm:Summary of the Summary}
The value functions $(\tilde{V}^{m})_{m\in\mathcal{N}}$ in~\eqref{eq:quadratic_finite}
together with the strategy mapping $\gamma^{*}$  in~\eqref{eq:mapping}
satisfy FPE~\ref{alg:Fixed-Point finite}.
\end{thm}
\begin{IEEEproof}
See Appendix~\ref{thm:Summary of the Summary_proof}.
\end{IEEEproof}

\section{Equilibrium Analysis}\label{sec:existence}
The convenient form of FPE~\ref{alg:Fixed-Point WY} allows us to analyze properties of the PBE and even to verify intuitive PBE solutions.  We first present an intermediate lemma which will be useful in proving subsequent results.

\begin{lem}\label{thm:totexp}
	The following are true for all solutions of FPE~\ref{alg:Fixed-Point WY}.
	\begin{itemize}
		\item 	For $\delta=1$, players with $x=1$ are indifferent between buying and waiting for $y\geq -1$ and all $w$ and $r$. Furthermore, players with $x=-1$ are indifferent between buying and waiting for $y+w\geq N$ and all $r$.
		\item For $\delta<1$,  players with $x=1$ prefer buying over waiting for $y\geq 0$ and all $w$ and $r$ and for $y=-1$, $r=1$, and are indifferent between buying and waiting for $y = -1$, $r=0$. Also,  players with $x=-1$  prefer buying over waiting for $y\geq 2$, $y+w\geq N$ and all $r$ and also for $y=1$, $w\geq N-1$ and $r=1$, and are indifferent for $y=1$, $r=0$ and $y=0$,  $r=1$.
		\item  For all $\delta\leq 1$, for $y<-1$, all $w$ and $r=0$ and for both values of $x$, players prefer to wait. Similarly, for $y<-2$, all $w$ and $r=1$ and both values of $x$, players prefer to wait. Finally, For $y=-2$, all $w$ and $r=1$ players with $x=1$ are indifferent between buying and waiting and players with $x=-1$ prefer to wait.
	\end{itemize}
	
\end{lem}

\begin{IEEEproof}
\optv{2col}{
The proof is omitted due to space limitations. It can be found in~\cite{BiHeAn19arxiv}.
}
\optv{arxiv}{
See Appendix~\ref{thm:totexp_proof}
}
\end{IEEEproof}

We comment at this point that the usefulness of the above lemma is the very fact that these statements are proved without explicitly solving FPE~\ref{alg:Fixed-Point WY}. Specifically, note that both the right hand side and the left hand side of equation \eqref{eq:fp_quadratica} depend on the solution $\gamma^*$. However, Lemma~\ref{thm:totexp} claims that for all of the solutions $\gamma^*$, the aforementioned properties hold.
Most of the remaining results of this section hinge on the above lemma.

The next theorem presents a solution of FPE	\ref{alg:Fixed-Point WY} for all values of $\delta\leq 1$ including $\delta=0$. For $\delta=0$, our scenario coincides with the original myopic scenario from \cite{Bikhchandani1992}, up to the fact that players who do not buy get another opportunity to play. Therefore, we refer to the strategy profile in Theorem \ref{thm:existence} as the myopic solution, even though it is a solution for all $\delta$.

	\begin{thm}[Existence] \label{thm:existence}
		The following strategy profile is a solution of FPE \ref{alg:Fixed-Point WY} for all $\delta$.\\
		For $r=0$, and all $w$,
\begin{subequations}
        \begin{equation}
        \gamma^*=\phi[r,y,w]=
            \begin{cases}
                \boldsymbol{1}, & y\geq 2 \\
                \boldsymbol{0}, & y \leq -2\\
                \boldsymbol{I}, & -1  \leq y \leq 1.
            \end{cases}
        \end{equation}
		For $r=1$, and all $w$,
        \begin{equation}
        \gamma^*=\phi[r,y,w]=
            \begin{cases}
                \boldsymbol{1}, & y\geq 2\\
                \boldsymbol{0}, & y\leq -2\\
                \boldsymbol{I}, & -1\leq y\leq 0. 
            \end{cases}
        \end{equation}
\end{subequations}
        Finally, for $r=1$, $y=1$, and all $w$, $\gamma^*=\phi[r,y,w]$ can be chosen appropriately, as a function of $\delta$ (it is either $\boldsymbol{1}$ or $\boldsymbol{I}$).
	\end{thm}
\begin{IEEEproof}
See Appendix~\ref{thm:existence_proof}.
\end{IEEEproof}

\begin{figure}[htbp]
	\begin{center}
		\includegraphics[width=0.49\columnwidth]{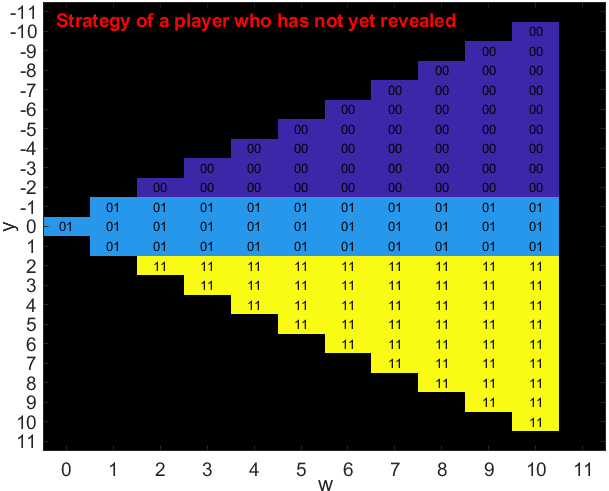}
		\includegraphics[width=0.49\columnwidth]{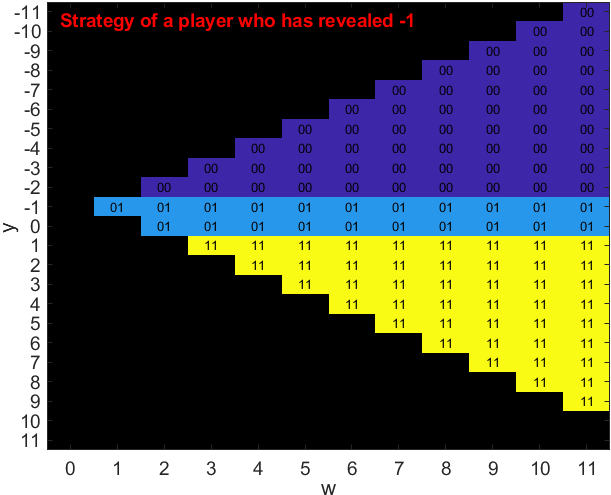}
	\end{center}
	\caption{Equilibrium strategies for $N=11$ and all $\delta\leq 1$, including $\delta=0$. ``00'', ``01'', and ``11'' denote strategies $\boldsymbol{0}$, $\boldsymbol{I}$, and $\boldsymbol{1}$, respectively. The strategies for $r=1$ and $y=1$ are specifically for $\delta=0$.}
	\label{fig:myopic1}
\end{figure}

This strategy profile is depicted in Fig.~\ref{fig:myopic1} (in all such figures we present the case for $r=0$ and the case for $r=1$ with $\tilde{x}=-1$, since if a player has revealed and her private information is 1 it means that she has bought the product already). Notice that it mostly consists of strategies $\gamma^*=\boldsymbol{1}$ and  $\gamma^*=\boldsymbol{0}$ which implies that players do not tend to reveal their private signal. Intuitively, if a player knows that others do not reveal their private signal, she does not gain from waiting for more information. Hence, revelation of private signals, which occur when $\gamma^*=\boldsymbol{I}$ is played, does not happen when both players  with $x=1$ and $x=-1$ have positive instantaneous reward. Therefore, for all values of $\delta$, acting myopically is always an equilibrium.

Although the strategy profile of Theorem \ref{thm:existence} is referred to as myopic, it captures the non-myopic aspect of the game too. For instance, at $y=1$, a player with $r=0$ and $x=-1$, does not buy the product because her value function is positive by not buying and therefore, she gains from waiting. But in the myopic case, her valuation is 0 for both buying and not buying so the player is indifferent between playing $a=1$ and $a=0$. This implies that if we change the apriori belief about $V$ from $Q(v=1)=0.5$ to $Q(v=1)=0.5+\epsilon$ for small enough $\epsilon$, this strategy profile is an equilibrium for $\delta\neq 0$ but not for $\delta=0$. This follows since a player with $x=-1$ at $y=1$ strictly prefers to buy at the myopic setting, while she still prefers to wait at the non-myopic setting if  $\epsilon$ is small enough.

We next investigate a solution for FPE~\ref{alg:Fixed-Point WY} for both $\delta=1$ and large enough $\delta<1$.

\begin{thm}\label{thm:nocas_st}
	The following strategy profile is a solution of FPE \ref{alg:Fixed-Point WY},
\begin{subequations}
	\begin{itemize}
		\item  For $\delta=1$,
\begin{equation}
  \gamma^*=\phi[r,y,w]=
        \begin{cases}
            \boldsymbol{0}, & y\leq-2 \\
            \boldsymbol{I}, & y\geq -1, w<N \\
            \boldsymbol{1}, & y\geq 1, w=N, r=1 \\
            \boldsymbol{I}, & y\in\{ 0,-1\}, w=N, r=1
        \end{cases}
\end{equation}

		\item For large enough $\delta<1$ (which depends on $N$ and other parameters of the game),
\begin{equation} \label{eq:fixedN_largedelta}
        \gamma^*=\phi[r,y,w]=
        \begin{cases}
            \boldsymbol{0}, & y\leq-2 \\
            \boldsymbol{I}, & y\geq -1, y+w<N \\
            \boldsymbol{I}, & y=1, w=N-1, r=0 \\
            \boldsymbol{I}, & y=0, w=N, r=1 \\
            \boldsymbol{1}, & y\geq 2, y+w\geq N \\
            \boldsymbol{1}, & y=1, w\geq N-1, r=1
        \end{cases}
\end{equation}
	\end{itemize}
\end{subequations}

\end{thm}

\begin{IEEEproof}
	\optv{2col}{
		The proof is omitted due to space limitations. It can be found in~\cite{BiHeAn19arxiv}.
	}
	\optv{arxiv}{
		See Appendix \ref{thm:nocas_st_proof}
	}
\end{IEEEproof}

\begin{figure}[htbp]
	\begin{center}
		\includegraphics[width=0.49\columnwidth]{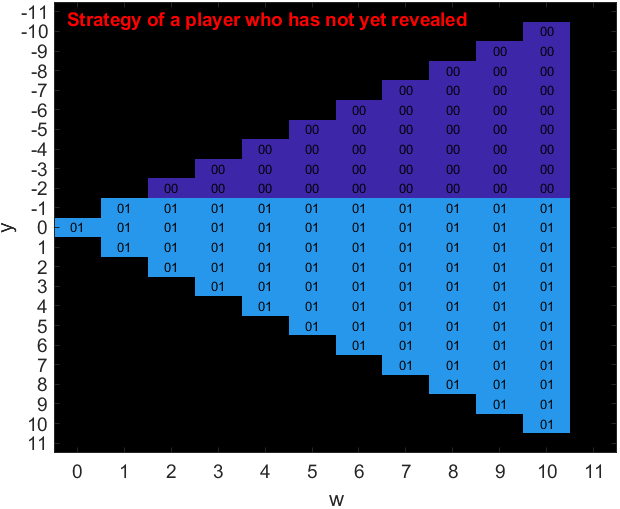}
		\includegraphics[width=0.49\columnwidth]{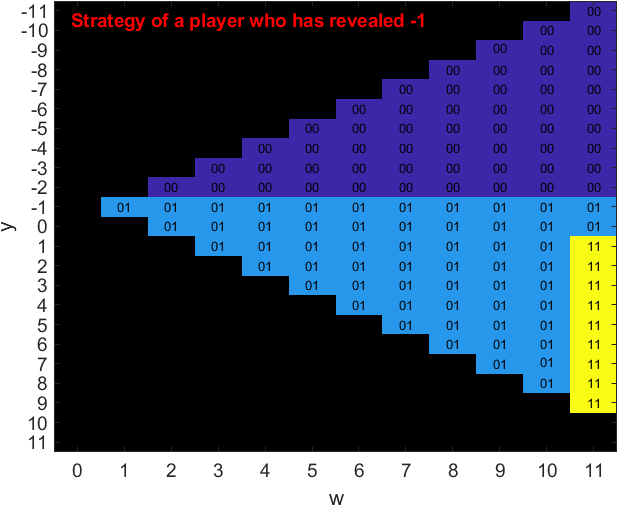}
	\end{center}
	\caption{Equilibrium strategies for $N=11$ and $\delta=1$. ``00'', ``01'', and ``11'' denote strategies $\boldsymbol{0}$, $\boldsymbol{I}$, and $\boldsymbol{1}$, respectively.}
	\label{fig:nocas1}
\end{figure}

\begin{figure}[htbp]
	\begin{center}
		\includegraphics[width=0.49\columnwidth]{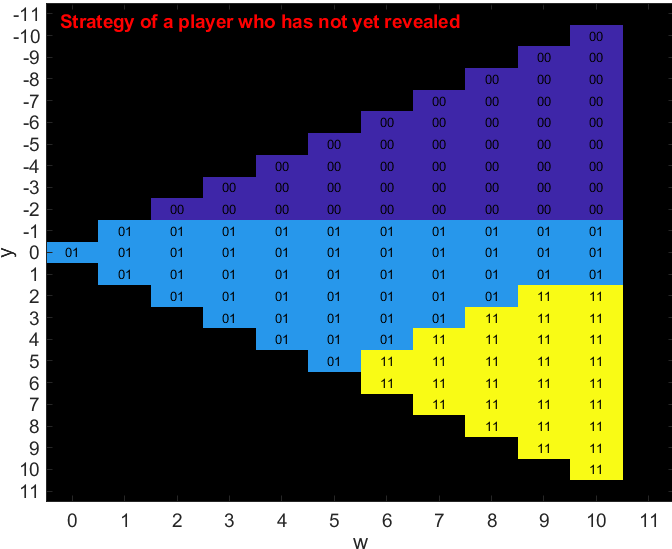}
		\includegraphics[width=0.49\columnwidth]{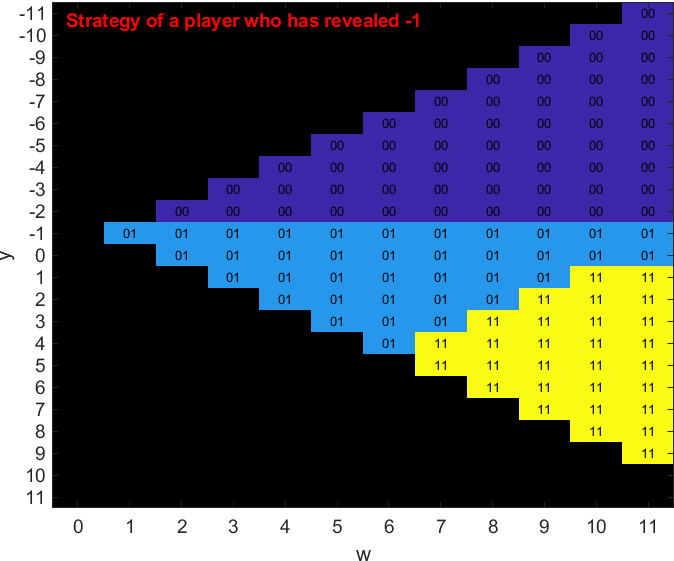}
	\end{center}
	\caption{Equilibrium strategies for $N=11$ and large enough $\delta<1$. ``00'', ``01'', and ``11'' denote strategies $\boldsymbol{0}$, $\boldsymbol{I}$, and $\boldsymbol{1}$, respectively.}
	\label{fig:nocas}
\end{figure}

The strategy profiles presented in Theorem \ref{thm:nocas_st} are depicted in Fig. \ref{fig:nocas1} and \ref{fig:nocas} for $N=11$ and $\delta=1$ and large enough $\delta<1$, respectively. Note that the strategy $\gamma^*=\boldsymbol{I}$ (denoted by $01$) is extended throughout all the states with $y\geq -1$ and $w<N$ for $\delta=1$.

\optv{arxiv}{
The FPE~\ref{alg:Fixed-Point WY} may exhibit more PBE than the PBE of Theorem \ref{thm:existence}. Nevertheless, all these potential PBE share similar structure, as the next theorem shows. We have also presented the existence results for the solutions that are threshold policies wrt $w$ and $y$ in the next theorem.

\begin{thm} \label{thm:wy:behavior}
		The following properties hold for the solutions of FPE \ref{alg:Fixed-Point WY} for $b=0$:
		\begin{itemize}
			\item All of the solutions of FPE \ref{alg:Fixed-Point WY} are either threshold policies (from $0$ to $1$) wrt $w$ or there exists a threshold policy wrt $w$ corresponding to a solution that is not of this type.
			\item For $\delta<1$, all of the solutions of FPE \ref{alg:Fixed-Point WY} that are threshold functions wrt $w$,  must be threshold  functions wrt $y$ for $r=0$, when all other parameters are fixed. This implies that if $\gamma^*(x)=\phi[0,y,w](x)=1$, then $\gamma^*(x)=\phi[0,y',w'](x)=1$ for $y'\geq y$ and $w'\geq w$. Further whenever the solution is threshold policy wrt $y$ for $r=0$, the solutions can also be threshold policy wrt $y$ for $r=1$.
		\end{itemize}
		Further, for all of the solutions of FPE \ref{alg:Fixed-Point WY}, we have the following properties:
		\begin{itemize}
			\item They are threshold functions wrt $y$ for $x=1$ and $r=0$, and the threshold is either $y=-1$ or $y=0$ for all $w$.
			
			\item They are such that $\gamma^*=\phi[r,y,w]=\boldsymbol{0}$ for $y \leq -3$ and all other parameters, and for $y=-2$ and $r=0$. Also, $\gamma^*=\phi[0,y,w]\neq\boldsymbol{0}$ for $y \geq 0$.
			
			\item We have $\gamma^*=\phi[0,0,w]=\boldsymbol{I}$.
			
			\item For $y\neq -1$, $\gamma^*=\phi[0,y,w]=\boldsymbol{0}$ for all $w$ (constant wrt $w$) or can only be either $\gamma^*=\phi[0,y,w]=\boldsymbol{I}$ or $\gamma^*=\phi[0,y,w]=\boldsymbol{1}$ for all $w$. It implies that by changing $w$ and fixing other parameters, the equilibrium strategies either do not change and are always $\boldsymbol{0}$, or they can change between $\boldsymbol{I}$ and $\boldsymbol{1}$.
			
			\item For $y=-1$, both $\gamma^*=\phi[0,-1,w]=\boldsymbol{I}$ and $\gamma^*=\phi[0,-1,w]=\boldsymbol{0}$ are always solutions for all $w$.
			
			\item For $y=-2$, both  $\gamma^*=\phi[1,-2,w]=\boldsymbol{I}$ and $\gamma^*=\phi[1,-2,w]=\boldsymbol{0}$ are always solutions for all $w$.
		\end{itemize}
\end{thm}
\begin{IEEEproof}
\optv{2col}{
The proof is omitted due to space limitations. It can be found in~\cite{BiHeAn19arxiv}.
}
\optv{arxiv}{
See Appendix~\ref{thm:wy:behavior_proof}.
}
\end{IEEEproof}

	The first two parts of this theorem imply that there exist solutions of FPE  \ref{alg:Fixed-Point WY} that by increasing $y$ or $w$,  the equilibrium strategies change from $\boldsymbol{0}$'s to $\boldsymbol{I}$'s and then to $\boldsymbol{1}$'s. This is evident in all of the solutions that we have proposed in this paper (Fig.~\ref{fig:myopic1}, \ref{fig:nocas1}, \ref{fig:nocas}). Other parts present more general statements about  the solutions. For instance, as we can see in the proposed solutions, the equilibrium strategies are $\gamma^*=\phi[r,y,w]=\boldsymbol{0}$ for $y\leq -2$, which is because the instantaneous reward of players is non-positive. One can also verify other parts of the theorem by the solutions proposed in this paper.

}

The boundary of $|y|=2$ is of special importance for the equilibria. The reason is that $y=2$ is the smallest $y$ for which the instantaneous reward is positive for all players, regardless of their private information. Similarly, $y=-2$ is the largest $y$ for which the instantaneous reward is negative for all players regardless of their private information. For the myopic scenario, these facts determine the equilibrium strategies at $y\geq 2$ and $y \leq -2$, and this is a possible PBE in our non-myopic scenario as well, as Theorem \ref{thm:existence} shows. However, in our non-myopic scenario, more intricate behaviors are also possible at equilibrium. For $y\leq -2$, waiting, which gives zero reward, is always better than buying. Therefore this side of the boundary behaves like the myopic scenario in all PBE. Nevertheless, for $y\geq 2$, players may choose to wait and not buy the product even if their instantaneous reward is positive. Therefore, we may have signaling strategies for $y\geq 2$ and hence, we observe different equilibrium strategies for these values of $y$ in Theorem \ref{thm:nocas_st}.

\section{Informational Cascades}\label{sec:cascades}

Our results from the previous section allow us
to evaluate PBE of the game by solving equations with a quadratic number (in $N$) of variables.
This methodology provides us with the necessary tools to investigate whether informational cascades occur in settings with large number of players.
\begin{defn}
An informational cascade is a sequence of turns in our game, starting from some $t_{0}\geq0$, such that  $\gamma_{t}\neq\boldsymbol{I}$ for all $t\geq t_{0}$. We say that an informational cascade is bad if it leads to the wrong decision:  users choose $\gamma = \boldsymbol{0}$ when $V=1$ or $\gamma = \boldsymbol{1}$ when $V=0$. 
\end{defn}

While the sequence of events in a realization of the game is random, given a PBE, we can identify the histories of the game at which an informational cascade occurs. Using Theorem \ref{thm:Summary of the Summary}, we can characterize these histories using only $w$ and $y$.

According to the definition above, an informational cascade can affect any number of players, from 1 to $N$. Obviously, informational cascades that affect more players are more significant. A natural question is then how much damage a bad informational cascade causes to the network. Our FPE \ref{alg:Fixed-Point WY} with its variables $(y,w)$ gives an easy way to tackle this question. If the cascade occurred at state $(y,w)$, then two things affect the damage done to the community: the probability that the cascade is bad, and the number of players that received the worst possible reward if the cascade is bad. Interestingly enough, both of these numbers are characterized by $w$. 

The players that participate in a bad cascade with $\gamma=\boldsymbol{0}$ receive 0 reward, which is the worst possible. The best reward is 1 up to the discounting in the first turn they get to act. The number of players that made the right decision and bought the product before the cascade is bounded from above by $w$.

The players that participate in a bad cascade with $\gamma=\boldsymbol{1}$ receive a reward of -1 up to the discounting in the first turn they get to act. This is the worst reward possible, while the best reward is 0. The number of players that initially made the right decision not to buy the product is bounded from above by $w$. Indeed, any such player must have played $\gamma=\boldsymbol{I}$ since otherwise, she would have started a $\gamma=\boldsymbol{0}$ cascade instead.

We conclude that in any bad cascade, at least $N-w$ players receive the worst reward possible. Hence, both the probability for a bad cascade and the damage it causes decrease with $w$. Using $w$, one can bound the system performance using a metric of choice (e.g., social welfare, or some notion of fairness). In the next section, we numerically evaluate the probability for a bad cascade as a function of $w$.


A direct consequence of our model, which induces players to be forward-looking instead of acting myopically,
is a multitude of equilibrium behaviours for the players.
This rich spectrum of behaviours includes the myopic strategies that have been reported in the literature and that lead to informational cascades, but also--and more importantly--includes more cooperative strategies that induce players to reveal their information with the potential of alleviating or even eliminating informational cascades.
The next two subsections explore these two extremes by proving conclusively the above claims.

\subsection{The case of $\delta<1$ and $N\rightarrow\infty$}

In this part, we employ the results of the previous section to conclude
that an informational cascade indeed happens with probability approaching 1 as the number of players approaches infinity even in a non-myopic scenario for a fixed $\delta<1$.

Our methodology consists of defining a Markov chain and studying its properties.
Specifically this Markov chain is not defined on absolute time $t$,
but on the random times when a new revelation happens (i.e., when a player plays strategy $\gamma_t=\boldsymbol{I}$).
Towards this goal we provide the following definition.

\begin{defn}
	Let $\phi[\cdot]$ be a solution to FPE~\ref{alg:Fixed-Point WY}. Define
	the random variables $(D_t)_{t\geq 0}$ with realization
	\begin{equation}
	d_{t}=\Biggl\{\begin{array}{cc}
	1, & \phi[r^{n_t}_{t},y_{t},w_{t}]=\boldsymbol{I}\text{ and }r^{n_t}_{t}=0 \text{ and } b^{n_t}_t=0\\
	0, & \text{else},
	\end{array}\label{eq:29}
	\end{equation}
which indicates if the player who acts at turn $t$ reveals her private information. 
Let $Y_t$ be the random aggregated state information at time $t$ (see \eqref{eq:y}). Let $T_i$ be the random time of the $i$-th revealing, so
$T_0=0$ and $T_i=\min \{t>T_{i-1}| D_t=1\}$ for $i\geq 1$.
	We also define the random process $(\bar{Y}_i)_{i\geq 0}$ with $\bar{Y}_i=Y_{T_i}$ when $T_i<\infty$ and $\bar{Y}_{i}=\bar{Y}_{i-1}$ otherwise.
\end{defn}
The next lemma characterizes the reason why cascades still occur in
a non-myopic scenario. 
\begin{lem}
	\label{lem:y Markov}
	Let $\phi[\cdot]$ be a solution to FPE \ref{alg:Fixed-Point WY}.
	The induced process $( \bar{Y}_i )_{i\geq 0}$ is a Markov
	chain where, for large enough $N$, there exist absorbing states $y_{R},y_{L}$ such that for all $y_{L}<y<y_{R}$,
	if $T_{i+1}<\infty$ then
	\begin{equation}
	\Pr\left(\bar{Y}_{i+1}=y'|\bar{Y}_i=y\right)=\left\{ \begin{array}{cc}
	\frac{p+\left(1-p\right)q^{y}}{q^{y}+1} & y'=y+1\\
	\frac{1-p+pq^{y}}{q^{y}+1} & y'=y-1
	\end{array}.\right.\label{eq:30}
	\end{equation}
\end{lem}
\begin{IEEEproof}
	First we show the Markovianity of $(\bar{Y}_i)_{i\geq 0}$
	\begin{subequations}
		\optv{2col}{
		\begin{align}
		&\Pr\left(\bar{Y}_{i+1}=y'|\bar{Y}_{0:i}=y_{0:i}\right) \nonumber \\
		&=\Pr\left(Y_{T_{i}}+X^{N_{T_{i}}}=y'\,|\,Y_{T_{0:i}}=y_{0:i}\right)\\
		&=\Pr\left(X^{N_{T_{i}}}=y'-y_{i}|Y_{T_{0:i}}=y_{0:i}\right)\\
		&=\frac{Q(y'-y_{i}|0)+Q(y'-y_{i}|1)q^{y_{i}}}{q^{y_{i}}+1}\\
		&=\Pr\left(\bar{Y}_{i+1}=y'|\bar{Y}_i=y_{i}\right).\label{eq:38}
		\end{align}}
		\optv{arxiv}{
		\begin{align}
		\Pr\left(\bar{Y}_{i+1}=y'|\bar{Y}_{0:i}=y_{0:i}\right) &=\Pr\left(Y_{T_{i}}+X^{N_{T_{i}}}=y'\,|\,Y_{T_{0:i}}=y_{0:i}\right)\\
		&=\Pr\left(X^{N_{T_{i}}}=y'-y_{i}|Y_{T_{0:i}}=y_{0:i}\right)\\
		&=\frac{Q(y'-y_{i}|0)+Q(y'-y_{i}|1)q^{y_{i}}}{q^{y_{i}}+1}\\
		&=\Pr\left(\bar{Y}_{i+1}=y'|\bar{Y}_i=y_{i}\right).\label{eq:38}
		\end{align}}
	\end{subequations}
	Now we characterize the absorbing states. For $\delta<1$ and $Y_{\max}=\left\lceil 1+\log_{q}(\frac{1+\delta}{1-\delta})\right\rceil<N$,
	we have
	\begin{equation}
	\frac{q^{Y_{\max}+r_t+x}-1}{q^{Y_{\max}+r_t+x}+1}>\delta>\delta U_{a}\left(x,r_{t+1},y_{t+1},w_{t+1}\right).\label{eq:39}
	\end{equation}
	So either $y_{R}=Y_{\max}$ is absorbing or there exists a $y_{R}<Y_{\max}$
	that is absorbing. In $Y_{t}=y_{R}$, all players, regardless of $x$,
	prefer to buy. Similarly, for $Y_{\min}=-2$ we have
	\begin{equation}
	\frac{q^{-1}-1}{q^{-1}+1}=2p-1<0<\delta U_{a}\left(x,r_{t+1},y_{t+1},w_{t+1}\right)\label{eq:40}
	\end{equation}
	So either $y_{L}=Y_{\min}=-2$ or $y_{L}=-1$ is absorbing. In $Y_{t}=y_{L}$,
	all players, regardless of $x$, prefer to wait. Hence, in $Y_{t}=y_{L}$
	or $Y_{t}=y_{R}$ no more revealings occur and $Y_{t}$ (and $\bar{Y}_i$)
	remains constant for all $t'>t$ with probability 1.
\end{IEEEproof}
The absorbing states of the Markov chain we defined above are informational
cascades. As a result, an informational cascade will occur with probability approaching 1 as $N$ increases as in the gambler's ruin problem.
However, an informational cascade that occurs after (almost) all players have revealed their private information is of
little concern. In such a case (almost) all available information about $V$ has been revealed, so $w$ is close to $N$ and the cascade affects only a few players and also has small probability to be bad. Unfortunately, for a fixed $\delta<1$, the following theorem shows that this is far from being the case, as an informational cascade occurs early on:
\begin{thm} \label{thm:Cascades}
	For $\delta<1$, the probability that an informational cascade occurs
	in finite time approaches 1 as $N\rightarrow\infty$.
	
	Furthermore, let $M_{N}$ be a sequence such that $\lim_{N\rightarrow\infty}\frac{M_{N}}{\sqrt{N}}=0$
	and $\lim_{N\rightarrow\infty}M_{N}=\infty$.
	\begin{enumerate}
		\item The probability that less than $M_{N}$ players have revealed their
		private information before the cascade occurred approaches 1 as $N\rightarrow\infty$.
		\item If, in addition, the solution is such that $\phi\left[r,y,w\right]=\boldsymbol{1}$
		implies $\phi\left[r,y,\hat{w}\right]=\boldsymbol{1}$ for all $\hat{w}>w$ (according to \optv{2col}{\cite[Th. 5]{BiHeAn19arxiv}}\optv{arxiv}{Theorem {\ref{thm:wy:behavior}}}, we know such solutions exist),
		then the cascade happens in less than $M_{N}$ turns with a probability
		that approaches 1 as $N\rightarrow\infty$.
	\end{enumerate}
\end{thm}
\begin{IEEEproof}
	See Appendix~\ref{thm:Cascades_proof}.
\end{IEEEproof}
Theorem \ref{thm:Cascades} implies that an informational cascade will occur at some finite time with probability approaching 1 as $N$ increases.
Secondly, the theorem implies that the cascade happens too early. This follows since when a cascade occurs, with high probability, less than $M_N$ players have revealed their information for any increasing sequence that grows slower than $\sqrt{N}$ (e.g., $M_N=\log N$). Hence, a minuscule amount of the available information about $V$ has been revealed before a cascade occurs (for large $N$). This is undesirable, since it means that the probability for a bad cascade can be significant, and that the cascade will affect almost all of the players.

\subsection{The case of $\delta=1$ or large enough $\delta<1$ and finite $N$}

In this subsection, we study informational cascades for a fixed $N$ and for either $\delta=1$ or large enough $\delta<1$.
We refer to these cases as infinitely patient and sufficiently patient players, respectively.
As it will be shown, a very surprising result emerges in this setting.
 For $\delta=1$ and $V=-1$, there exists a PBE that completely avoids bad information cascades. For $V=-1$ and with large enough $\delta<1$,  there exists a PBE that has a vanishing probability (in $N$) for a bad information cascade, since it is guaranteed that at least half of the players will reveal their private information. The next two theorems formalize these results.

	
\begin{thm}
For $\delta=1$, there exists a PBE in which there is no bad informational cascade for $V=-1$.
\label{thm:infocas1}
\end{thm}
\begin{IEEEproof}
Consider the strategy profile of Theorem \ref{thm:nocas_st} for $\delta=1$ (depicted in Fig.~\ref{fig:nocas1}). There is no strategy $\gamma^*=\phi[r=0,y,w]=\boldsymbol{1}$. This means that for $V=-1$, bad informational cascades never happen for this strategy profile.
\end{IEEEproof}	
Although Theorem~\ref{thm:infocas1} states that bad informational cascades can be avoided for $V=-1$, they will always happen for $V=1$ with positive probability due to the strategies $\gamma^*=\phi[r=0,y,w]=\boldsymbol{0}$ that are played for $y\leq -2$ and all $w$.
\begin{thm}
For sufficiently large $\delta<1$ (which depends on $N$) there exists a PBE for which bad informational cascades for $V=-1$ happen only when at least half of the players have revealed their private information. Consequently, in this PBE, the probability that a bad informational cascade for $V=-1$ happens is bounded from above by $e^{-{\frac{N}{4}(1-2p)^{2}}}$.
\label{thm:infocas}
\end{thm}
\begin{IEEEproof}
Assume that $\delta<1$ is large enough such that the strategy profile of the second part of Theorem~\ref{thm:nocas_st} (depicted in Fig.~\ref{fig:nocas}) is a PBE. This strategy profile consists of strategies $\gamma^*=\phi[0,y,w]=\boldsymbol{1}$ for $y\geq 2$ and $y+w\geq N$ (yellow cells in Fig.~\ref{fig:nocas}). This implies that for $V=-1$, a bad informational cascade happens only when $y\geq 2$ and $y+w\geq N$. This in turn means that a bad informational cascade happens when at least $w=\frac{N}{2}$. Since the initial value of $y$ is 0 and the strategies played before reaching $y\geq 2$ and $y+w\geq N$, are all $\gamma^*=\phi[r,y,w]=\boldsymbol{I}$, then $w$ is equal to the number of players who have revealed. Therefore, a bad cascade can only happen after at least $\frac{N}{2}$ players have revealed their private information.

\optv{2col}{
Let $\mathcal{T}$ be the set of turns when players revealed their private information throughout the game. Let $R=\left|\mathcal{T}\right|$. Let $Y_{\infty}$ be the random value of $y$ when an information cascade occurs, such that $Y_{\infty}=\infty$ if it does not occur. Let $E$ be the error event, in which a bad information cascade happens. 
Then, using that $p<\frac{1}{2}$ we have
\begin{multline}
\Pr\left(E\right)\overset{(a)}{\leq}\Pr(Y_{\infty}=\sum_{t\in \mathcal{T}}x^{n_{t}}\leq0\mid V=1)\Pr\left(V=1\right)\\+\Pr(Y_{\infty}=\sum_{t\in \mathcal{T}}x^{n_{t}}\geq0\mid V=-1)\Pr\left(V=-1\right)\\
=\frac{1}{2}\Pr(\sum_{t\in \mathcal{T}}x^{n_{t}}-\left(1-2p\right)R\leq-\left(1-2p\right)R\mid V=1)\\
+\frac{1}{2}\Pr(\sum_{t\in \mathcal{T}}x^{n_{t}}+\left(1-2p\right)R\geq\left(1-2p\right)R\mid V=-1)\\
\overset{(b)}{\leq}e^{-\frac{R}{2}\left(1-2p\right)^{2}}\overset{(c)}{\leq}e^{-\frac{N}{4}\left(1-2p\right)^{2}}\end{multline}
where (a) follows since a bad information cascade can only occur if $VY_{\infty}$ is non-positive, (b) is Hoeffding's inequality for bounded random variables, and (c) uses that $R\geq \frac{N}{2}$.
}%
\optv{arxiv}{
Let $\mathcal{T}$ be the set of turns when players revealed their private information throughout the game. Let $R=\left|\mathcal{T}\right|$. Let $Y_{\infty}$ be the random value of $y$ when an information cascade occurs, such that $Y_{\infty}=\infty$ if it does not occur. Let $E$ be the error event, in which a bad information cascade happens. 
Then, using that $p<\frac{1}{2}$ we have
\begin{multline}
\Pr\left(E\right)\overset{(a)}{\leq}\Pr(Y_{\infty}=\sum_{t\in \mathcal{T}}x^{n_{t}}\leq0\mid V=1)\Pr\left(V=1\right)+\Pr(Y_{\infty}=\sum_{t\in \mathcal{T}}x^{n_{t}}\geq0\mid V=-1)\Pr\left(V=-1\right)\\
=\frac{1}{2}\Pr(\sum_{t\in \mathcal{T}}x^{n_{t}}-\left(1-2p\right)R\leq-\left(1-2p\right)R\mid V=1)
+\frac{1}{2}\Pr(\sum_{t\in \mathcal{T}}x^{n_{t}}+\left(1-2p\right)R\geq\left(1-2p\right)R\mid V=-1)\\
\overset{(b)}{\leq}e^{-\frac{R}{2}\left(1-2p\right)^{2}}\overset{(c)}{\leq}e^{-\frac{N}{4}\left(1-2p\right)^{2}}\end{multline}
where (a) follows since a bad information cascade can only occur if $VY_{\infty}$ is non-positive, (b) is Hoeffding's inequality for bounded random variables, and (c) uses that $R\geq \frac{N}{2}$.
}
\end{IEEEproof}

\section{Numerical Results}\label{sec:numerical}

In this section, we present numerical results for the solution of FPE~\ref{alg:Fixed-Point WY}.
The results were obtained as follows. First an iterative algorithm was used to solve the FPE, much like the value iteration algorithm used in the solution of Markov Decision Processes. The iterative process was run until the value functions converged numerically. In order to verify without a doubt that this solution is an equilibrium, a second step was followed.
At the second step, the equilibrium strategy obtained by this iterative process was fixed and a linear system of equations was formulated with the unknowns being all value functions. This system was solved using infinite precision arithmetic (through rational number representation) and the exact value functions were obtained corresponding to this strategy profile.
The final step involved checking if sequential rationality is satisfied for the obtained value functions, i.e., if all inequalities in~\eqref{eq:fp_quadratic} are satisfied.

\begin{figure}[htbp]
\begin{center}
\includegraphics[width=0.49\columnwidth]{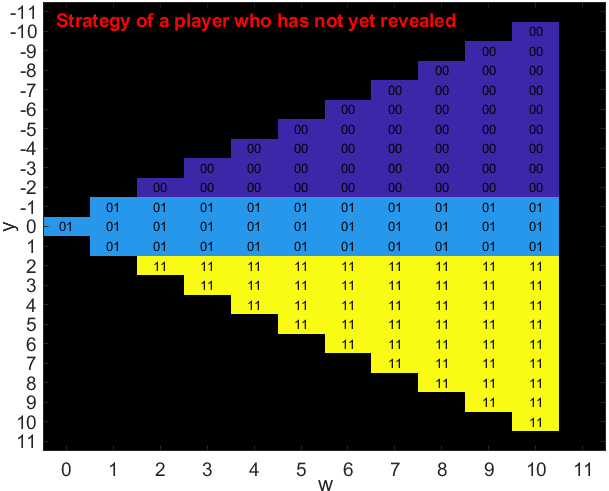}
\includegraphics[width=0.49\columnwidth]{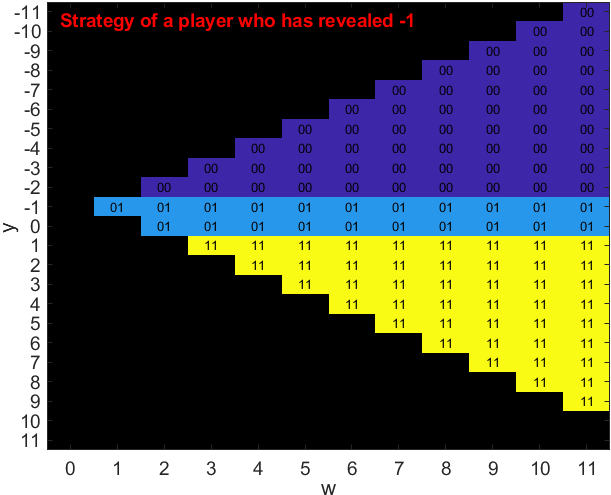}
\end{center}
\caption{Equilibrium strategies for $N=11$, $p=0.1$, $\delta=0$. ``00'', ``01'', and ``11'' denote strategies $\boldsymbol{0}$, $\boldsymbol{I}$, and $\boldsymbol{1}$, respectively.}
\label{fig:myopic}
\end{figure}

In the following we present results for $N=11$, $p=0.1$ and three different cases: $\delta=0$, $\delta=0.999$, and $\delta=1$.
The first case ($\delta=0$) is essentially the case of myopic players and the results in Fig.~\ref{fig:myopic} confirm the ones in~\cite{Bikhchandani1992}. Regardless of the value of $w$, players who have not yet revealed their information, wait for $y\leq -2$, buy for $y\geq 2$ and reveal their information for $-1\leq y \leq 1$. Note that for $y=1$ a non-revealing player is indifferent between $\gamma=\boldsymbol{I}$ and $\gamma=\boldsymbol{1}$, and similarly for $y=-1$.
We resolve the tie by assuming that the player always reveals. In addition, for $y=0$ a player who has already revealed is indifferent between any action, and we resolve this ambiguity by assuming that she always reveals.

\begin{figure}[htbp]
\begin{center}
\includegraphics[width=0.49\columnwidth]{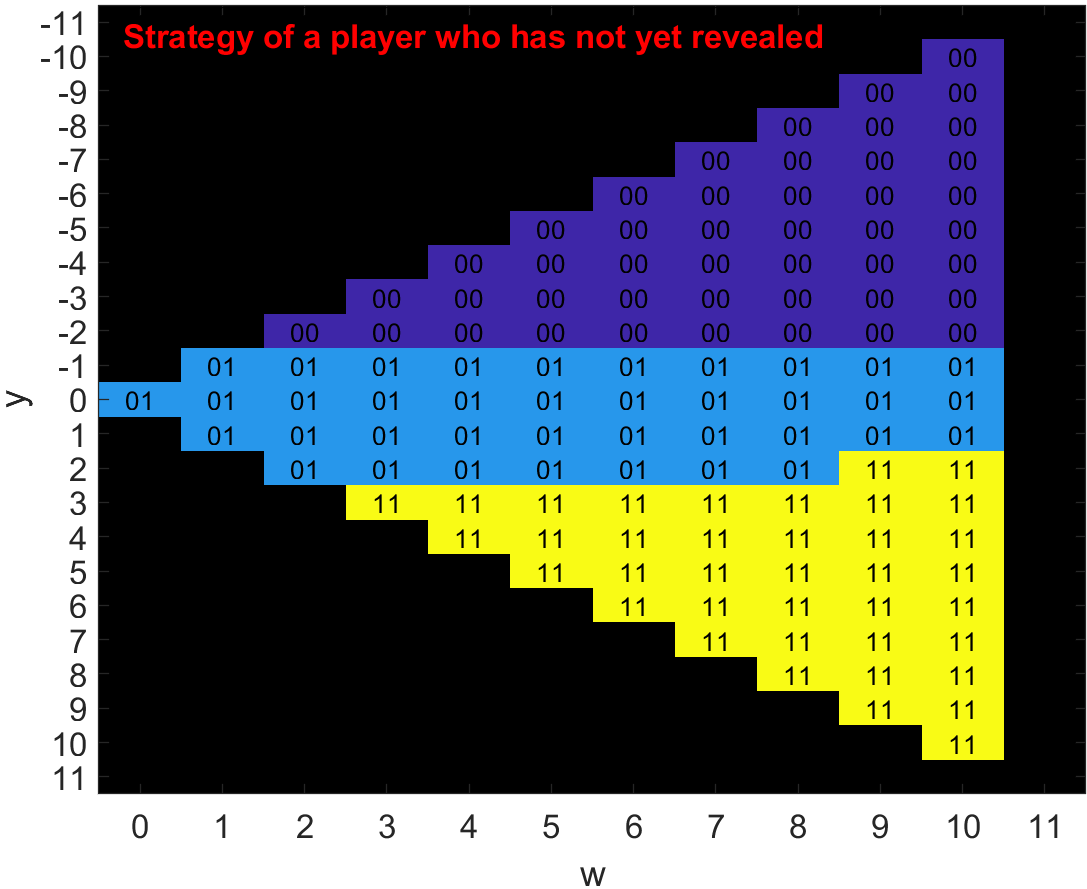}
\includegraphics[width=0.49\columnwidth]{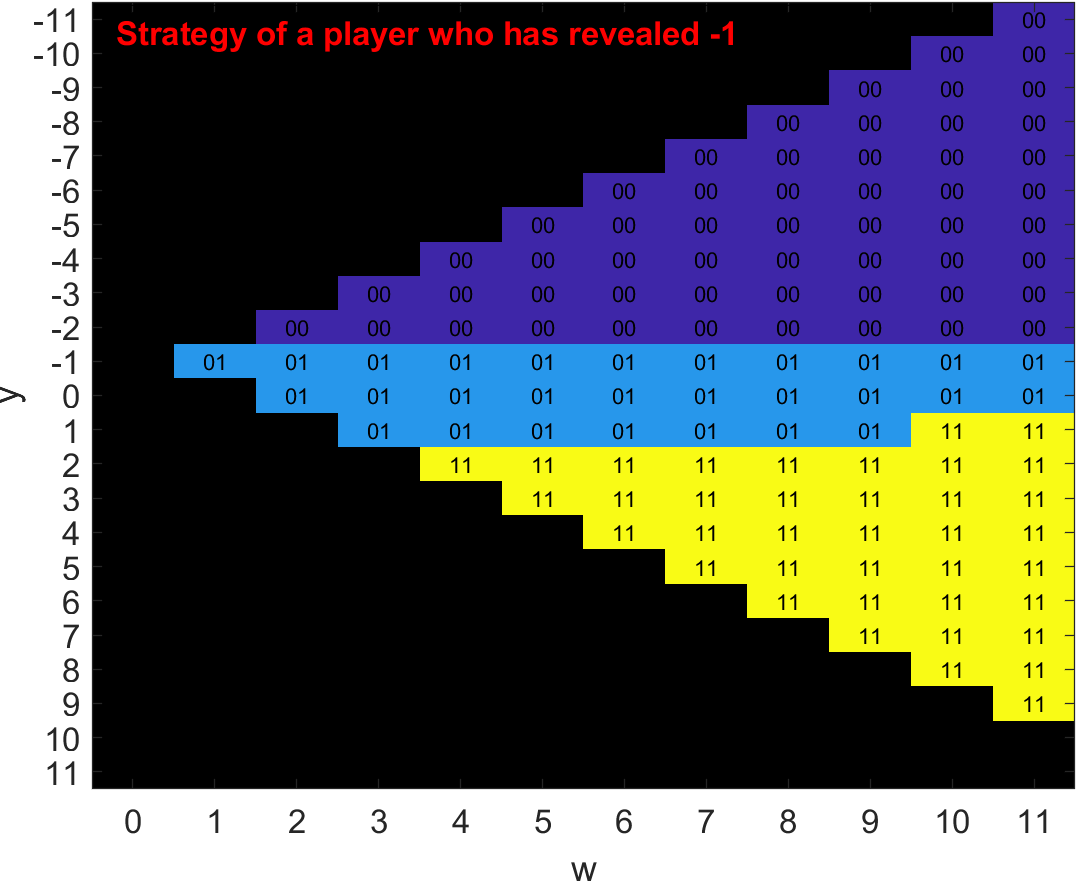}
\end{center}
\caption{Equilibrium strategies for $N=11$, $p=0.1$, $\delta=0.999$. ``00'', ``01'', and ``11'' denote strategies $\boldsymbol{0}$, $\boldsymbol{I}$, and $\boldsymbol{1}$, respectively.}
\label{fig:morepatient}
\end{figure}

The second case ($\delta=0.999$) studies more patient players and the results are depicted in Fig.~\ref{fig:morepatient}.
Not surprisingly, players are willing to wait more before committing to a buying decision. In fact, for values of $w=2$ to $w=5$ and with a
believed product quality of $y=2$ a player is not committing to buy (i.e., to play $\gamma=\boldsymbol{1}$) but the equilibrium strategy is to reveal her information ($\gamma=\boldsymbol{I}$). Similarly, with a
believed product quality of $y=2$ a player who has already revealed her private information $X^n=-1$ chooses to wait ($\gamma=\boldsymbol{0}$).

\begin{figure}[htbp]
\begin{center}
\includegraphics[width=0.49\columnwidth]{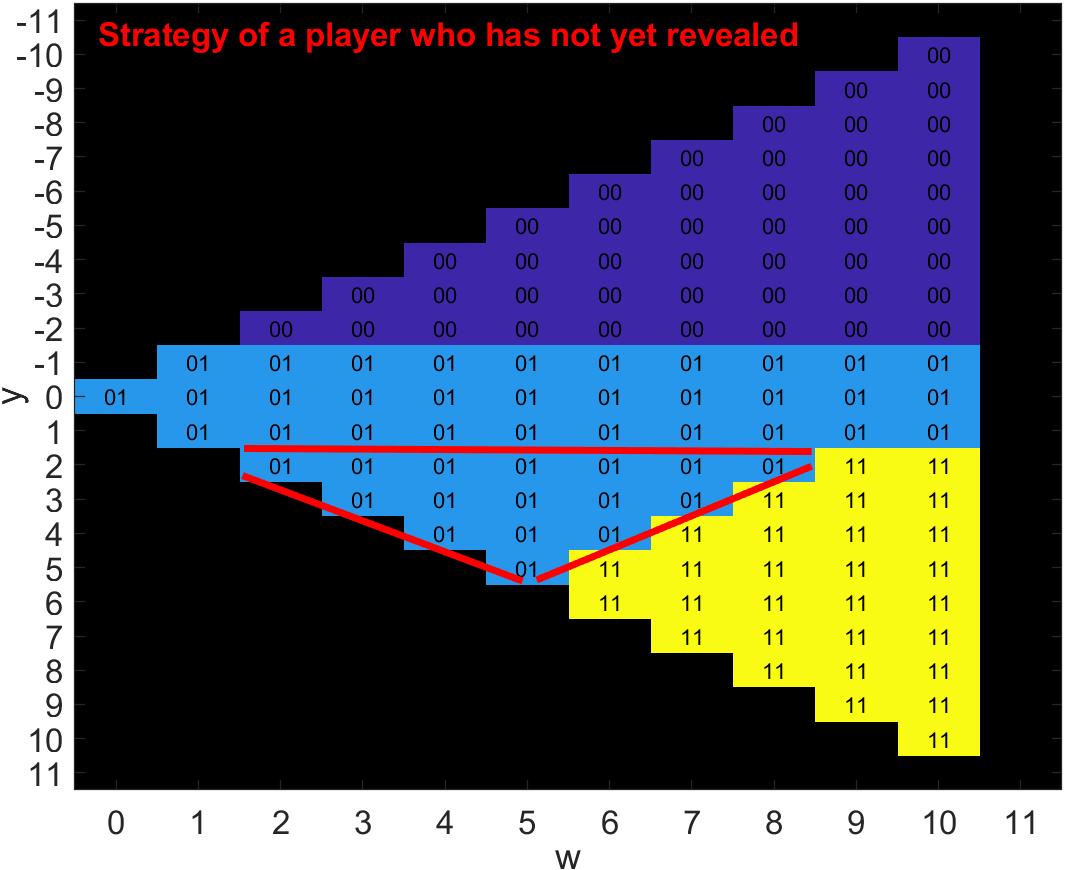}
\includegraphics[width=0.49\columnwidth]{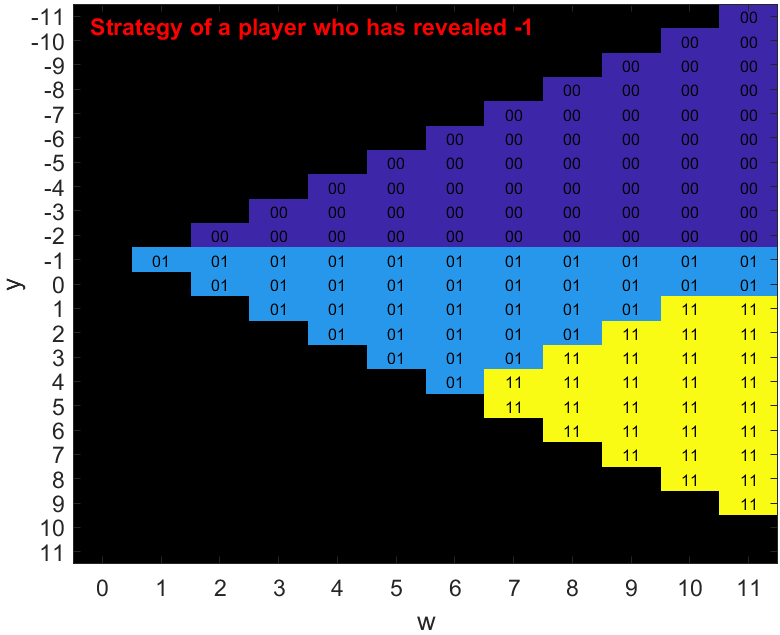}
\end{center}
\caption{Equilibrium strategies for $N=11$, $p=0.1$, $\delta=1$. ``00'', ``01'', and ``11'' denote strategies $\boldsymbol{0}$, $\boldsymbol{I}$, and $\boldsymbol{1}$, respectively.}
\label{fig:patient}
\end{figure}

The third case ($\delta=1$) studies infinitely patient players and the results are depicted in Fig.~\ref{fig:patient}.
As intuition suggests, players are willing to wait more before committing to a buying decision. In fact, for $w=5$ and with a
believed product quality of $y=5$ a player is not committing to buy (i.e., to play $\gamma=\boldsymbol{1}$) but the equilibrium strategy is to reveal her information ($\gamma=\boldsymbol{I}$). Similarly, for $w=6$ and with a
believed product quality of $y=4$ a player who has already revealed her private information $X^n=-1$ chooses to wait ($\gamma=\boldsymbol{0}$). Clearly, as $w$ increases and we are approaching the end of the game, players become more aggressive, as there is less information to be learnt by waiting, and at $w=N$ the equilibrium strategies for $\delta=0$ and $\delta=1$ coincide. Nevertheless, in the case of patient players a more cooperative equilibrium emerges (see strategies indicated in the red triangle in Fig.~\ref{fig:patient}) where players are willing to help each other learn the unknown state $V$ by revealing their private information.

We remark that these results are not inconsistent with Theorem~\ref{thm:nocas_st} since the theorem claims existence of specific solutions to the FPE but not uniqueness. Indeed, although this is the case of $\delta=1$ our numerical algorithm converges to the equilibrium described in~\eqref{eq:fixedN_largedelta} and also depicted in Fig.~\ref{fig:nocas}.

\begin{figure}[htbp]
\begin{center}
\includegraphics[width=0.49\columnwidth]{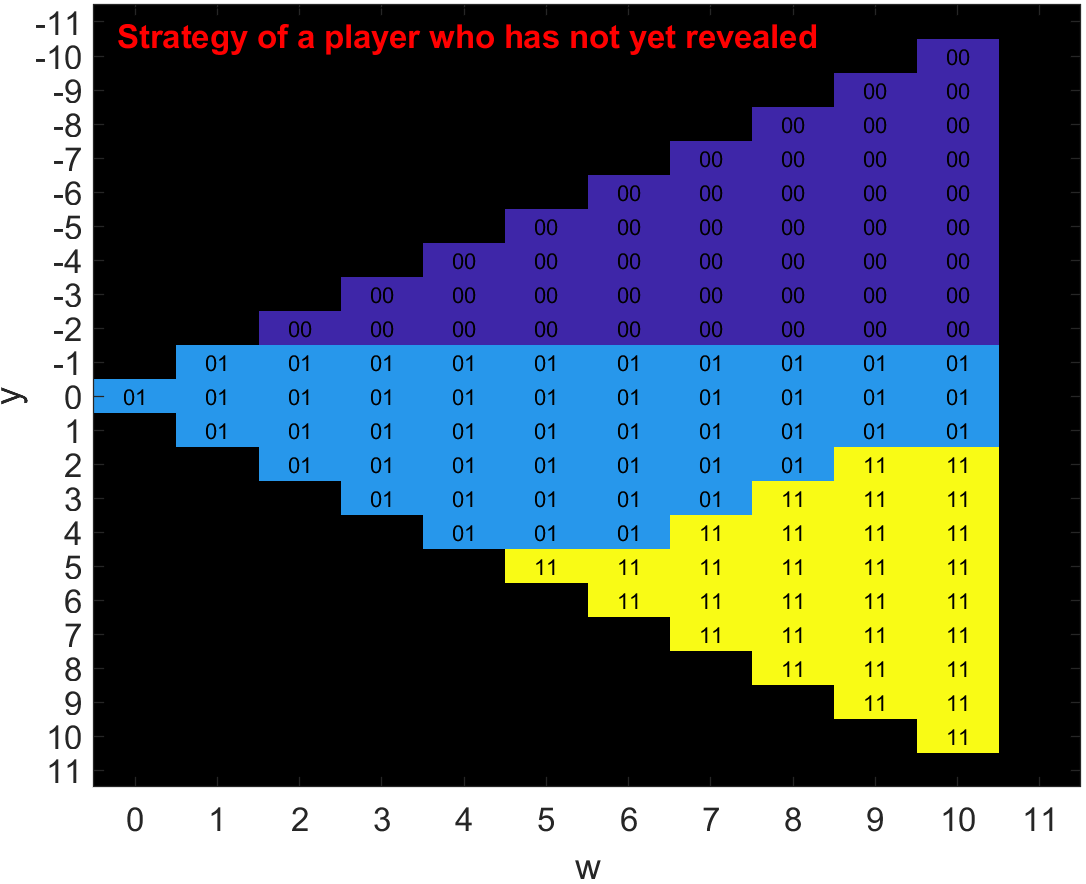}
\includegraphics[width=0.49\columnwidth]{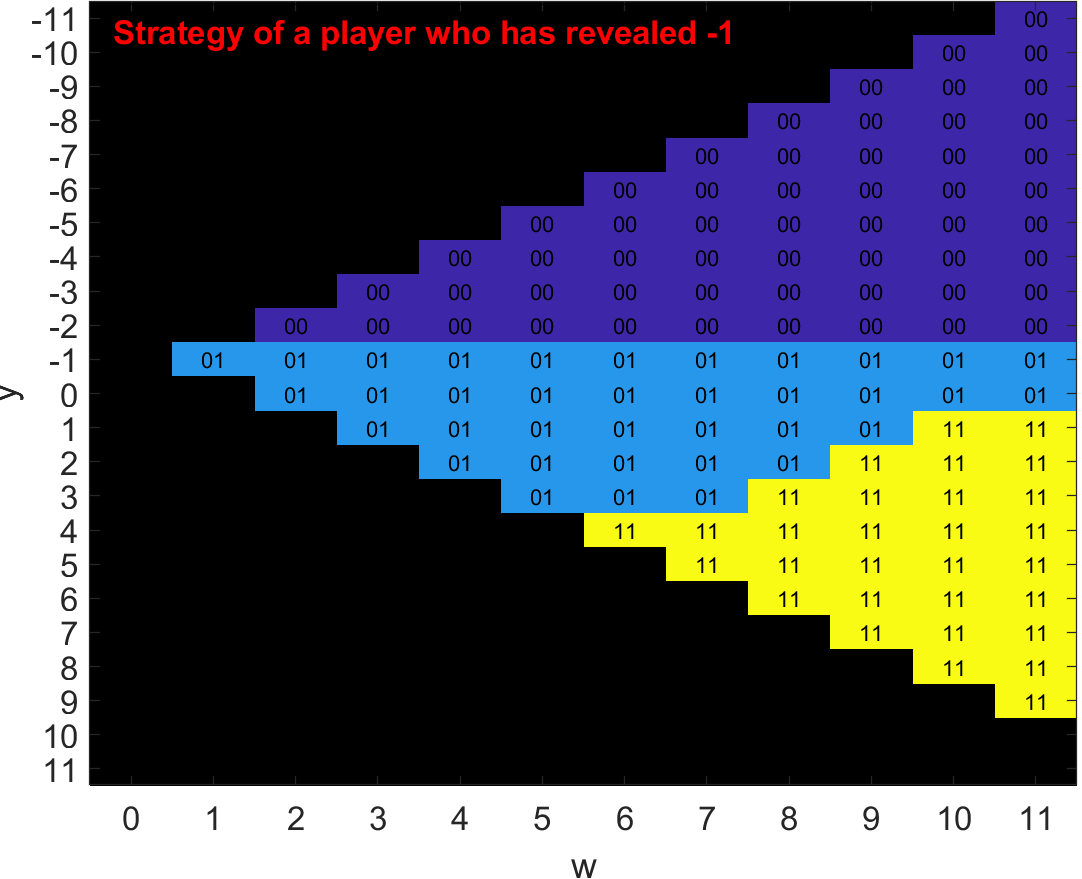}
\end{center}
\caption{Equilibrium strategies for $N=11$, $p=0.4$, $\delta=0.999$. ``00'', ``01'', and ``11'' denote strategies $\boldsymbol{0}$, $\boldsymbol{I}$, and $\boldsymbol{1}$, respectively.}
\label{fig:morepatient4}
\end{figure}

The next set of figures shows the effect of the quality of information. In Fig.~\ref{fig:morepatient4} the equilibrium for the case of $\delta=0.999$ and $p=0.4$ is depicted. This is a much noisier private observation compared to the one depicted in Fig.~\ref{fig:morepatient}. As a result, equilibrium behavior is ``softer'': players are willing to wait more and reveal their information, since a single observation is now of lower quality than before.

The last figure shows the probability of a bad cascade for the two different values of $V$, for different values of $p$ and for a larger number of users $N=21$.
We further disaggregate this probability according to the value of $W$ when a cascade occurs. We depict this information as cumulative bad cascade probabilities with $W\leq w$ for $w\in\{0,\ldots,N\}$ in Fig.~\ref{fig:bad_cascade}.
The figure shows that cascading behaviour is significantly asymmetric for the values $V=1$ and $V=-1$ and it is more severe for $V=1$, i.e., when the product is good and players opt to not buy it.
This is due to the asymmetry of the sets of values of $(y,w)$ for which the equilibrium is $\gamma=\boldsymbol{0}$ vs. that for which the equilibrium is $\gamma=\boldsymbol{1}$.
\optv{2col}{
\begin{figure}[htbp]
\begin{center}
\includegraphics[width=0.95\columnwidth]{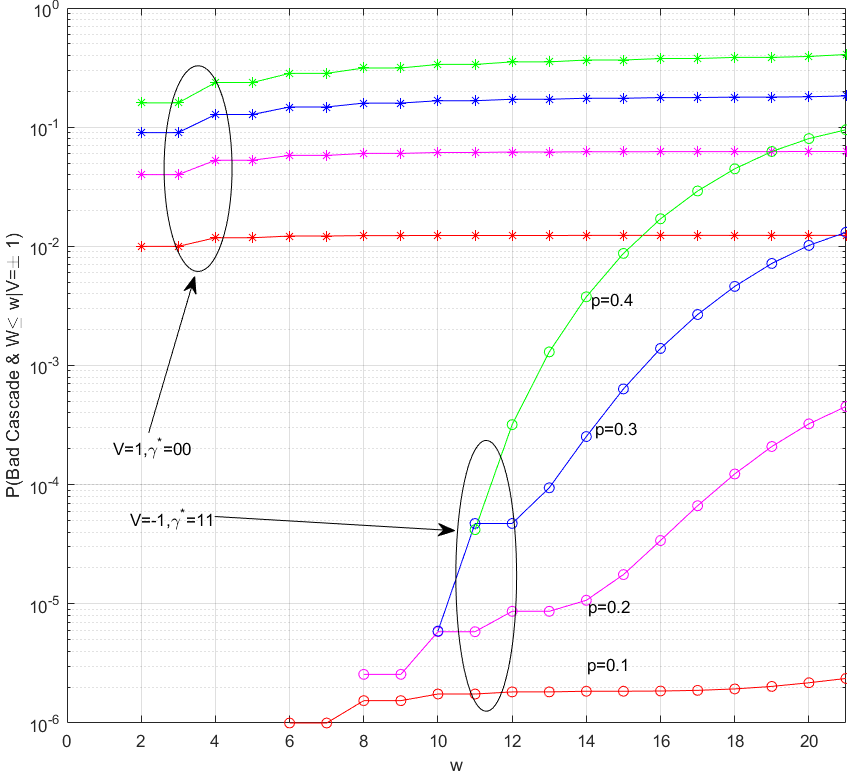}
\end{center}
\caption{Bad cascade probability for $N=21$, $\delta=0.999999$, $p\in\{0.1,0.2,0.3,0.4\}$.}
\label{fig:bad_cascade}
\end{figure}}
\optv{arxiv}{
	\begin{figure}[htbp]
		\begin{center}
			\includegraphics[width=0.6\columnwidth]{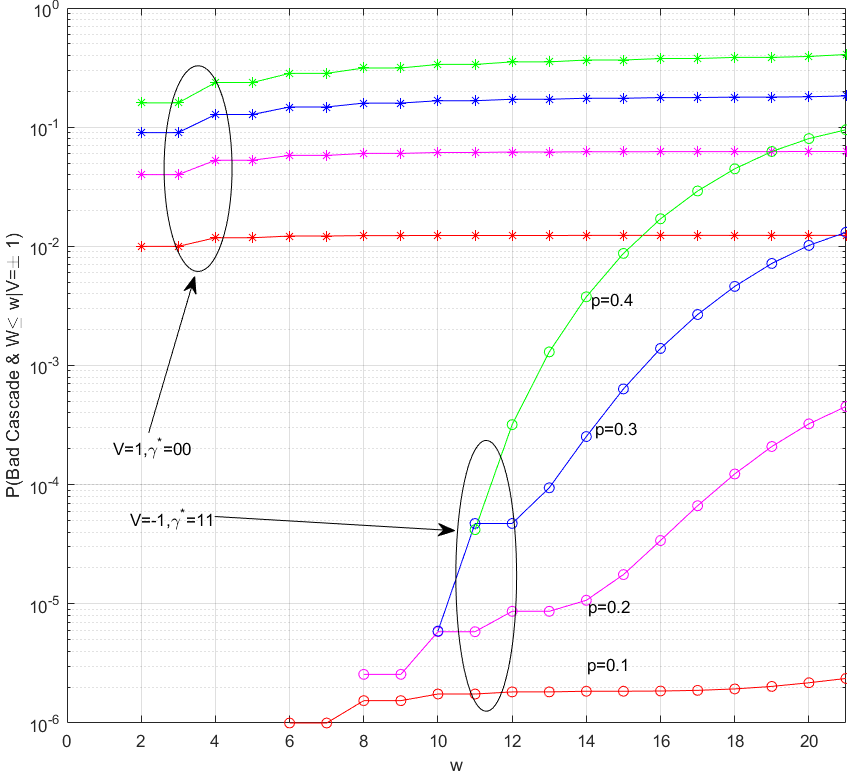}
		\end{center}
		\caption{Bad cascade probability for $N=21$, $\delta=0.999999$, $p\in\{0.1,0.2,0.3,0.4\}$.}
		\label{fig:bad_cascade}
\end{figure}}

\section{Conclusions }\label{sec:conclusions}

We studied a Bayesian learning scenario with non-myopic players. Our model generalizes the classic myopic and sequential one-shot scenario where informational cascades were first reported. In order to analyze information cascades in this scenario, an intricate analysis of the PBE of the dynamic game was needed. To that end, we first constructed FPEs that involve value functions defined on a finite domain. By further exploiting the structure of the model, we constructed FPEs with intuitive interpretations where the value functions has domain that is only quadratic in the number of players $N$.
Building on the tractability of these equations, we investigated their solutions in two regimes. The first
is for a fixed $\delta<1$ and asymptotically large $N$. The second is for a fixed $N$ and $\delta=1$ or asymptotically approaching 1.
For the first regime, we proved that an informational cascade eventually happens with high probability for large $N$. In these informational cascades, only a small portion of the information has been revealed, with high probability, making these cascades inefficient. For the second regime we proved that, surprisingly, infinitely patient players can completely avoid bad cascades when the product is bad. Furthermore, for sufficiently patient players when a bad cascade occurs (for a bad product) at least half of the players have already revealed their private information, which guarantees an error probability that vanishes with $N$.
%
Numerical solutions of the developed FPEs show that players exhibit a non-myopic behavior that is much more intricate than in the myopic case we generalized.

We were able to compress the fixed point equation based on the symmetry and structure of the problem. It could be interesting to apply the techniques introduced here to when the observation model $Q\left(x^{n}|v\right)$ is different between players. Extending FPE \ref{alg:Fixed-Point finite} is relatively straightforward. Extending FPE \ref{alg:Fixed-Point WY} is possible if there is a discrete set of available $Q\left(x^{n}|v\right)$. Then we can add a pair of $w,y$ variables to count players that have the same $Q\left(x^{n}|v\right)$. As expected, the dimension of the FPE would then increase with the complexity of the scenario.


\appendices{}

\section{Proof of Lemma \ref{lem:BeliefDecomposition}}\label{lem:BeliefDecomposition_proof}
\begin{IEEEproof}
The proof follows by induction. For $t=0$ we have
$\pi_0(\boldsymbol{x},v)=\Pr^s(\boldsymbol{x},v|n_0)=Q(v)\prod_{m=1}^N Q(x^m|v)$.
Assuming that $\pi_{t-1}(\boldsymbol{x},v)=\pi_{t-1}(v)\prod_{m=1}^N \pi_{t-1}(x^m|v)$ we have
\begin{subequations}
	\optv{2col}{
\begin{align}\label{eq:joint}
&\pi_{t}(\boldsymbol{x},v)= \Pr^s(\boldsymbol{x},v|\boldsymbol{a}_{0:t-1},n_{0:t}) \\
 &= \frac{\Pr^s(\boldsymbol{x},v,\boldsymbol{a}_{t-1},n_{t}|\boldsymbol{a}_{0:t-2},n_{0:t-1})}{\Pr^s(\boldsymbol{a}_{t-1},n_{t}|\boldsymbol{a}_{0:t-2},n_{0:t-1})} \\
 &= \frac{(1/N)\Pr^s(\boldsymbol{a}_{t-1}|\boldsymbol{x},v,\boldsymbol{a}_{0:t-2},n_{0:t-1}) \Pr^s(\boldsymbol{x},v|\boldsymbol{a}_{0:t-2},n_{0:t-1})}{\Pr^s(\boldsymbol{a}_{t-1},n_{t}|\boldsymbol{a}_{0:t-2},n_{0:t-1})} \\
 &=\frac{(1/N) \left(\prod_{m=1}^N \textbf{1}_{\gamma_{t-1}^m(x^m)}(a^m_{t-1}) \right) \pi_{t-1}(\boldsymbol{x},v)}
        {\sum_{x,v}(1/N) \left(\prod_{m=1}^N \textbf{1}_{\gamma_{t-1}^m(x^m)}(a^m_{t-1}) \right) \pi_{t-1}(\boldsymbol{x},v)}\\
 &=\frac{\left(\prod_{m=1}^N \textbf{1}_{\gamma_{t-1}^m(x^m)}(a^m_{t-1}) \right) \pi_{t-1}(v)\prod_{m=1}^N \pi_{t-1}(x^m|v)}
        {\sum_{x,v}\left(\prod_{m=1}^N \textbf{1}_{\gamma_{t-1}^m(x^m)}(a^m_{t-1}) \right) \pi_{t-1}(v)\prod_{m=1}^N \pi_{t-1}(x^m|v)}\\
 &=\frac{\left(\prod_{m=1}^N \textbf{1}_{\gamma_{t-1}^m(x^m)}(a^m_{t-1}) \pi_{t-1}(x^m|v) \right) \pi_{t-1}(v)}
        {\sum_{v}\left(\prod_{m=1}^N \sum_{x^m }\textbf{1}_{\gamma_{t-1}^m(x^m)}(a^m_{t-1}) \pi_{t-1}(x^m|v) \right) \pi_{t-1}(v) }.
\end{align}}
	\optv{arxiv}{
	\begin{align}\label{eq:joint}
	\pi_{t}(x,v) &= \Pr^s(x,v|\boldsymbol{a}_{0:t-1},n_{0:t})\\&= \frac{\Pr^s(x,v,\boldsymbol{a}_{t-1},n_{t}|\boldsymbol{a}_{0:t-2},n_{0:t-1})}{\Pr^s(\boldsymbol{a}_{t-1},n_{t}|\boldsymbol{a}_{0:t-2},n_{0:t-1})} \\
	&= \frac{(1/N)\Pr^s(\boldsymbol{a}_{t-1}|x,v,\boldsymbol{a}_{0:t-2},n_{0:t-1}) \Pr^s(x,v|\boldsymbol{a}_{0:t-2},n_{0:t-1})}{\Pr^s(\boldsymbol{a}_{t-1},n_{t}|\boldsymbol{a}_{0:t-2},n_{0:t-1})} \\
	&=\frac{(1/N) \left(\prod_{m=1}^N \textbf{1}_{\gamma_{t-1}^m(x^m)}(a^m_{t-1}) \right) \pi_{t-1}(x,v)}
	{\sum_{x,v}(1/N) \left(\prod_{m=1}^N \textbf{1}_{\gamma_{t-1}^m(x^m)}(a^m_{t-1}) \right) \pi_{t-1}(x,v)}\\
	&=\frac{\left(\prod_{m=1}^N \textbf{1}_{\gamma_{t-1}^m(x^m)}(a^m_{t-1}) \right) \pi_{t-1}(v)\left(\prod_{m=1}^N \pi_{t-1}(x^m|v)\right)}
	{\sum_{x,v}\left(\prod_{m=1}^N \textbf{1}_{\gamma_{t-1}^m(x^m)}(a^m_{t-1}) \right) \pi_{t-1}(v)\left(\prod_{m=1}^N \pi_{t-1}(x^m|v)\right)}\\
	&=\frac{\left(\prod_{m=1}^N \textbf{1}_{\gamma_{t-1}^m(x^m)}(a^m_{t-1}) \pi_{t-1}(x^m|v) \right) \pi_{t-1}(v)}
	{\sum_{v}\left(\prod_{m=1}^N \sum_{x^m }\textbf{1}_{\gamma_{t-1}^m(x^m)}(a^m_{t-1}) \pi_{t-1}(x^m|v) \right) \pi_{t-1}(v) }.
	\end{align}}
\end{subequations}
The conditional distribution of $X$ given $V$ and $\boldsymbol{h}^c_t$ can now be written as
\begin{subequations}
\begin{align}
\pi_{t}(\boldsymbol{x}|v)
 &=\frac{\prod_{m=1}^N \textbf{1}_{\gamma_{t-1}^m(x^m)}(a^m_{t-1}) \pi_{t-1}(x^m|v)  }
        {\sum_x \left(\prod_{m=1}^N \textbf{1}_{\gamma_{t-1}^m(x^m)}(a^m_{t-1}) \pi_{t-1}(x^m|v) \right) } \\
 &=\prod_{m=1}^N  \frac{\textbf{1}_{\gamma_{t-1}^m(x^m)}(a^m_{t-1}) \pi_{t-1}(x^m|v)  }
        { \sum_{x^m}  \textbf{1}_{\gamma_{t-1}^m(x^m)}(a^m_{t-1}) \pi_{t-1}(x^m|v) } \\
 &=\prod_{m=1}^N  \pi_{t}(x^m|v), \label{eq:cond_update}
\end{align}
\end{subequations}
where the second equality follows since given $V$,  $\{x^{m}\}$ are independent, so the expectation of the product is the product of the expectations. This completes the induction step proving that  $X^1,\ldots,X^N$
are conditionally independent given $v,\boldsymbol{h}_{t}^{c}$, which gives~\eqref{eq:cond_ind}.
Furthermore,~\eqref{eq:cond_update} provides an update equation for the conditional beliefs as
\begin{subequations}
\begin{align}
\pi_{t}&(x^m|v)
 =\frac{\textbf{1}_{\gamma_{t-1}^m(x^m)}(a^m_{t-1}) \pi_{t-1}(x^m|v)  }
        { \sum_{x^m}  \textbf{1}_{\gamma_{t-1}^m(x^m)}(a^m_{t-1}) \pi_{t-1}(x^m|v) } \nonumber \\
 &=\begin{cases}
 \pi_{t-1}(x^m|v), & m\neq n_{t-1} \text{ or } \gamma^m_{t-1}\neq \boldsymbol{I} \\
 \textbf{1}_{\frac{x^m+1}{2}}(a^m_{t-1}), & m = n_{t-1} \text{ and } \gamma^m_{t-1}= \boldsymbol{I}.
 \end{cases}
\end{align}
\end{subequations}
Consequently, if player $m$ has not yet revealed her information up to time $t$, then
$\pi_{t}(x^m|v)=\cdots=\pi_0(x^m|v)=Q(x^m|v)$.
Alternatively, if player $m$ has revealed her information before time $t$, we have $\pi_{t}(x^m|v)=\textbf{1}_{\tx^m}(x^m)$, thus proving~\eqref{eq:conditional}.

Now, marginalizing~\eqref{eq:joint} w.r.t. $\boldsymbol{x}$ we have
\begin{subequations}
	\optv{2col}{
\begin{align}
&\frac{\pi_{t+1}(1)}{\pi_{t+1}(-1)} \nonumber 
 =\frac{\prod_{m=1}^N \sum_{x^m}\textbf{1}_{\gamma_{t}^m(x^m)}(a^m_{t}) \pi_{t}(x^m|1) }
       {\prod_{m=1}^N \sum_{x^m} \textbf{1}_{\gamma_{t}^m(x^m)}(a^m_{t}) \pi_{t}(x^m|-1)  }
   \frac{\pi_t(1)}{\pi_t(-1)}\\
 &=\frac{\sum_{x^{n_t}}\textbf{1}_{\gamma_{t}(x^{n_t})}(a^{n_t}_{t}) \pi_{t}(x^{n_t}|1) }
       {\sum_{x^{n_t}} \textbf{1}_{\gamma_{t}(x^{n_t})}(a^{n_t}_{t}) \pi_{t}(x^{n_t}|-1)  }
   \frac{\pi_t(1)}{\pi_t(-1)},
\end{align}}
	\optv{arxiv}{
	\begin{align}
	\frac{\pi_{t+1}(1)}{\pi_{t+1}(-1)}
	&=\frac{\prod_{m=1}^N \sum_{x^m}\textbf{1}_{\gamma_{t}^m(x^m)}(a^m_{t}) \pi_{t}(x^m|1) }
	{\prod_{m=1}^N \sum_{x^m} \textbf{1}_{\gamma_{t}^m(x^m)}(a^m_{t}) \pi_{t}(x^m|-1)  }
	\frac{\pi_t(1)}{\pi_t(-1)}\\
	&=\frac{\sum_{x^{n_t}}\textbf{1}_{\gamma_{t}(x^{n_t})}(a^{n_t}_{t}) \pi_{t}(x^{n_t}|1) }
	{\sum_{x^{n_t}} \textbf{1}_{\gamma_{t}(x^{n_t})}(a^{n_t}_{t}) \pi_{t}(x^{n_t}|-1)  }
	\frac{\pi_t(1)}{\pi_t(-1)},
	\end{align}}
where the last equality is due to the fact that for all non-acting players $m \neq n_t$ we have $\gamma_{t}^m=\boldsymbol{0}$. Hence, for $m \neq n_t$ 
we always have  $\textbf{1}_{\gamma_{t}(x^{m})}(a^{m}_{t})=1$, and then $\sum\pi_{t}\left(x^{m}\mid\pm1\right)=1$ so these terms have no effect on the products in the numerator and denominator.
Furthermore, if $\gamma_t\neq \boldsymbol{I}$ or the acting player has already revealed her information, the multiplicative factor reduces to 1. Else, the factor becomes
\optv{2col}{\label{eq:37}
\begin{align}
\frac{\sum_{x^{n_t}}\textbf{1}_{\frac{x^{n_t}+1}{2}}(a^{n_t}_{t}) Q(x^{n_t}|1) }
     {\sum_{x^{n_t}}\textbf{1}_{\frac{x^{n_t}+1}{2}}(a^{n_t}_{t}) Q(x^{n_t}|-1)  }
 &=\frac{Q(2a^{n_t}_t-1|1) }{Q(2a^{n_t}_t-1|-1)  } \\
  &= q^{2a^{n_t}_t-1},
\end{align}}
\optv{arxiv}{
	\begin{align}\label{eq:37}
	\frac{\sum_{x^{n_t}}\textbf{1}_{\frac{x^{n_t}+1}{2}}(a^{n_t}_{t}) Q(x^{n_t}|1) }
	{\sum_{x^{n_t}}\textbf{1}_{\frac{x^{n_t}+1}{2}}(a^{n_t}_{t}) Q(x^{n_t}|-1)  }
	=\frac{Q(2a^{n_t}_t-1|1) }{Q(2a^{n_t}_t-1|-1)  } = q^{2a^{n_t}_t-1},
	\end{align}}
\end{subequations}
which gives~\eqref{eq:marginal_up}. We derive \eqref{eq:beliefv} by repeating the substitution of \eqref{eq:37} for all $n$, and using $Q(1)=Q(-1)=\frac{1}{2}$.
\end{IEEEproof}

\section{Proof of Theorem \ref{thm:PBE}}\label{thm:PBE_proof}
\begin{IEEEproof}
For clarity, we first prove the result with $\pi$ replacing $\boldsymbol{\tilde{x}}$ all throughout FPE \ref{alg:Fixed-Point finite}, and then use the fact that $\pi$ can be computed from $\boldsymbol{\tilde{x}}$.
Let us assume that all players other than player $n$ play according to $\gamma^*_t=\theta[n_t,\pi_t,\boldsymbol{b}_t]$, i.e., so if at time $t$ we have $n_{t}\neq n$ then 
$a^{n_t}_t = \gamma^*_t(x^{n_t}) = \theta[n_t,\pi_t,\boldsymbol{b}_t](x^{n_t})$.
Let us further assume that the update of the belief $\pi_t$ is fixed to $\pi_{t+1}=F(\pi_t,\gamma^*_t,a^{n_t}_t,n_t)=F(\pi_t,\theta[n_t,\pi_t,\boldsymbol{b}_t],a^{n_t}_t,n_t)=:F^{\theta}(\pi_t,n_t,a^{n_t}_t,\boldsymbol{b}_t)$.
We will show that the optimization problem faced by player $n$ can be formulated as a Markov decision process (MDP).
For this we will define a state, action, and instantaneous reward of a dynamical system as follows.
The state of the system is defined as $\lambda_t=(x^n,n_t,\pi_t,\boldsymbol{b}_t)$. Further, the action space is defined according to  \eqref{eq:actionset}, where at each time $t$, player $n$ takes the action $a^n_t \in \mathcal{A}^n(b^n_t,n_t)$ and receives an expected instantaneous reward of $R(\lambda_t,a^n_t)=a^n_t\sum_{v}v \mu^{n}(v)$.

We first show that $(\Lambda_t)_t$ is a controlled Markov process with actions $a^n_t$, i.e.,
\begin{align} \label{eq:cmc}
\Pr(\Lambda_{t+1}|\Lambda_{1:t},a^n_{1:t})=\Pr(\Lambda_{t+1}|\Lambda_t,a^n_t).
\end{align}
Indeed,
\begin{subequations}
	\optv{2col}{
\begin{align}
 &\Pr(\Lambda_{t+1}|\Lambda_{1:t},a^n_{1:t}) \nonumber\\
 &=\Pr(\bar{x}^n,n_{t+1},\pi_{t+1},\boldsymbol{b}_{t+1}|x^n,n_{1:t},\pi_{1:t},\boldsymbol{b}_{1:t},a^n_{1:t}) \\
 &=\textbf{1}_{x^n}(\bar{x}^n)\frac{1}{N}  Q_b(\boldsymbol{b}_{t+1}|x^n,n_t,\pi_t,\boldsymbol{b}_t,a^n_t)\\
 &\qquad Q_{\pi}(\pi_{t+1}|x^n,n_t,\pi_t,\boldsymbol{b}_t,a^n_t),
\end{align}}
	\optv{arxiv}{
	\begin{align}
	\Pr(\Lambda_{t+1}|\Lambda_{1:t},a^n_{1:t}) &=\Pr(\bar{x}^n,n_{t+1},\pi_{t+1},\boldsymbol{b}_{t+1}|x^n,n_{1:t},\pi_{1:t},\boldsymbol{b}_{1:t},a^n_{1:t}) \\
	&=\textbf{1}_{x^n}(\bar{x}^n)\frac{1}{N}  Q_b(\boldsymbol{b}_{t+1}|x^n,n_t,\pi_t,\boldsymbol{b}_t,a^n_t)Q_{\pi}(\pi_{t+1}|x^n,n_t,\pi_t,\boldsymbol{b}_t,a^n_t),
	\end{align}}
\end{subequations}
where the kernels $Q_b$ and $Q_{\pi}$ are defined through
\begin{subequations}
		\optv{2col}{
\begin{align}
 &Q_b(\boldsymbol{b}_{t+1}|x^n,n_t,\pi_t,\boldsymbol{b}_t,a^n_t) \nonumber\\
 &=Q_{b^n}(b^n_{t+1}|b^n_t,a^n_t)\hspace{-0.2cm}\prod_{m=1, m \neq n}^{N}\hspace{-0.2cm}Q_{\boldsymbol{b}^{-n}}(b^m_{t+1}|x^n,n_t,\pi_t,\boldsymbol{b}_t)
\end{align}}
	\optv{arxiv}{
	\begin{align}
	Q_b(\boldsymbol{b}_{t+1}|x^n,n_t,\pi_t,\boldsymbol{b}_t,a^n_t) =Q_{b^n}(b^n_{t+1}|b^n_t,a^n_t)\prod_{m=1, m \neq n}^{N}Q_{\boldsymbol{b}^{-n}}(b^m_{t+1}|x^n,n_t,\pi_t,\boldsymbol{b}_t)
	\end{align}}
with
\begin{align}\label{eq:kernel_b}
Q_{b^n}(b^n_{t+1}=1|b^n_t,a^n_t)=
\left\{\begin{array}{ll}
1, & b_t^n=1, \text{ or } a^n_t=1 \\
0, & \text{else}
\end{array}\right.
\end{align}
\optv{2col}{
\begin{align}
&Q_{\boldsymbol{b}^{-n}}(b^m_{t+1}=1|x^n,n_t,\pi_t,\boldsymbol{b}_t)= \nonumber \\
& \begin{cases}
\textbf{1}_{m}(n_t)\sum_{x^{m}} \mu^{n}(x^{m})\textbf{1}_{\theta[n_t,\pi_t,\boldsymbol{b}_t](x^{m})}(1), & b_t^m=0 \\
1, & b_t^m =1
\end{cases}
\end{align}}
\optv{arxiv}{
	\begin{align}
	Q_{\boldsymbol{b}^{-n}}(b^m_{t+1}=1|x^n,n_t,\pi_t,\boldsymbol{b}_t)= \begin{cases}
	\textbf{1}_{m}(n_t)\sum_{x^{m}} \mu^{n}(x^{m})\textbf{1}_{\theta[n_t,\pi_t,\boldsymbol{b}_t](x^{m})}(1), & b_t^m=0 \\
	1, & b_t^m =1
	\end{cases}
	\end{align}}
and
\optv{2col}{
	\begin{align}
	&Q_{\pi}(\pi_{t+1}|x^n,n_t,\pi_t,\boldsymbol{b}_t,a^n_t)= \nonumber \\
	&\begin{cases}
	%
	\sum\limits_{x^{n_t}} \mu^{n}(x^{n_t})
	\textbf{1}_{F^{\theta}(\pi_t,n_t,\theta[n_t,\pi_t,\boldsymbol{b}_t](x^{n_t}),\boldsymbol{b}_t)}(\pi_{t+1}),
	& n_t\neq n\\
	\textbf{1}_{F^{\theta}(\pi_t,n_t,a^{n}_t,\boldsymbol{b}_t)}(\pi_{t+1}),
	& n_t= n.
	\end{cases}
	\end{align}}
\optv{arxiv}{
\begin{align}
Q_{\pi}(\pi_{t+1}|x^n,n_t,\pi_t,\boldsymbol{b}_t,a^n_t)=\begin{cases}
%
\sum\limits_{x^{n_t}} \mu^{n}(x^{n_t})
    \textbf{1}_{F^{\theta}(\pi_t,n_t,\theta[n_t,\pi_t,\boldsymbol{b}_t](x^{n_t}),\boldsymbol{b}_t)}(\pi_{t+1}),
    & n_t\neq n\\
\textbf{1}_{F^{\theta}(\pi_t,n_t,a^{n}_t,\boldsymbol{b}_t)}(\pi_{t+1}),
    & n_t= n.
\end{cases}
\end{align}}
\label{kernels}
\end{subequations}
It is exactly the above equation that reveals why the belief update has to be fixed in order to prove that player $n$ faces an MDP. If that were not the case, the above equation would require that the belief is updated through an expression of the form $\pi_{t+1}=F(\pi_t,\gamma_t,a^{n_t}_t,n_t)$ which would require to include the partial function $\gamma_t$ in the action space for the case $n_t=n$ as opposed to only including the action $a^n_t$.
We have now proved~\eqref{eq:cmc}. Hence, the state process $(\Lambda_t)_t$ with the reward $R(\Lambda_t,a^n_t)$ form an infinite horizon MDP and so the optimal pure strategy can be derived from the following FPE for the state $\Lambda=(x^n,n_a,\pi,\boldsymbol{b})$,
\begin{subequations}
\optv{2col}{
\begin{align}
 &a^{*n}=\gamma^*(x^n) = \argmax_{a^n \in \mathcal{A}^n(b^n,n_a)} \{ a^n\sum_{v}v \mu^{n}(v)+ \nonumber \\
 &\qquad  \delta \mathbb{E}[V^n(x^n,N_a',\Pi,B)|x^n,n_a,\pi,\boldsymbol{b},a^n]\},
\end{align}}
\optv{arxiv}{
	\begin{align}
	a^{*n}=\gamma^*(x^n) = \argmax_{a^n \in \mathcal{A}^n(b^n,n_a)} \left\{ a^n\sum_{v}v \mu^{n}(v)+  \delta \mathbb{E}\left[V^n(x^n,N_a',\Pi,B)|x^n,n_a,\pi,b,a^n\right]\right\},
	\end{align}}
where $n_a$ denotes the acting player and $N_a'$, $\Pi$ and $B$ are random variables for the next state elements and the expectation is according to the transition kernels \eqref{kernels}. Furthermore,
\optv{2col}{
\begin{align}
 & V^n(x^n,n_a,\pi,\boldsymbol{b})=\max_{a^n \in \mathcal{A}^n(b^n,n_a)}\{ a^n\sum_{v}v \mu^n(v)+\nonumber \\
 &\qquad  \delta \mathbb{E}[ V^n(x^n,N_a',\Pi,B)|x^n,n_a,\pi,\boldsymbol{b},a^n]\}.
\end{align}}
\optv{arxiv}{
	\begin{align}
	V^n(x^n,n_a,\pi,b)=\max_{a^n \in \mathcal{A}^n(b^n,n_a)}\left\{ a^n\sum_{v}v \mu^n(v)+ \delta \mathbb{E}\left[ V^n(x^n,N_a',\Pi,B)|x^n,n_a,\pi,b,a^n\right]\right\}.
	\end{align}}
\label{fp13-2}
\end{subequations}
Next, we need to show that the above FPE is equivalent to FPE~\ref{alg:Fixed-Point finite}.
We first show that $V^n(x^n,n_a,\pi,b^n=1,\boldsymbol{b}^{-n})=0$ for all $x^n,n,\pi,\boldsymbol{b}^{-n}$. According to the action space defined in \eqref{eq:actionset}, if $b^n=1$,  $\mathcal{A}^n(b^n,n_a)=\{0\}$. This means that the instantaneous reward at this state is $0$. On the other hand, according to the transition kernel of $\boldsymbol{b}$ in~\eqref{kernels}, this state is absorbing in terms of $b^n$, which means that $b^n=1$ for all future states too. This will cause player $n$ to have 0 rewards in all of the upcoming states and so $V^n(x^n,n_a,\pi,b^n=1,\boldsymbol{b}^{-n})=0$.
The above implies that player $n$ faces a stopping time problem.

If $n$ is the acting player ($n=n_a$), then FPE~\eqref{fp13-2} is indeed choosing between buying and getting the instantaneous reward $\sum_{v}v \mu^{n}(v)$, or waiting and getting \optv{2col}{$\delta\mathbb{E}\left[V^n(x^n,N_a',\Pi,B)|x^n,n_a,\pi,\boldsymbol{b},a^n\right]
=\frac{\delta}{N}\sum_{n_a'=1}^{N}V^{n}\left(x^{n},n_a',F\left(\pi,\gamma^{*},0,n\right),\boldsymbol{b}\right)$,}\optv{arxiv}{$$\delta\mathbb{E}\left[V^n(x^n,N_a',\Pi,B)|x^n,n_a,\pi,\boldsymbol{b},a^n\right]
=\frac{\delta}{N}\sum_{n_a'=1}^{N}V^{n}\left(x^{n},n_a',F\left(\pi,\gamma^{*},0,n\right),\boldsymbol{b}\right)$$} using the transition kernels in \eqref{kernels}. Hence, for $n=n_a$, FPE~\eqref{fp13-2} is equivalent to \eqref{eq:fpe2a} and the first three cases of \eqref{eq:fpe2b}.

If $n$ is not the acting player ($n\neq n_a$), since $\mathcal{A}^n(b^n,n_a)=\{0\}$ then \optv{2col}{$V^n(x^n,n_a,\pi,\boldsymbol{b})=\delta \mathbb{E}\left[ V^n(x^n,N_a',\Pi,B)|x^n,n_a,\pi,\boldsymbol{b},a^n\right]$.}\optv{arxiv}{$$V^n(x^n,n_a,\pi,\boldsymbol{b})=\delta \mathbb{E}\left[ V^n(x^n,N_a',\Pi,B)|x^n,n_a,\pi,\boldsymbol{b},a^n\right].$$}
According the transition kernels~\eqref{kernels}, \optv{2col}{$\delta \mathbb{E}\left[ V^n(x^n,N_a',\Pi,B)|x^n,n_a,\pi,\boldsymbol{b},a^n\right]=\frac{\delta}{N}\sum_{n_a'=1}^{N}\mathbb{E}\left[ V^n(x^n,n_a',\Pi,(B^{n_a},\boldsymbol{b}^{-n_a}))|x^n,n_a,\pi,\boldsymbol{b},a^n\right]$.}\optv{arxiv}{$$\delta \mathbb{E}\left[ V^n(x^n,N_a',\Pi,B)|x^n,n_a,\pi,\boldsymbol{b},a^n\right]=\frac{\delta}{N}\sum_{n_a'=1}^{N}\mathbb{E}\left[ V^n(x^n,n_a',\Pi,(B^{n_a},\boldsymbol{b}^{-n_a}))|x^n,n_a,\pi,\boldsymbol{b},a^n\right],$$}
and $\Pi=F(\pi,\gamma^*,\gamma^*(X^{n_a}),n_a)$ with probability 1. Thus,
\optv{2col}{
\begin{align*}
&V^n(x^n,n_a,\pi,\boldsymbol{b})
 =\frac{\delta}{N}\sum_{n_a'=1}^{N}
  \mathbb{E}\{ V^n(x^n,n_a',\nonumber \\
&\qquad F(\pi,\gamma^*,\gamma^*(X^{n_a}),n_a),(B^{n_a},\boldsymbol{b}^{-n_a}))|x^n,n_a,\pi,\boldsymbol{b},a^n\},
\end{align*}}
\optv{arxiv}{
	\begin{align*}
	V^n(x^n,n_a,\pi,\boldsymbol{b})
	=\frac{\delta}{N}\sum_{n_a'=1}^{N}
	\mathbb{E}\{ V^n(x^n,n_a', F(\pi,\gamma^*,\gamma^*(X^{n_a}),n_a),B^{n_a}\boldsymbol{b}^{-n_a})|x^n,n_a,\pi,\boldsymbol{b},a^n\},
	\end{align*}}
which is the fourth case of \eqref{eq:fpe2b}.
Next, note that the transition kernel of $B^{n_a}$ in~\eqref{eq:kernel_b} is the same is in~\eqref{eq:fpe2c}.
It is now a simple task to construct the PBE by the forward algorithm in~\eqref{eq:forward} following each information set recursively (we are also using the fact that the private variables $X^1,\ldots,X^N$ are independent conditioned on $V$, as  shown in Lemma~\ref{lem:BeliefDecomposition}).
\optv{2col}{
The proof is completed by showing that $\pi$ can be computed using $\boldsymbol{\widetilde{x}}$.
In particular, using \eqref{eq:beliefv} in Lemma~\ref{lem:BeliefDecomposition},  and~\eqref{eq:private} in~\eqref{eq:fpe2a} we substitute
\begin{equation}
\pi(1\,|\,x^{n})=\frac{q^{\sum_{m}\widetilde{x}^{m}-\widetilde{x}^n+x^{n}}}{1+q^{\sum_{m}\widetilde{x}^{m}-\widetilde{x}^n+x^{n}}}.
\end{equation}
Similarly, using \eqref{eq:conditional} in Lemma~\ref{lem:BeliefDecomposition}, in~\eqref{eq:fpe2e} we get~\eqref{eq:fpe2d}.}\optv{arxiv}{
The proof is completed by showing that $\pi$ can be computed using $\boldsymbol{\widetilde{x}}$.
In particular, using \eqref{eq:beliefv} in Lemma~\ref{lem:BeliefDecomposition},  and~\eqref{eq:private} in~\eqref{eq:fpe2a} we substitute
\begin{equation}
\pi(1\,|\,x^{n})=\frac{q^{\sum_{m}\widetilde{x}^{m}-\widetilde{x}^n+x^{n}}}{1+q^{\sum_{m}\widetilde{x}^{m}-\widetilde{x}^n+x^{n}}}.
\end{equation}
Similarly, using \eqref{eq:conditional} in Lemma~\ref{lem:BeliefDecomposition}, in~\eqref{eq:fpe2e} we get~\eqref{eq:fpe2d}.}
\end{IEEEproof}

\optv{arxiv}{

\section{Computing a PBE though a polynomial-dimensional FPE}\label{sec:PBE_polynomial}

Owing to the symmetry of the problem we define
the set $\mathcal{K}=\{00,-10,01,-11,+11\}$ where the elements of this set are all possible values that the pair $\tx ^i b^i$ can take for each player $i$. Note that $+10$ can never happen under any strategy so it is not included in the set.
So players are grouped into 5 groups according to their value of the pair $\tx ^i b^i$.
We define the joint type (scaled empirical distribution), $t_{\boldsymbol{\tx} \boldsymbol{b}}$ of the sequence $(\boldsymbol{\tx} ,\boldsymbol{b})$ as
\begin{equation}
t_{\boldsymbol{\tx} \boldsymbol{b}}(k) = 
  \sum_{i=1}^N \textbf{1}_{\tx ^i b^i}(k) , \qquad \forall k\in\mathcal{K}.
\end{equation}
Clearly for every type $\boldsymbol{t}$,  $t(k)\geq 0$ and $\sum_{k\in\mathcal{K}} t(k)=N$, so there are exactly ${N+4\choose 4} \sim N^4$ such possible types.

Note that with the above definition, the aggregate state information $y=\sum_{i=1}^N \tilde{x}_i$ equals to $y=t(+11)-t(-10)-t(-11)$.

We define the following functions $U_a: \mathcal{X} \times\mathcal{K} \times \mathcal{T} \rightarrow \mathbb{R}$, and $U_{na}^l: \mathcal{X} \times \mathcal{K} \times \mathcal{T} \rightarrow \mathbb{R}$ for all $l\in\mathcal{K}$.
The meaning of these functions is as follows. $U_a(x,k,\boldsymbol{t})$ denotes the value function of the acting player $n$ whose private information $x^n=x$, her pair $\tx ^n b^n = k$ (and so she belongs to group $k$) and the joint type of the sequence $(\boldsymbol{\tx} , \boldsymbol{b})$ is $t$. Similarly, $U_{na}^l(x,k,\boldsymbol{t})$ denotes the value function of a non-acting player $m$ whose private information $x^m=x$, her pair $\tx ^m b^m = l$ (and so she belongs to group $l$), with an acting player $n$ whose pair $\tx ^n b^n = k$ (i.e., belonging to group $k$), and the joint type of the sequence $(\boldsymbol{\tx} , \boldsymbol{b})$ is $\boldsymbol{t}$.

Finally we define the update functions $g^x$, $g^b$, and $g^t$ as follows
\begin{subequations}
\begin{align}
 g^x(k_x,\gamma,a)  &=
 \left\{\begin{array}{ll}
 2a-1 & \text{, if } k_x= 0 \text{ and } \gamma = \boldsymbol{I}\\
  k_x  & \text{, else}
 \end{array} \right., \\
  g^b(k_b,a)  &=
 \left\{\begin{array}{ll}
 a & \text{, if } k_b=0 \\
  k_b & \text{, else}
 \end{array} \right., \\
 g^{\boldsymbol{t}}(k,\boldsymbol{t},\gamma,a)(k') &=
 \left\{\begin{array}{ll}
 t(k')-1  &  \text{, if } k'=k \text{ and } g^{xb}(k,\gamma,a)\neq k\\
 t(k')+1  &  \text{, if } k'=g^{xb}(k,\gamma,a) \text{ and } g^{xb}(k,\gamma,a)\neq k\\
 t(k') & \text{, else}
 \end{array} \right.
\end{align}
\end{subequations}
where we use the notation $k=k_x k_b$ to decompose the two parts of the $k$ index, and
with the understanding that we also use the notation $g^{efg}$ to denote $(g^e,g^f,g^g)$ for any $e,f,g \in \{x,b,t\}$.

We consider the following FP equation in FPE~\ref{alg:Fixed-Point POLY}.

\begin{algo}[Polynomial dimension]\label{alg:Fixed-Point POLY}

For every $k=k_x k_b\in\mathcal{K}$, $\boldsymbol{t}\in\mathcal{T}$ we evaluate $\gamma^*=\phi[k,\boldsymbol{t}]$ as follows.
\begin{itemize}
\item If $k_b=1$ then $\gamma^*=\textbf{0}$.
\item If $k_b=0$ then $\gamma^*$ is the solution of the following system of equations
\begin{subequations}
\label{eq:fp_polynomial}
\begin{align}
& \gamma^*(x) = \arg\max \{
     \underbrace{A }_{0=\text{don't buy}},
     \underbrace{\frac{q^{y+x\textbf{1}_{0}(k_x)}-1}
       {q^{y+x\textbf{1}_{0}(k_x)}+1}}_{1=\text{buy}}
     \}   \quad \forall x\in\mathcal{X}, \\&  \text{ where} \nonumber \\
   &  A =\frac{\delta}{N} U_a(x,g^{xb\boldsymbol{t}}(k,\boldsymbol{t},\gamma^*,0)) +\frac{\delta}{N} \sum_{k'\in \mathcal{K}}
      [t(k')-\textbf{1}_k(k')]
      U_{na}^{g^{xb}(k,\gamma^*,0)}(x,k',g^{\boldsymbol{t}}(k,\boldsymbol{t},\gamma^*,0))] \label{eq:fp_polynomial_A}
\end{align}
where the value functions satisfy
\begin{align}
 U_a&(x,k,\boldsymbol{t}) = \left\{\begin{array}{ll}
  0, & \text{if } k_b=1 \\
  A, & \text{if } k_b=0, \gamma^*(x)=0\\
  \frac{q^{y+x\textbf{1}_{0}(k_x)}-1}
       {q^{y+x\textbf{1}_{0}(k_x)}+1}, & \text{if } k_b=0,  \gamma^*(x)=1
 \end{array} \right.,
\end{align}
and for all $l=l_x l_b\in \mathcal{K}$
\begin{align} \label{eq:fp_polynomial_d}
U_{na}^l(x,k,\boldsymbol{t}) =
 \left\{\begin{aligned}
& 0, \hspace{5.5cm}  \hspace{5.35cm}\text{if } l_b=1, \\
   &\frac{\delta}{N} \mathbb{E}[ U_a(x,l,g^{\boldsymbol{t}}(k,\boldsymbol{t},\gamma^*,\gamma^*(X^n)))]
   + \frac{\delta}{N} \mathbb{E}[ U_{na}^{l}(x,g^{xb\boldsymbol{t}}(k,\boldsymbol{t},\gamma^*,\gamma^*(X^n)))] \\
&+\frac{\delta}{N} \sum_{k'\in\mathcal{K}}
   [ t(k') -\textbf{1}_{k}(k') -\textbf{1}_{l}(k') ] \mathbb{E}[U_{na}^{l}(x,k',g^{\boldsymbol{t}}(k,\boldsymbol{t},\gamma^*,\gamma^*(X^{n})))], \hspace{1cm} \text{if } l_b=0,
 \end{aligned} \right.,
\end{align}
where expectation in the last equation is wrt the RV $X^{n}$  where
\begin{align}
P(X^{n}=x^{n}|l,x,k,\boldsymbol{t})  = \left\{
\begin{array}{ll}
 \textbf{1}_{k_x}(x^n) & \text{, if } k_x\neq 0\\
 \frac{Q(x^n|-1)+Q(x^n|1)q^{y+x\textbf{1}_0(l_x)}}
 {q^{y+x\textbf{1}_0(l_x)}+1} & \text{, else.}
\end{array}
\right.
\end{align}
\end{subequations}%
\hfill $\blacksquare$
\end{itemize}

\end{algo}

We will now show that if the above FPE  has a solution $U^*$, then the original FPE  has a solution $V^*$ where $V^*$ can be readily derived from $U^*$.

Given the solution $U^*$ of the above FP equation (together with the strategy $\phi$) we construct the following strategies and value functions.
\begin{subequations}
\begin{align}
 \gamma^*=\theta[n,\boldsymbol{\tx} , \boldsymbol{b}]= \phi[\tx ^n b^n,\boldsymbol{t}_{\boldsymbol{\tx} ,\boldsymbol{b}}]
\end{align}
\begin{align}
 \tV^m(\cdot,n,\boldsymbol{\tx} ,\boldsymbol{b})= \left\{\begin{array}{ll}
  U_a(\cdot,\tx ^n b^n,\boldsymbol{t}_{\tx ,b}),  & \text{if } m=n \\
  U_{na}^{\tx ^m b^m}(\cdot,\tx ^n b^n ,\boldsymbol{t}_{\boldsymbol{\tx} ,\boldsymbol{b}}), & \text{if } m\neq n
 \end{array} \right..
\end{align}
\end{subequations}
We will show that these value functions are solutions of the original FPE~\ref{alg:Fixed-Point finite}.

\begin{thm}\label{thm:polynomial}
The value functions $(\tV^m)_{m\in\mN}$ together with the strategy mapping $\gamma^*=\phi[\cdot]$
satisfy  FPE~\ref{alg:Fixed-Point finite}.
\end{thm}


\begin{IEEEproof}
Fix $n$, $\boldsymbol{\tx}$, and $\boldsymbol{b}$ that result in a type $\boldsymbol{t}$ with accumulated state $y$.
The acting player $n$ belongs to a group $k=k_x k_b=\tx^n b^n$.
If $b_n=1$ then $k_b=1$ and $\gamma^*=\textbf{0}$.
If $b_n=0$ then it is clear that the second term in~\eqref{eq:fpe2a} becomes $\frac{q^{y+x^n\textbf{1}_{0}(k_x)}-1}
       {q^{y+x^n\textbf{1}_{0}(k_x)}+1}$, which is exactly the same as the second term in~\eqref{eq:fp_polynomial_A} (with $x^n=x$).
Consider the first term in~\eqref{eq:fpe2a}.
The new group of the acting player $n$ is $\hat{k}=\left(f(\tx^{n},\gamma^{*},0),0\right)=g^{xb}(k,\gamma^*,0)$ and
the new value for the overall type will change to $\boldsymbol{\hat{t}}=g^{\boldsymbol{t}}(k,\boldsymbol{t},\gamma^*,0)$. The implication of the above is that the first term in~\eqref{eq:fpe2a} will be
 \begin{equation}
 \begin{aligned}
 &\sum_{n'=1}^N \tV^n(x^n,n',\left(\ensuremath{\boldsymbol{\tx}^{-n},f(\tx^{n},\gamma^{*},0)}\right),(\boldsymbol{b}^{-n},0))  \\
  &= \tV^n(x^n,n,\left(\ensuremath{\boldsymbol{\tx}^{-n},f(\tx^{n},\gamma^{*},0)}\right),(\boldsymbol{b}^{-n},0))+\sum_{n'=1,n'\neq n}^N \tV^n(x^n,n',\left(\ensuremath{\boldsymbol{\tx^{-n}},f(\tx^{n},\gamma^{*},0)}\right),(\boldsymbol{b}^{-n},0)) \\
  &= U_a(x^n,\hat{k},\boldsymbol{\hat{t}})
   + \sum_{n'=1,n'\neq n}^N U_{na}^{\hat{k}}(x^n,\tx^{n'} b^{n'} ,\boldsymbol{\hat{t}}) \\
   &= U_a(x^n,\hat{k},\boldsymbol{\hat{t}})
   + \sum_{k'\in\mK}\sum_{n'=1,n'\neq n, \tx^{n'} b^{n'}=k'}^N \hspace{-0.2cm} U_{na}^{\hat{k}}(x^n,\tx^{n'} b^{n'} ,\boldsymbol{\hat{t}}) \\
   &= U_a(x^n,g^{xb\boldsymbol{t}}(k,\boldsymbol{t},\gamma^*,0))+ \sum_{k'\in\mK} [ t(k')-\textbf{1}_k(k') ] U_{na}^{g^{xb}(k,\gamma^*,0)}(x^n,k' ,g^{\boldsymbol{t}}(k,\boldsymbol{t},\gamma^*,0)),
   \end{aligned}
 \end{equation}
where the term $ t(k')-\textbf{1}_k(k')$ enumerates all players $n'\neq n$ in the vector $\left(\ensuremath{\boldsymbol{\tx^{-n}},f(\tx^{n},\gamma^{*},0)}\right),(\boldsymbol{b}^{-n},0)$ which are given by the original type $\boldsymbol{t}$ subtracting one from the group of the acting player. This is exactly the expression in~\eqref{eq:fp_polynomial_A} and thus~\eqref{eq:fpe2a} is satisfied.

Now consider~\eqref{eq:fpe2b}. Fix $m$ and denote the group of the $m$-th player by $l=l_x l_b=\tx^m b^m$.
The first three branches of this equation are obviously satisfied.
Regarding the fourth branch we know that the new group of the acting player $n$ will be $\hat{K}=f(\tx^n,\gamma^*,\gamma^*(X^n)),B^{\prime n}=g^b(k_b,\gamma^*(X^n))$ and the new type will be $\hat{T}=g^{\boldsymbol{t}}(k,\boldsymbol{t},\gamma^*,\gamma^*(X^n))$. The left-hand side of~\eqref{eq:fpe2b} becomes $U_{na}^l(x^m,k,\boldsymbol{t})$ with $l_b=0$. The right-hand side becomes
 \begin{equation}
 \begin{aligned}
 \sum_{n'=1}^{N}&\mathbb{E}\left[V^{m}\left(x^{m},n',\left(\boldsymbol{\tx^{-n}},f\left(\tilde{x}^{n},\gamma^{*},\gamma^{*}\left(X^{n}\right)\right)\right),(\boldsymbol{b}^{-n},B^{\prime n})\right)\right]\\
 &= \mathbb{E}\left[V^{m}\left(x^{m},m,\left(\boldsymbol{\tx^{-n}},f\left(\tilde{x}^{n},\gamma^{*},\gamma^{*}\left(X^{n}\right)\right)\right),(\boldsymbol{b}^{-n},B^{\prime n})\right)\right] \\&+
 \mathbb{E}\left[V^{m}\left(x^{m},n,\left(\boldsymbol{\tx^{-n}},f\left(\tilde{x}^{n},\gamma^{*},\gamma^{*}\left(X^{n}\right)\right)\right),(\boldsymbol{b}^{-n},B^{\prime n})\right)\right]  \\ &+\sum_{n'=1,n'\neq m,n}^{N}\hspace{-0.3cm}\mathbb{E}\left[V^{m}\left(x^{m},n',\left(\boldsymbol{\tx^{-n}},f\left(\tilde{x}^{n},\gamma^{*},\gamma^{*}\left(X^{n}\right)\right)\right),(\boldsymbol{b}^{-n},B^{\prime n})\right)\right] \\
   &=\mathbb{E}[ U_a(x^m,l,\hat{T})] +
     \mathbb{E}[ U_{na}^l(x^m,\hat{K},\hat{T})] +
     \sum_{k'\in\mK}\sum_{n'=1,n'\neq n,m,\ \tx^{n'} b^{n'}=k'}^N  \mathbb{E}[U_{na}^{l}(x^m,\tx^{n'} b^{n'} ,\hat{T})] \\
 &  = \mathbb{E}[ U_a(x^m,l,g^{\boldsymbol{t}}(k,\boldsymbol{t},\gamma^*,\gamma^*(X^n)))] +
     \mathbb{E}[ U_{na}^l(x^m,g^{xb\boldsymbol{t}}(k,\boldsymbol{t},\gamma^*,\gamma^*(X^n)))]  \\
     &+\sum_{k'\in\mK} [ t(k') -\textbf{1}_{k}(k') -\textbf{1}_{l}(k') ] \mathbb{E}[ U_{na}^{l}(x^m,k',g^{\boldsymbol{t}}(k,\boldsymbol{t},\gamma^*,\gamma^*(X^n)))].
\end{aligned}
 \end{equation}
This is exactly the expression in~\eqref{eq:fp_polynomial_d} and thus~\eqref{eq:fpe2b} is satisfied.
\end{IEEEproof}

We remark at this point that this method can be generalized for heterogeneous players with different values of $\delta$. All is needed is to consider joint types of the vectors $\boldsymbol{\tx},\boldsymbol{b},\delta$. The corresponding dimensionality of the FP equation will be $\sim N^{4 K_{\delta}}$ where
$K_{\delta}$ is the number of different types of $\delta$.

}


\section{Proof of Theorem \ref{thm:Summary of the Summary}}\label{thm:Summary of the Summary_proof}
\begin{IEEEproof}
Fix $n$, $\boldsymbol{\tx}$ that results in population parameters $y$ and $w$.
The acting player has either not revealed her information ($\tx^n=0$)
or she has revealed a negative signal ($\tx^n=-1$), since otherwise she would have already bought the product and must play $a^{n}=0$.
This implies that $\tx^n=-r$.
It is then clear that the first term in \eqref{eq:fpe2a} becomes
$\frac{q^{y+r+x^n}-1}
      {q^{y+r+x^n}+1}$,
which is exactly the same as the first term in~\eqref{eq:fp_quadratica} (with $x^n=x$).
Consider the second term in~\eqref{eq:fpe2a}.
The new parameter of the acting player $n$ is $\hat{r}=\left|f(\tilde{x}^{n},\gamma^{*},0)\right|=G^{r}(r,\gamma^{*})$. Define the new  population parameters by $\hat{y}=G^{y}\left(r,y,\gamma^{*},0\right)$ and
$\hat{w}=G^{w}\left(r,w,\gamma^{*},0\right)$. The
implication of the above is that the second term in \eqref{eq:fpe2a}
will be (apart for the $\delta/N$ factor)
\begin{subequations}
	\optv{2col}{
\begin{align}
&\sum_{n'=1}^{N}\tilde{V}^{n}\left(x^{n},n',(\boldsymbol{\tilde{x}}^{-n},f(\tilde{x}^{n},\gamma^{*},0)),(\boldsymbol{b}^{-n},0)\right) \nonumber \\
&=\tilde{V}^{n}\left(x^{n},n,(\boldsymbol{\tilde{x}}^{-n},f\left(\tilde{x}^{n},\gamma^{*},0\right)),(\boldsymbol{b}^{-n},0)\right) \nonumber \\
 &\ \ + \hspace{-0.2cm}
\sum_{n'=1,n'\neq n}^{N}\hspace{-0.3cm} \tilde{V}^{n}\left(x^{n},n',(\boldsymbol{\tilde{x}}^{-n},f\left(\tilde{x}^{n},\gamma^{*},0\right)),(\boldsymbol{b}^{-n},0)\right)  \\
&=U_{a}\left(x^{n},\hat{r},\hat{y},\hat{w}\right)
 +\sum_{n'=1,n'\neq n}^{N}U_{na}^{\hat{r}}\left(x^{n},z^{n'},\hat{y},\hat{w}\right)  \\
&=U_{a}\left(x^{n},\hat{r},\hat{y},\hat{w}\right)
 +\sum_{n'=1,n'\neq n,z^{n'}=0}^{N}U_{na}^{\hat{r}}\left(x^{n},0,\hat{y},\hat{w}\right)
 \nonumber  \\
 &\quad +\sum_{n'=1,n'\neq n,z^{n'}=1}^{N}U_{na}^{\hat{r}}\left(x^{n},1,\hat{y},\hat{w}\right)  \\
&=U_{a}\left(x^{n},\hat{r},\hat{y},\hat{w}\right)
 +\left(N-w-1+r\right)U_{na}^{\hat{r}}\left(x^{n},0,\hat{y},\hat{w}\right) \nonumber
 \\
 &\quad +\left(w-r\right)U_{na}^{\hat{r}}\left(x^{n},1,\hat{y},\hat{w}\right).  \\
%
\end{align}}
	\optv{arxiv}{
	\begin{align}
&	\sum_{n'=1}^{N}\tilde{V}^{n}\left(x^{n},n',\boldsymbol{\tilde{x}}^{-n}f(\tilde{x}^{n},\gamma^{*},0),(\boldsymbol{b}^{-n},0)\right)
\\	&=\tilde{V}^{n}\left(x^{n},n,\boldsymbol{\tilde{x}}^{-n}f\left(\tilde{x}^{n},\gamma^{*},0\right),(\boldsymbol{b}^{-n},0)\right) +
	\sum_{n'=1,n'\neq n}^{N}\tilde{V}^{n}\left(x^{n},n',\boldsymbol{\tilde{x}}^{-n}f\left(\tilde{x}^{n},\gamma^{*},0\right),(\boldsymbol{b}^{-n},0)\right)  \\
	&=U_{a}\left(x^{n},\hat{r},\hat{y},\hat{w}\right)
	+\sum_{n'=1,n'\neq n}^{N}U_{na}^{\hat{r}}\left(x^{n},z^{n'},\hat{y},\hat{w}\right)  \\
	&=U_{a}\left(x^{n},\hat{r},\hat{y},\hat{w}\right)
	+\sum_{n'=1,n'\neq n,z^{n'}=0}^{N}U_{na}^{\hat{r}}\left(x^{n},0,\hat{y},\hat{w}\right)
 +\sum_{n'=1,n'\neq n,z^{n'}=1}^{N}U_{na}^{\hat{r}}\left(x^{n},1,\hat{y},\hat{w}\right)  \\
	&=U_{a}\left(x^{n},\hat{r},\hat{y},\hat{w}\right)
	+\left(N-w-1+r\right)U_{na}^{\hat{r}}\left(x^{n},0,\hat{y},\hat{w}\right) \nonumber
+\left(w-r\right)U_{na}^{\hat{r}}\left(x^{n},1,\hat{y},\hat{w}\right)  \\
	&=U_{a}\left(x^{n},G^{ryw}(r,y,w,\gamma^{*},0)\right)  +\left(N-w-1+r\right)U_{na}^{G^{r}(r,\gamma^{*})}\left(x^{n},0,G^{yw}(r,y,w,\gamma^{*},0)\right)  \nonumber  \\
	&\quad +\left(w-r\right)U_{na}^{G^{r}(r,\gamma^{*})}\left(x^{n},1,G^{yw}(r,y,w,\gamma^{*},0)\right).
	\label{eq:52}
	\end{align}}
\end{subequations}
This is exactly the expression in~\eqref{eq:fp_quadraticb} so~\eqref{eq:fpe2a} is satisfied.
\optv{2col}{
The proof for~\eqref{eq:fpe2b} is quite similar and is omitted due to space limitations.
It can be found in~\cite{BiHeAn19arxiv}.}
\optv{arxiv}{
Now consider~\eqref{eq:fpe2b}. Fix $m$ and denote the parameter of the $m$-th player by $\tilde{r}=|\tx^m|$.
The first three branches of this equation are obviously satisfied.
Regarding the fourth branch we know that the new parameter of the acting player $n$ will be $\hat{Z}=G^{z}(z,\gamma^*,\gamma^*(X^n))$ and the new population parameters will be $(\hat{Y},\hat{W})=G^{yw}(z,y,w,\gamma^*,\gamma^*(X^n))$.
The left-hand side of~\eqref{eq:fpe2b} becomes $U_{na}^{\tilde{r}}(x^m,z,y,w)$. The right-hand side becomes
\begin{align}
\sum_{n'=1}^{N}&\mathbb{E}\left[V^{m}\left(x^{m},n',\tx^{-n}f\left(\tilde{x}^n,\gamma^{*},\gamma^{*}\left(X^{n}\right)\right),(\boldsymbol{b}^{-n},B^{\prime n})\right)\right] \nonumber \\
&=\mathbb{E}\left[V^{m}\left(x^{m},m,\tx^{-n}f\left(\tilde{x}^n,\gamma^{*},\gamma^{*}\left(X^{n}\right)\right),(\boldsymbol{b}^{-n},B^{\prime n})\right)\right]  +\mathbb{E}\left[V^{m}\left(x^{m},n,\tx^{-n}f\left(\tilde{x}^n,\gamma^{*},\gamma^{*}\left(X^{n}\right)\right),\boldsymbol{b}^{-n}B^{\prime n}\right)\right] \nonumber  \\
  &+ \sum_{n'=1,n'\neq m,n}^{N}\hspace{-0.5cm}\mathbb{E}\left[V^{m}\left(x^{m},n',\tx^{-n}f\left(\tilde{x}^n,\gamma^{*},\gamma^{*}\left(X^{n}\right)\right),(\boldsymbol{b}^{-n},B^{\prime n})\right)\right]\nonumber \\
&=\mathbb{E}[ U_a(x^m,\tilde{r},\hat{Y},\hat{W})] +
     \mathbb{E}[ U_{na}^{\tilde{r}}(x^m,\hat{Z},\hat{Y},\hat{W})] +\sum_{n'=1,n'\neq n,m,\ z^{n'}=1}^N  \mathbb{E}[U_{na}^{\tilde{r}}(x^m,1 ,\hat{Y},\hat{W})] \nonumber \\
     & +
     \sum_{n'=1,n'\neq n,m,\ z^{n'}=0}^N  \mathbb{E}[U_{na}^{\tilde{r}}(x^m,0 ,\hat{Y},\hat{W})] \nonumber  \\
&=\mathbb{E}[ U_a(x^m,\tilde{r},G^{yw}(z,y,w,\gamma^*,\gamma^*(X^n)))] +
     \mathbb{E}[ U_{na}^{\tilde{r}}(x^m,G^{zyw}(z,y,w,\gamma^*,\gamma^*(X^n)))]   \nonumber  \\
     &+(w-z-\tilde{r})  \mathbb{E}[U_{na}^{\tilde{r}}(x^m,1 ,G^{yw}(z,y,w,\gamma^*,\gamma^*(X^n)))] \nonumber \\
     &+(N-w-2+z+\tilde{r})
     \mathbb{E}[U_{na}^{\tilde{r}}(x^m,0 ,G^{yw}(z,y,w,\gamma^*,\gamma^*(X^n)))].
\end{align}
This is exactly the expression in~\eqref{eq:fp_quadraticd} and thus~\eqref{eq:fpe2b} is satisfied.
}
\end{IEEEproof}


\optv{arxiv}{

\section{Proof of Lemma  \ref{thm:totexp}\label{thm:totexp_proof}}

First, we show that whenever $\gamma^*=\phi[0,y,w]=\boldsymbol{0}$, the valuation functions are all 0 and we must have $\gamma^*=\phi[1,y,w]=\boldsymbol{0}$. According to FPE \ref{alg:Fixed-Point WY}, at the state $\left(x,r,y,w\right)$ we have
\begin{align*}
A=\frac{\delta}{N}U_{a}\left(x,r,y,w\right)
+\frac{\delta}{N}\left(N-w-1+z\right)U_{na}^{r}\left(x,0,y,w\right)+
\frac{\delta}{N}\left(w-z\right)U_{na}^{r}\left(x,1,y,w\right),
\end{align*}
where for both $\tilde{z}=0,1$,
\begin{align*}
U_{na}^{r}\left(x,\tilde{z},y,w\right)=\frac{\delta}{N}U_{a}\left(x,r,y,w\right)
+\frac{\delta}{N}\left(N-w-1+z\right)U_{na}^{r}\left(x,0,y,w\right)+
\frac{\delta}{N}(w-z)U_{na}^{r}\left(x,1,y,w\right).
\end{align*}
and since $\gamma^{*}=\phi[r,y,w]=\boldsymbol{0}$, we should have $U_{a}\left(x,r,y,w\right)=A$. Therefore, we can solve for  $U_{a}\left(x,r,y,w\right)$, $U_{na}^{r}\left(x,0,y,w\right)$, $U_{na}^{r}\left(x,1,y,w\right)$ and  $A$ in above equations. It is easy to see that the solution for all of these quantities is 0 and hence, $A=0$. Therefore, $U_{a}\left(x,r,y,w\right)=0$ and it is obvious that for $y<-2$, players strictly prefer to wait since the expected value of instantaneous reward is negative and they prefer to get $A$, which is 0. Further, for $y=-2$, players with $r=0$ or $r=1$ and $x=-1$ strictly prefer to wait. A player with $r=1$ and $x=1$ is indifferent between buying and not buying. The reason is that a player with $r=1$ that gets to play again, must have revealed $x=-1$. Therefore, if $x=1$, the player is at an off-equilibrium point and according to equation \eqref{eq:private}, she forms her true belief by canceling out what she has revealed and then augmenting the belief by her private information. In terms of $y$, this is translated into using $y-\tilde{x}+x$  to form the belief over $V$. For $y=-2$, a player with $r=1$ and $x=1$  uses $y-(-1)+1=0$ to form her belief over $V$. Thus, the expected value of her instantaneous reward is 0. It completes the proof of the third part of the theorem.

Now consider $\delta=1$.  Assume that $\gamma^*=\phi[r,y,w]$ is a solution of FPE \ref{alg:Fixed-Point WY}. According to $\gamma^*$, define player $n$'s terminating states to be the $(r,y,w)$ values for which player $n$ decides to either buy the product (and leave the game) or to not buy it ever after, i.e., playing $\gamma^*=\boldsymbol{0}$ when everyone else is (which means that the player practically leaves the game). At each state of the game, $\gamma^*$ imposes a probability distribution on the future terminating states of the game that are reached by not buying decision of the acting player.  So at each state, a player compares the expected value of $V$ with the expected valuation she can get in future by not buying, which is an average between the value of $V$ at the terminating states where she buys the product, and zero, corresponding to the terminating states she decides not to buy the product ever.
\begin{lem}\label{lem:indif_all_buy}
	Assume that according to $\gamma^*=\phi[r,y,w]$, the acting player will buy the product in all of the future terminating states (the ones with positive probability to happen if the acting player decided not to buy). Then, the player is indifferent between buying and not buying the product for $\delta=1$. Otherwise, she strictly prefers to wait for $\delta=1$.
\end{lem}
	\begin{IEEEproof}
According to FPE \ref{alg:Fixed-Point WY}, the acting player compares the expected value of $v$ with the average of expected values of $v$  in future terminating states with positive probability to happen if the acting player decided not to buy. More formally, if we denote the current state by $s$ and the future terminating states that happen with positive probability if the player does not buy with  $s_1,...,s_k$, then we have
\begin{align}
\gamma^*(x)=\arg\max\left\{ \sum_{j=1}^k\mathbb{E}[v|s_j]p(s_j|s),\mathbb{E}[v|s]\right\}
\label{deci}
\end{align}

By the law of total expectation, we know that the above terms are always equal to each other, no matter what the states $s_1,...s_k$ are and with what probability they happen. The only requirement is that at all of the states $s_1,...s_k$ the player decides to buy the product. Therefore, if a player finds herself in a state $s$ that could lead to terminating states $s_1,...s_k$ (by not buying), in all of which she will decide to buy the product (according to $\gamma^*$), she is in fact indifferent between buying and not buying at the current state for $\delta=1$.

Next, assume that at one of the terminating states $s_1,...s_k$, let's say $s_j$, the player strictly prefer not to buy the product, she will receive zero valuation at $s_j$ and therefore, the expected value of $v$ should have been negative at $s_j$. Hence, by substituting zero instead of $\mathbb{E}[v|s_j]$ in equation \eqref{deci}, we get a term that is greater than $\mathbb{E}[v|s]$. It means that the expected valuation of not buying is greater than the expected value of $v$ at the current state $s$ which implies that the player strictly prefers not to buy the product. The same argument holds if in more than one future terminating states the player strictly prefers not to buy.
\end{IEEEproof}

\begin{lem}\label{lem:dif_notall_buy}
	 Assume that according to $\gamma^*$, we know that for state $s$, there exists at least one future terminating state $s_j$ (with positive probability to happen if the acting player does not buy the product), at which the acting player strictly prefers not to buy the product, then she strictly prefers to wait at the state $s$ for large enough $\delta\leq1$.
\end{lem}
\begin{IEEEproof}
	According to the proof of the second part of Lemma \ref{lem:indif_all_buy}, for $\delta=1$, the acting player strictly prefers to wait. Hence, there exists large enough $\delta<1$ for which the acting player still strictly prefers to wait.
\end{IEEEproof}


Next, we prove the first two parts of the theorem.
We first characterize the equilibrium strategies for $w=N$. It is evident what the equilibrium is for $w=N$, because all of the states are absorbing and players act based on the expected instantaneous reward. For any value of $\delta$, we have $\gamma^*=\phi[1,y,N]=\boldsymbol{0}$ for\footnote{Note that we can also have $\gamma^*=\phi[1,y,N]=\boldsymbol{I}$ for $y= -2$ due to the tie for the player with $x=1$.} $y\leq -2$ , $\gamma^*=\phi[1,y,N]=\boldsymbol{I}$ for $y=-1$ and $\gamma^*=\phi[1,y,N]=\boldsymbol{1}$ for $y\geq 0$. We also know that $\gamma^*=\phi[r,y,w]=\boldsymbol{0}$ for $y\leq -2$. In order to prove the theorem, we investigate the terminating states for all states of the game with $y\geq -1$. Since $\gamma^*=\boldsymbol{0}$ is played at $y=-2$, all states with $y=-2$ are absorbing. Hence, no state with $y<-2$ is reachable from states with $y \geq -1$. Therefore, all of the terminating states have $y\geq -2$.  On the other hand, if $\gamma^*=\boldsymbol{I}$ at the current state, the acting player with $x=1$ can reach her terminating states only when she has $r=1$. A player with $x=1$ and $r=1$ is indifferent between buying and not buying at $y=-2$ (she is on an off-equilibrium path) and prefers to buy at all states with $y\geq -1$. Therefore, at all of the terminating states the player prefers to buy and according to Lemma \ref{lem:indif_all_buy}, she is indifferent between buying and not buying at the current state. Furthermore, if $\gamma^*=\boldsymbol{1}$ at the current state, the acting player should be indifferent between buying and not buying according to Lemma \ref{lem:indif_all_buy} (the player should either be indifferent or strictly prefer to wait. The latter is impossible due to the strategy $\gamma^*=\boldsymbol{1}$).
It means that a player with $x=1$ is indifferent between buying and not buying for all states with $y\geq-1$ and all $w$.   It implies that for $\delta<1$ a player with $x=1$ strictly prefers to buy if her instantaneous reward is positive, i.e., $y\geq0$ and is indifferent if her instantaneous reward is 0, i.e., $y=-1$.

Next consider the players with $x=-1$. If at all of the terminating states of a player with $x=-1$  we have $y\geq 1$ or $y=0$ and $w=N$ (the states in which she prefers to buy the product), then this player should be indifferent between buying and not buying at the current state. Assume that we have  $\gamma^*=\phi[r,y,w]=\boldsymbol{I}$ for every $r,y,w$ (this strategy profile shows us the biggest set of approachable terminating states from each state $s$, although it may not be the solution). It is evident that for $y+w\geq N$ and all $r$, all of the terminating states that are approachable have $y\geq1$ or $y=0$ and $w=N$ (at each state $(r,y,w)$, the player can move to $y-1$ and $w+1$ by playing $\gamma^*=\boldsymbol{I}$). Hence, for $\delta=1$, a player with $x=-1$ is indifferent between buying and not buying  for $y+w\geq N$ and all $r$. It implies that for $\delta<1$, a player with $x=-1$ strictly prefers to buy for $y+w\geq N$ if her instantaneous reward is positive, i.e., $y\geq2$ or $y=1$ and $r=1$, and is indifferent if her instantaneous reward is 0, i.e., $y=1$ and $r=0$ or $y=0$ and $r=1$.

}



\section{Proof of Theorem \ref{thm:existence}}\label{thm:existence_proof}
	We prove this theorem by referring to Lemma \ref{thm:totexp}. The strategy profile proposed for $y\leq -2$ is an evident solution of FPE \ref{alg:Fixed-Point WY} due to the fact that both types of players with $r=0$ prefer not to buy and hence they play $\gamma^*=\boldsymbol{0}$. This implies that both types of players with $r=1$ will also play $\gamma^*=\boldsymbol{0}$.  For $y\geq -1$, a player with $x=1$ is either indifferent or prefers to buy for all $\delta\leq1$ and therefore, she can decide to buy for $-1\leq y\leq 1$. Furthermore, a player with $x=-1$ and $r=0$ always prefers to wait or is indifferent for $-1\leq y \leq 1$ (her expected instantaneous reward is either negative or zero) and so she can decide to wait for $-1\leq y\leq 1$. The same argument holds for a player with $x=-1$ and $r=1$ for $-1 \leq y \leq 0$. For $y=1$, a player with $x=-1$ and $r=1$ has positive expected instantaneous reward and so whether she prefers to wait or to buy depends on $\delta$. Since we know that the action of a player with $x=1$ and $r=1$ at $y=1$ is buying, the strategy at $y=1$ and $r=1$ should be either of $\boldsymbol{1}$ or $\boldsymbol{I}$.
Notice that this strategy does not affect the decision of players at other states
(it can not be reached from states with $y>1$ and the solution for $y<1$ does not depend on what is played at $y=1$, as we just proved). Hence, it can be determined independently based on FPE~\ref{alg:Fixed-Point WY}.
We next have to prove that the strategy $\gamma^*=\phi[r,y,w]=\boldsymbol{1}$ is a solution for $y\geq 2$ and all $w$ and $r$. According to Lemma \ref{thm:totexp}, if the strategy profile is $\gamma^*=\phi[r,y,w]=\boldsymbol{1}$ for some state of the game $s=(r,y,w)$ and we have $\gamma^*=\phi[r,y,w']=\boldsymbol{1}$ for all $w'>w$ (which is the case in the suggested strategy profile in this theorem), then in all of the terminating states (see the proof of Lemma \ref{thm:totexp}\optv{2col}{ in \cite{BiHeAn19arxiv}}) that are reachable from $s$,  the acting player buys the product. Hence, the player is either indifferent ($\delta=1$) or strictly prefers to buy ($\delta<1$) and it completes the proof of the strategy $\gamma^*=\phi[r,y,w]=\boldsymbol{1}$ being a solution for $y\geq 2$, all $w$, $r$ and all $\delta\leq 1$.

\optv{arxiv}{
	\section{Proof of Theorem \ref{thm:nocas_st}\label{thm:nocas_st_proof}}
	According to Lemma \ref{thm:totexp}, the solution is evident for $y\leq-2$, for both $\delta=1$ and $\delta<1$. We also know the solution for $w=N$ and all $\delta$ according to the proof of Lemma \ref{thm:totexp}. For $w=N$, which implies that $r=1$, we must have $\gamma^*=\phi[r,y,w]=\boldsymbol{1}$ for $y\geq1$. Further, since for $y=0,\-1$, the expected instantaneous reward for players with $x=1$ and $x=-1$ is positive and non-positive, respectively, we can have $\gamma^*=\phi[r,y,w]=\boldsymbol{I}$ as the solution. This proves the first, third and fourth part of $\delta=1$ case and first and fifth part ($w=N$) of $\delta<1$ case.
	
	Next, consider the second part of $\delta=1$ case. According to  Lemma \ref{thm:totexp}, a player with $x=1$ and all $r$ and $w<N$ is indifferent between buying and not buying for $y\geq-1$ and hence, she can decide to buy. On the other hand, a player with $x=-1$ is indifferent for $y+w\geq N$ and so she can decide not to buy for these states. If the proposed strategy is the solution of FPE \ref{alg:Fixed-Point WY}, then from the states with $y+w< N$, a player with $x=-1$ can reach the terminating states with negative $y$ (-1 or -2), in which she strictly prefers not to buy (this is evident by tracing the states that can be reached by going from $y,w$ to $y-1,w+1$ by each revelation through the strategy $\gamma^*=\phi[r,y,w]=\boldsymbol{I}$). Therefore, for $\delta=1$ a player with $x=-1$ strictly prefers to wait for $y+w<N$ and so strategy $\gamma^*=\phi[r,y,w]=\boldsymbol{I}$ can be a solution for $y\geq-1$ and $w<N$. This completes the proof of the $\delta=1$ case.
	
	Now consider $\delta<1$. With the same arguments as in the $\delta=1$ case, since a player with $x=1$ is indifferent for $y\geq -1$ and $\delta=1$, she strictly prefers to buy if $\delta<1$ (she is losing valuation by waiting and is not gaining anything). In the same manner, a player with $x=-1$ strictly prefers to buy for $y+w\geq N$ and $w<N$. Therefore, the strategy $\gamma^*=\phi[r,y,w]=\boldsymbol{1}$ could be a solution for $y+w\geq N$ and $w<N$. For the rest of the states which are $y\geq -1$, $w<N$ and $y+w<N$, a player with $x=-1$ strictly prefers to wait for $\delta=1$ and hence, there exists large enough $\delta<1$ such that this player still prefers to wait and therefore, $\gamma^*=\phi[r,y,w]=\boldsymbol{I}$ can be a solution for $y\geq -1$, $w<N$ and $y+w<N$ when $\delta<1$ is large enough. Further, for $w=N-1$, $y=1$ and $r=1$, both types prefer buying over waiting therefore $\gamma^*=\phi[r,y,w]=\boldsymbol{1}$ can be a solution. This completes the proof of this theorem.
	
}


\optv{arxiv}{

\section{Proof of Theorem \ref{thm:wy:behavior}}\label{thm:wy:behavior_proof}
	
%
We first prove the fourth part of the theorem.  For $y \leq -2$, the instantaneous reward is negative for $r=0$ and both $x=1$ and $x=-1$. On the other hand, according to the proof of Lemma \ref{thm:totexp}, the value functions are 0 when $\gamma^*=\phi(r,y,w)=\boldsymbol{0}$. Therefore, the equilibrium strategy is not buying for both values of $x$ and so $\gamma^*=\phi(r,y,w)=\boldsymbol{0}$ is the only solution for $y \leq -2$. For $y \geq 0$, the instantaneous reward is positive for $x=1$ and therefore, $\gamma^*=\phi(0,y,w)=\boldsymbol{0}$ (so that the value functions are all 0) can not be an equilibrium strategy.

The fifth part is obvious due to the fact that at $y=0$ the reward is negative for $x=-1$ and it is positive for $x=1$. Hence, neither  $\gamma^*=\phi[0,0,w]=\boldsymbol{0}$ nor $\gamma^*=\phi[0,0,w]=\boldsymbol{1}$ can be solution of FPE \ref{alg:Fixed-Point WY}. Therefore, if a solution exists, which we know it does, we must have $\gamma^*=\phi(0,0,w)=\boldsymbol{I}$.

Now we prove the sixth part. If for some equilibrium strategy and some $w$ and $y$, $\gamma^*=\phi[0,y,w]=\boldsymbol{I}$ or $\gamma^*=\phi[0,y,w]=\boldsymbol{1}$, we can not have $\gamma^*=\phi[0,y,w]=\boldsymbol{0}$ for $w'\neq w$ and $y \neq -1$. The reason is that if for $w' \neq w$, we have $\gamma^*=\phi[r,y,w']=\boldsymbol{0}$, the valuation function $U_a(x,0,y,w)=0$ as proved in the proof of Lemma \ref{thm:totexp}. On the other hand, since $\gamma^*=\phi[0,y,w]=\boldsymbol{I}$ or $\gamma^*=\phi[0,y,w]=\boldsymbol{1}$, we know that $\frac{q^{y+1}-1}{q^{y+1}+1}> 0$ for $y \neq -1$. Hence the instantaneous reward for a player with $x=1$ at $r=0,y,w'$ is positive and therefore, $\gamma^*=\phi[0,y,w']=\boldsymbol{0}$ can not be an equilibrium strategy. Hence, $\gamma^*=\phi[0,y,w]=\boldsymbol{I}$ or $\gamma^*=\phi[0,y,w]=\boldsymbol{1}$ can not happen with $\gamma^*=\phi[0,y,w']=\boldsymbol{0}$ for the same $y$. Therefore, for a fixed $y$, we either have $\gamma^*=\phi[0,y,w]=\boldsymbol{0}$  for all $w$ or a combination of   $\gamma^*=\phi[0,y,w]=\boldsymbol{I}$ or $\gamma^*=\phi[0,y,w]=\boldsymbol{1}$  for different  $w$.

The seventh part is evident by using the fourth part and Lemma \ref{thm:totexp}. As we saw in  Lemma \ref{thm:totexp}, a player with $x=1$ is indifferent between buying and waiting for $y\geq -1$, which includes $y=-1$. It means that she can always decide to buy for $y\geq -1$. On the other hand, a player with $x=-1$ has negative instantaneous reward and she should not buy at $y=-1$. Hence, since the expected reward of the player with $x=1$ is 0, both  $\gamma^*=\phi[0,-1,w]=\boldsymbol{I}$ and $\gamma^*=\phi[0,-1,w]=\boldsymbol{0}$ can be solutions for all $w$.

We can prove the eighth part in a similar way. A player with $x=1$ and $r=1$ is always indifferent between buying and not buying at $y=-2$ since the instantaneous reward is 0 (she is on an off-equilibrium path). On the other hand, a player with $x=-1$ and $r=1$ prefers to wait at $y=-2$ since her instantaneous reward is negative and hence, both  $\gamma^*=\phi[1,-2,w]=\boldsymbol{I}$ and $\gamma^*=\phi[1,-2,w]=\boldsymbol{0}$ are the solutions.

The third part is a direct consequence of fourth and seventh parts.

In order to prove the first part, it is sufficient to show that if $\gamma^*=\phi[r,y,w]=\boldsymbol{I}$ and the solution is a threshold policy wrt $w$ for $y'<y$, then  $\gamma^*=\phi[r,y,w']=\boldsymbol{I}$ is a solution for $w'<w$ (Note that it might not be the only case, and we are arguing about existence. So if a solution is not of this type, we can construct a solution of this type, as we explain later on).

Assume that for the state $s=(x,r,y,w)$, we have $\gamma^*=\phi[r,y,w]=\boldsymbol{I}$. It means that the instantaneous reward for $x=-1$ at $y$ has been no more than the expected valuation of not buying, which is the average of rewards, at those terminating states that the player will buy, and $0$'s, for those in which the player decides not to buy the product (see the proof of Lemma \ref{thm:totexp}). The  more likely the final states with not buying decision are, the larger the difference between the instantaneous reward and expected valuation of not buying is. So for two different states with the same instantaneous reward, i.e., the same $y$, we can compare their terminating states to get a sense of how the player decides in these two states. Consider $s'=(r,y,w')$ for $w'<w$. Since the solution is a threshold policy wrt $w$ for $y'<y$, it is clear that for each terminating state $s_j$ for the state $s$ that the player decides not to buy the product, there is a corresponding state $s_j'$ for the state $s'$ which is at least as likely to happen as $s_j$ (there are more players that can reveal their private signal and change the state). At $s_j'$, the player has the opportunity to decide not to buy the product, or decide later on if it is beneficial for her (which implies that $s_j'$ may not be a terminating state for $s'$). In both cases, the valuation of not buying at $s'$ is at least as good as $s$ and hence, if the valuation of not buying at $s$ is not less than the instantaneous reward, it has to be true for $s'$ too. Therefore, if $\gamma^*=\phi[r,y,w]=\boldsymbol{I}$, we can have $\gamma^*=\phi[r,y,w']=\boldsymbol{I}$ for $w'<w$. If a solution is not of this type, we can construct such strategy as follows. According to the other parts of this theorem, we know the solution can be  $\gamma^*=\phi[r,y,w]=\boldsymbol{0}$ for $y\leq -2$, all $r$ and $w$, and also $\gamma^*=\phi[r,y,w]=\boldsymbol{I}$ for $r=0$, $-1 \leq y\leq 0$ and all $w$ and for $r=1$, $-1 \leq y\leq 0$ and all $w$. So whatever else is the solution, we can change it to the mentioned strategy profile. Next, we start at $y=1$ and we know that the solution is a threshold policy for $y'<y$. Starting at $w=N$ and going back step by step for both $r=0$ and $r=1$, we can change all the solutions $\gamma^*=\phi[r,y,w']=\boldsymbol{1}$ to $\gamma^*=\phi[r,y,w']=\boldsymbol{I}$ for all $w'<w$ such that the solution is $\gamma^*=\phi[r,y,w]=\boldsymbol{I}$.  In this way we construct a strategy profile that is a threshold policy wrt $w$ and is  solution to FPE \ref{alg:Fixed-Point WY}.

Now we restrict our attention to the equilibrium strategies that are threshold policies wrt $w$ and prove that whenever $\gamma^*=\phi[0,y,w]=\boldsymbol{1}$, then we must have $\gamma^*=\phi[0,y',w]=\boldsymbol{1}$ for all $y'>y$ and whenever $\gamma^*=\phi[0,y,w]=\boldsymbol{I}$ then we must have $\gamma^*=\phi[0,y',w]\neq \boldsymbol{0}$ for all $y'>y$.

Similar to the arguments in the proof of sixth part of this theorem, whenever $\gamma^*=\phi[0,y,w]=\boldsymbol{I}$, the instantaneous reward is positive for $x=1$ and if we have  $\gamma^*=\phi[0,y',w]= \boldsymbol{0}$, the valuation will be 0 while the instantaneous reward for $y'$ is greater than $y$ and so it is positive. Hence, we can not have  $\gamma^*=\phi[0,y',w]= \boldsymbol{0}$ as a solution.

In order to prove that whenever $\gamma^*=\phi[0,y,w]=\boldsymbol{1}$, then we must have $\gamma^*=\phi[0,y',w]=\boldsymbol{1}$ for all $y'>y$, we assume this is not true and hence, we have a case where $\gamma^*=\phi[0,y,w]=\boldsymbol{1}$ and $\gamma^*=\phi[0,y+1,w]=\boldsymbol{I}$. In this case the player with $x=-1$  at $y+1$ is choosing not buying over buying which means that
\begin{multline}
\frac{q^{y}-1}{q^{y}+1} \leq \frac{\delta}{N}U_{a}\left(-1,1,y,w+1\right)
 +\frac{\delta}{N}\left(N-w-1\right)U_{na}^{1}\left(-1,0,y,w+1\right)+
  \frac{\delta}{N}wU_{na}^{1}\left(-1,1,y,w+1\right),
  \label{Fpi}
 \end{multline}
since $\gamma^*=\phi[0,y,w]=\boldsymbol{1}$, we know that $\gamma^*=\phi[0,y,w+1]=\boldsymbol{1}$ and hence, according to the proof of Theorem \ref{thm:existence}, $U_{a}\left(-1,1,y,w+1\right)=\frac{q^{y}-1}{q^{y}+1}$, $U_{na}^{1}\left(-1,0,y,w+1\right)\leq \frac{q^{y}-1}{q^{y}+1}$ and $U_{na}^{1}\left(-1,1,y,w+1\right) \leq \frac{q^{y}-1}{q^{y}+1}$ which means that for $\delta <1$, $\frac{q^{y}-1}{q^{y}+1} <\frac{q^{y}-1}{q^{y}+1} $ and it is a contradiction.

Next consider $r=1$. We first prove a relation between a player's decision in states $s=(x,0,y,w)$ and $s'=(x,1,y-1,w+1)$.
Assume that $\gamma^*=\phi[0,y,w]=\boldsymbol{I}$. It means that
\begin{multline}
\frac{q^{y-1}-1}{q^{y-1}+1} \leq \frac{\delta}{N}U_{a}\left(-1,1,y-1,w+1\right)
+\frac{\delta}{N}\left(N-w-1\right)U_{na}^{1}\left(-1,0,y-1,w+1\right)+
\frac{\delta}{N}wU_{na}^{1}\left(-1,1,y-1,w+1\right),
\label{Fpi2}
\end{multline}
on the other hand, if we write the fixed point equation for $x=-1,\ r=1,\ y-1, \ w+1$, we have
\begin{multline}
\gamma^*(-1)=\arg\max \left\{ \frac{\delta}{N}U_{a}\left(-1,1,y-1,w+1\right)
+\frac{\delta}{N}\left(N-w-1\right)U_{na}^{1}\left(-1,0,y-1,w+1\right) \right. \\ \left. +
\frac{\delta}{N}wU_{na}^{1}\left(-1,1,y-1,w+1 \right),\frac{q^{y-1}-1}{q^{y-1}+1}\right\}.
\label{Fp1}
\end{multline}
According to \eqref{Fpi2}, we can say that the solution of the above fixed point equation can be not buy. Hence, whenever $\gamma^*=\phi[0,y,w]=\boldsymbol{I}$, we can have $\gamma^*=\phi[1,y-1,w+1]=\boldsymbol{I}$.  Also, according to Lemma  \ref{thm:totexp}, whenever $\gamma^*=\phi[0,y,w]=\boldsymbol{1}$ and  $\gamma^*=\phi[0,y,w']=\boldsymbol{1}$ for $w'>w$, we must have $\gamma^*=\phi[1,y,w]=\boldsymbol{1}$. This all means that whenever we have a solution that is a threshold policy wrt $y$ for $r=0$, the solution can also be a threshold policy wrt $y$ for $r=1$.


}


\section{Proof of Theorem \ref{thm:Cascades}}\label{thm:Cascades_proof}
If $Y_{t}$ remains constant with probability one for all $t'>t$,
then it is an informational cascade by definition since $y_{t}$
sums all the revealed private information. Hence, the absorbing states
of $\bar{Y}_{i}$ are informational cascades. We have shown that some
$\bar{Y}_{i}=y_{L}\geq Y_{\min}$ and $\bar{Y}_{i}=y_{R}\leq Y_{\max}$
are absorbing. The values of both $y_{L},y_{R}$ are independent of
$N$. The transition probabilities of $\bar{Y}_{i}$ are $\frac{p+\left(1-p\right)q^{y}}{q^{y}+1}$
for moving right and $\frac{1-p+pq^{y}}{q^{y}+1}$ for moving left,
so they are also independent of $N$. We conclude that the distribution
(specifically, expectation and variance) of the absorption time is
independent of $N$. Hence, for large enough $N$, the probability
that the absorption time is larger than $M_{N}$ vanishes to zero.
This absorption time is counted in the number of revealings $i$.
We conclude that the probability that a cascade occurs before $M_{N}$
revealings occur approaches 1 as $N\rightarrow\infty$.

Now assume that $\phi\left[r,y,w\right]=\boldsymbol{1}$ implies
that $\phi\left[r,y,\hat{w}\right]=\boldsymbol{1}$ for $\hat{w}>w$.
Denote the number of turns up to turn $t=M_{N}$ where the acting player
$n_{t}$ has $r^{n_{t}}=1$ or $b^{n_{t}}=1$ by $\overline{R}\left(M_{N}\right)$,
which is stochastically dominated by a binomial distributed variable
with $p=\frac{M_{N}}{N}$ and $M_{N}$ trials since $\frac{w_{t}}{N}\leq\frac{M_{N}}{N}$.

Hence, for all $N>0$,
\begin{equation}
\Pr\left(\overline{R}\left(M_{N}\right)\geq 1\right)\leq 1-\left(1-\frac{M_{N}}{N}\right)^{M_{N}}.\label{eq:54}
\end{equation}
Since by assumption $\frac{M_{N}^{2}}{N}\rightarrow0$ as $
N\rightarrow\infty$, then $1-\left(1-\frac{M_{N}}{N}\right)^{M_{N}}\rightarrow1-e^{-\frac{M_{N}^{2}}{N}}\rightarrow0$. We conclude that with high probability, at least $M_{N}-1$
of the first turns are of players with $z^{n}=0$, so we either have at least  $M_{N}-1$ revealings during these turns, or that the acting player at $t<M_{N}-1$ did not reveal her private information. But then:
\begin{itemize}
\item If she waited, then $w_{t+1}=w_{t}$ and $y_{t+1}=y_{t}$. The next
player with $z^{n}=0$ will also wait since she uses the same strategy
$\gamma^{*}=\phi\left[r,y,w\right]$.
\item If she bought, then $w_{t+1}=w_{t}+1$ and $y_{t+1}=y_{t}$. The next
acting player with $z^{n}=0$ will also buy for $x^{n}=-1,1$ (and
not reveal) since $w_{t+1}>w_{t}$.
\end{itemize}
The same occurs to all subsequent players with $z^{n}=0$, and by definition to players with $z^{n}=1$, so a cascade occurred. 


\bibliographystyle{IEEEtran}
\input{tac20_bha_r1.bbl}



\optv{2col}{
\vspace*{-0.5cm}

\input{ilai_bio}

\vspace*{-0.5cm}
\input{nas_bio_ieee}

\vspace*{-0.5cm}

\input{achilleas_bio}

}

\end{document}

%% file: tac20_bha_r1.bbl

%% file: ilai_bio.tex
\begin{IEEEbiography}[{\includegraphics[width=1in,height=1.25in,clip,keepaspectratio]{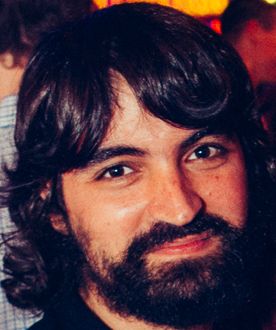}}]{Ilai Bistritz}
received his B.Sc. in 2012 (magna cum laude) and his M.Sc in 2016 (summa cum laude), both in Electrical Engineering and both from Tel-Aviv University, Israel. He is currently working toward a Ph.D. in Electrical Engineering at Stanford University, Stanford, CA, USA. His main research interests are game theory, decision-making over networks and multiagent learning.

\end{IEEEbiography}
\vfill

%% file: nas_bio_ieee.tex
\begin{IEEEbiography}[{\includegraphics[width=1in,height=1.25in,clip,keepaspectratio]{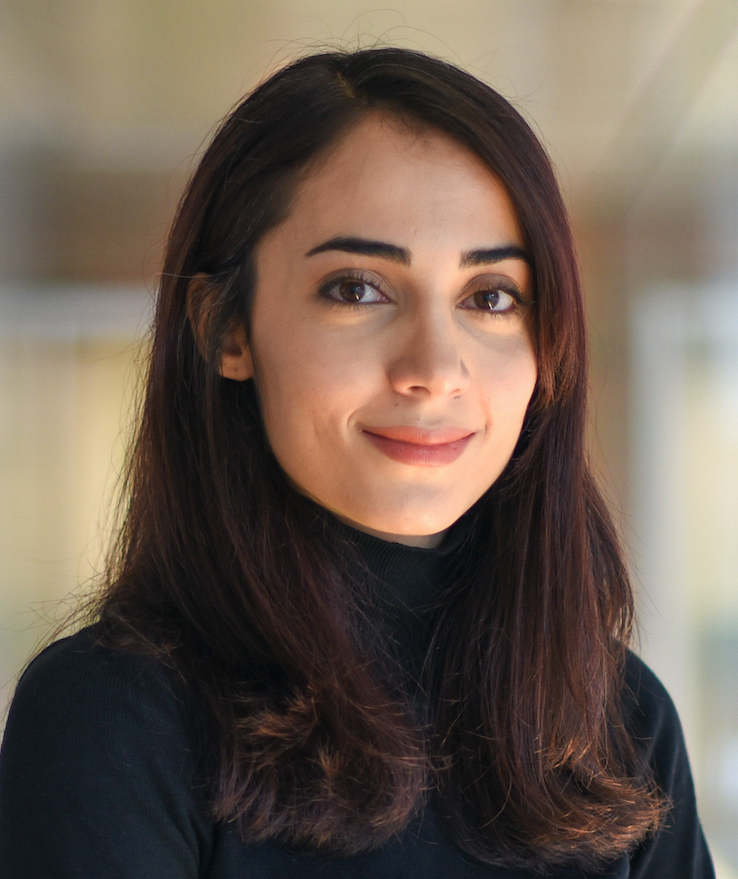}}]{Nasimeh Heydaribeni}
	received her B.S. and M.S. degrees in Electrical Engineering at Sharif University of Technology, Tehran, Iran, in 2015 and 2017, respectively. She is currently pursuing her Ph.D. degree in Electrical Engineering and Computer Science at University of Michigan, Ann Arbor. Her research interests are game theory and its applications in networked systems with emphasis on mechanism design, dynamic games with asymmetric information and information design.
\end{IEEEbiography}
\vfill

%% file: achilleas_bio.tex
\begin{IEEEbiography}[{\includegraphics[width=1in,height=1.25in,clip,keepaspectratio]{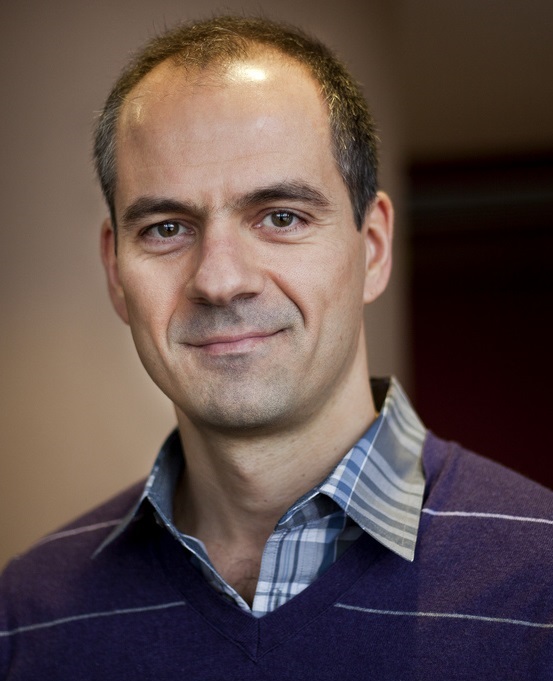}}]{Achilleas Anastasopoulos}
(S'97-M'99-SM'13) was born in Athens, Greece in 1971. He received the Diploma
in Electrical Engineering from the National Technical University of
Athens, Greece in 1993, and the M.S.\ and Ph.D.\ degrees in Electrical
Engineering from University of Southern California in 1994 and 1999,
respectively. He is currently an Associate Professor at the University of
Michigan, Ann Arbor, Department of Electrical Engineering and Computer
Science.

His research interests lie in the general area of communication and information theory,
with emphasis in channel coding and multi-user channels;
control theory with emphasis in decentralized stochastic control and its
connections to communications and information theoretic problems;
analysis of dynamic games and mechanism design for resource allocation on networked systems.

He is the co-author
of the book \emph{Iterative Detection: Adaptivity, Complexity Reduction,
and Applications,} (Reading, MA: Kluwer Academic, 2001).

Dr.\ Anastasopoulos is the recipient of the ``Myronis Fellowship'' in
1996 from the Graduate School at the University of Southern California,
the NSF CAREER Award in 2004,
and was a co-author for the paper that received the best student paper award in ISIT 2009.
He served as a technical program committee member for 
ICC 2003, 2015--2018; 
Globecom 2004, 2012; 
VTC 2007, 2014, 2015; 
ISIT 2015,
SPAWC 2018,
and is currently serving as the TPC co-Chair for the Communication Theory Symposium, ICC'21.
He was an associate editor for the IEEE Transactions on Communications in 2003--2008.
%

%
\end{IEEEbiography}
\vfill